\pdfoutput=1
\documentclass[fleqn,usenatbib]{mnras}

\usepackage{newtxtext,newtxmath}

\usepackage[T1]{fontenc}
\usepackage{ae,aecompl}

\usepackage{graphicx}	
\usepackage{amsmath}	
\usepackage{amssymb}	
\usepackage{adjustbox}
\usepackage{pdflscape}
\usepackage{hyperref}

 \newcommand{\mgii}{\ion{Mg}{II} } 
 \newcommand{\ang}{\AA $ $ }

\title[The comoving mass density of MgII from $z\sim2$ to $5.5$]{The comoving mass density of MgII from $z\sim2$ to $5.5$}

\author[A. Codoreanu et al]{
Alex Codoreanu,$^{1, 2}$\thanks{E-mail: acodoreanu@swin.edu.au}
Emma V. Ryan-Weber,$^{1, 2}$
Neil H.M. Crighton,$^{1}$
George Becker,$^{3}$ 
\newauthor
Max Pettini,$^{4}$
Piero Madau,$^{5}$
and Bram Venemans$^{6}$
\\
$^{1}$Centre for Astrophysics and Supercomputing, Swinburne University of Technology, Hawthorn, Victoria 3122, Australia\\
$^{2}$ARC Centre of Excellence for All-sky Astrophysics (CAASTRO)\\
$^{3}$Department of Physics Astronomy, University of California, Riverside, 900 University Avenue, Riverside, CA 92521, USA\\
$^{4}$Institute of Astronomy, Madingley Road, Cambridge, CB3 0HA, UK\\
$^{5}$Department of Astronomy $\&$ Astrophysics, University of California, 1156 High Street, Santa Cruz, CA 95064, US \\
$^{6}$Max-Planck Institute for Astronomy, K$\ddot{o}$nigstuhl 17, D-69117 Heidelberg, Germany \\
}

\date{Accepted XXX. Received YYY; in original form ZZZ}

\pubyear{2017}

\begin{document}
\label{firstpage}
\pagerange{\pageref{firstpage}--\pageref{lastpage}}
\maketitle

\begin{abstract}

We present the results of a survey for intervening \mgii absorbers in the redshift range $z$ $\simeq$2-6 in the foreground of four high redshift quasar spectra, 5.79$\le$$z_{\textrm{em}}\le$6.133, obtained with the ESO Very Large Telescope X-shooter. We visually identify 52 \mgii absorption systems and perform a systematic completeness and false positive analysis. We find 24 absorbers at $\ge$5$\sigma$ significance in the equivalent width range 0.117$\le$$W_{2796}$$\le$3.655\AA{} with the highest redshift absorber at $z$=4.89031$\pm$4$\times$10$^{-5}$. For weak ($W_{2796}$<0.3\AA) systems, we measure an incidence rate $dN/dz$=1.35$\pm$0.58 at <$z$>=2.34 and find that it almost doubles to $dN/dz$=2.58$\pm$0.67 by <$z$>=4.81. The number of weak absorbers exceeds the number expected from an exponential fit to stronger systems ($W_{2796}$>0.3\AA). We find that there must be significant evolution in the absorption halo properties of \mgii absorbers with $W_{2796}$>0.1\AA{} by <$z$>=4.77 and/or that they are associated with galaxies with luminosities beyond the limits of the current luminosity function at z $\sim$5. We find that the incidence rate of strong \mgii absorbers ($W_{2796}$>1.0\AA) can be explained if they are associated with galaxies with $L$$\ge$$0.29L_{\ast}$ and/or their covering fraction increases. If they continue to only be associated with galaxies with $L$$\ge$$0.50$$L_{\ast}$ then their physical cross section ($\sigma_{phys}$) increases from 0.015 Mpc$^2$ at z=2.3 to 0.041 Mpc$^2$ at <$z$>=4.77. We measure $\Omega_{\ion{Mg}{II}}$=2.1$^{+6.3}_{-0.6}$$\times$$10^{-8}$, 1.9$^{+2.9}_{-0.2} \times10^{-8}$, 3.9$^{+7.1}_{-2.4}\times10^{-7}$ at <$z$>=2.48, 3.41, 4.77, respectively. At <$z$>=4.77, $\Omega_{\ion{Mg}{II}}$ exceeds the value expected from $\Omega_{\ion{H}{I}}$ estimated from the global metallicity of DLAs at $z$ $\simeq$4.85 by a factor of $\sim$44 suggesting that either \mgii absorbers trace both ionised and neutral gas and/or are more metal rich than the average DLA at this redshift.
 
\end{abstract}


\section{Introduction}
Quasar (QSO) absorption systems provide a luminosity independent view of the universe and can trace the enrichment and ionisation state of the inter-galactic medium (IGM) and circum-galactic medium (CGM) (see \citealt{BECKER2015} for a {recent} review). The {$\ion{Mg}{II}\lambda\lambda$2796.3542 2803.5314 doublet ($\ion{Mg}{II}\lambda\lambda$2796 2803)} has long been used as a tracer of enriched and ionised cool gas surrounding galaxies (\citealt{CHURCHILL1999, CHURCHILL2013, WEINER2009, CHEN2010B, CHEN2010A, LOVEGROVESIMCOE2011, KACPRZAK2011, KORNEI2012, MATEJECK2012} and others) and the strength of the $\lambda$2796 feature, $W_{2796}$, has been used to associate absorption systems with {the properties of} star-forming galaxies \citep{MENARDCHELOUCH2009, MENARD2011, MATEJECK2012}, galaxy groups \citep{GAUTHIER2013}, $L_{*}$ galaxies \citep{BERGERONBOISSE1991, STEIDELDICKINSONPERSSON1994, MENARDCHELOUCH2009} as well as low-surface brightness galaxies \citep{CHURCHILL1999}. 

The Sloan Digital Sky Survey (SDSS) provides a large collection of QSO spectra which has been used to {investigate intervening \mgii absorbers.} Given the wavelength coverage {($\sim $0.38-0.92} $\mu$m), resolution ($\sim$150 kms$^{-1}$) and typical signal-to-noise (S/N) of SDSS {spectra,} such investigations are complete for only strong \mgii systems ($W_{2796} \ge 1$\AA)$ $ in the redshift window 0.4$\le$z$\le$2.3. Their incidence rate, $dN/dz$, can be well described {by a single power-law function}, $dN/dz=N^{*}(1+z)^{\beta}$ \citep{NESTOR2005, PROCHETER2006, PROCHETER2006B}. {These strong \mgii systems are associated with the haloes of star-forming galaxies \citep{MARTIN2009, MENARD2011, RUBIN2014, ZHUMENARD2013} and \citet{SEYFFERT2013} estimate their physical cross section ($\sigma_{phys}$) by connecting their incidence rate with that of B-band-selected galaxies. They find that their $\sigma_{phys}$ increases from 0.005 to 0.015 $Mpc^2$ {from z=0.4 to z=2.3}, which corresponds to a projected radius of $\sim$40 to 70 kpc. This is in agreement with non-SDSS studies on the connection between \mgii absorbers and galaxy halo gas \citep{BORDOLOI2011, NIELSEN2013a, NIELSEN2013b}. {This increase in cross section from redshift 0.4 to 2.3 is then driven by the strong outflows from star-forming galaxies whose star formation rate density increases and peaks around redshift $\sim$2 (see \citealt{MADAU2014} for a recent review)}.  }

{While SDSS studies are complete for only $\textit{strong}$ systems, they can identify \mgii absorbers down to $W_{2796}\sim$0.3\AA$ $. In order to push the detection limit below 0.3\AA, \citet{NESTOR2006} follow-up 381 SDSS QSOs with the MMT telescope. They identify 140 \mgii absorption systems with 0.1\AA$\le W_{2796} \le$3.2\AA{} and 0.1973$\le z \le$0.9265. {They find that the full equivalent width frequency distribution, $dN/dW$, can not be well described by a single exponential function ($dN/dW=(N^{*}/W^{*})e^{(-W_{2796}/W^{*})}$). The fit breaks down due to an excess of absorbers with $W_{2796}$<0.3\AA$ $. Thus, using data from a single survey, they identify an inflection point in the $dN/dW$ at $W_{2796}$ $\simeq$ 0.3\AA$ $. This transition {suggests} that the full range of \mgii absorbers is composed of systems enriched through physically distinct processes. }}

{\citet{CHURCHILL1999} also investigate weak \mgii absorption systems ($W_{2796}$$\le0.3$\AA). They perform} a spectroscopic survey using HIRES on Keck I and {find} 30 such systems in the range 0.4$\le$z$\le$1.4. \citet{NARAYANAN2007} {increase the search range to z $\sim$2.4, by }using the UVES spectrograph on VLT and {identify} a further 112 weak \mgii absorption systems. Both surveys are $\sim$80$\%$ complete down to $W_{2796}$ $\sim$0.02\AA $ $ and {also} identify an excess of {weak \mgii absorbers}. Given {their high incidence rates, weak \mgii absorbers } have been identified as possible tracers of sub-Lyman limit systems (sub-LLS; defined as $N_{HI}$<10$^{17.3}$ \citealt{WOLFE2005}) enriched by stellar processes \citep{CHURCHILL1998, CHURCHILL1999} {and/or} {CGM/halo gas \citep{NARAYANAN2007}}.

\citet{RAO2006} connect \mgii absorbers to HI ($z<1.65$) by {selecting SDSS QSOs with \mgii absorbers and {follow them} up with UV spectroscopy using the $\textit{Hubble Space Telescope}$. They find that damped $Ly \alpha$} systems (DLAs; defined as $N_{HI}$$\ge$$10^{20.3}cm^{-2}$ \citealt{WOLFE2005}) are exclusively drawn from systems with $W_{2796}$$\ge$0.6\AA $ $. If the selection criteria {are} expanded to include the presence of Fe transitions with $W_{FeII\lambda2600}$$\ge$0.5\AA $ $ then half of those systems are found to be DLAs.

 {Identifying \mgii absorbers {beyond} z $\sim$2.6 requires observations in the near infrared and \citet{MATEJECK2012} (MS12 from here on) as well as \citet{CHEN2016} (C16 from here on) undertook such a survey using the Folded-port InfraRed Echellette (FIRE, \citealt{SIMCOE2008}) on the Magellan Baade Telescope. They observe 100 QSOs in the redshift range 1.9<$z$<7.08 with five systems beyond redshift 6. They are unable to investigate the evolution of weak \mgii systems as MS12 is only complete down to $\sim$0.337\AA$ $. They find that $dN/dz$ is consistent with no evolution for {intermediate/medium} systems (0.3<$W_{2796}$<1\AA). For strong \mgii systems ($W_{2796}$>1\AA$ $) they find an increase towards z $\sim$3 {followed by} a decrease towards the early universe, {reminiscent of the evolution in cosmic star formation rate density {\citep{MADAU2014}}. }}

{\citet{MATEJECK2013} (MS13 from here on) investigate the metallicity and associated HI {column densities} of the \mgii absorbers presented in MS12. They find that the incidence rate and metallicity of absorbers with $W_{2796}$>0.4\ang are consistent with the hypothesis that they are DLAs. They also find that \mgii systems associated with sub-DLAs (defined as 10$^{19}$cm$^{-2}$<N$_{HI}$<10$^{20.3}$cm$^{-2}$ \citealt{DESSAUGES2003}) are metal rich when Fe, Si and Al transitions are considered and that the incidence rate of {classic  (as defined in \citealt{CHURCHILL2000})} intermediate/medium {\mgii systems does not evolve with redshift. They also separate strong \mgii absorbers into DLA/HI-rich and double systems (as defined in \citealt{CHURCHILL2000})} and find that only the incidence rate of strong double system rises from redshift 2 to 3 and then drops towards redshift 5 {while the incidence rate of strong \mgii DLA/HI-rich absorbers decreases from redshift 2 to 5}. They suggest that further considerations, beyond just the strength of the $W_{2796}$ feature, are needed when connecting these systems to star-forming galaxies. }

\mgii has proven itself an effective probe of the enrichment of the CGM as it traces low-ionisation, metal-enriched outflowing \citep{MENARDCHELOUCH2009, RUBIN2010, WEINER2009} and accreting gas \citep{KACPRZAK2012, RUBIN2012}. It has provided observational constraints for testing metal transport and ionisation in cosmological simulations. The high frequency of weak systems, the lack of evolution of intermediate/medium systems and the connection between strong  systems and global star formation history suggests that the total cross section of \mgii absorbers is influenced by a range of physical effects {such as} covering fractions and metallicity. Furthermore, accurately modelling the chemical evolution of the universe is problematic as it depends on density field variations as well as the strength and shape of the global UV ionising background \citep{OPPENHEIMER2009}. Metal enrichment is also dependent on the outflow models \citep{OPPENHEIMER2006}, the initial mass function of Population III stars \citep{PALLOTINI2014} and it requires high-resolution in order to identify low mass self-shielded regions \citep{BOLTONHAEHNELT2013}.

Given the difficulty of modelling \mgii systems{, especially those arising in self-shielded low mass systems at the mass resolution limit of current cosmological simulations}, only one attempt has been presented in the literature. \citet{KEATING2016} test the impact of different feedback schemes, choice of hydrodynamic code and UV background in four different simulations, including the Illustris and Sherwood simulations. These simulations are unable to reproduce the incidence rate of strong absorbers, under-predict the incidence rate of intermediate/medium systems and {do not address} the evolution of weak systems.

{Greater consistency is achieved with simulations of the high ionisation metal transition \ion{C}{IV}$\lambda\lambda$1548 1550. \citet{OPPENHEIMER2006}, \citet{OPPENHEIMER2009} and \citet{CEN2011} are able to reproduce the decline in the comoving mass density of \ion{C}{IV}, $\Omega_{\ion{C}{IV}}$, observed past redshift 5 \citep{SIMCOE2006, SIMCOE2011, RYANWEBER2006, BECKER2009, RYANWEBER2009, DODORICO2013}. This drop by a factor of 2 to 4 {from redshift 5 to 6} suggests either a rapid decrease in the enrichment or the ionisation {state of the IGM} \citep{BECKER2015}, a degeneracy that {could} be broken by measuring the comoving mass density of low ionisation ions.}

 {\citet{MATHES2017} explore the evolution of the comoving mass density of \ion{Mg}{ii} ($\Omega_{\mgii}$). They investigate 602 QSO spectra and find that $\Omega_{\mgii}$ increases from $\Omega_{\mgii} \simeq$0.9$\times$10$^{-8}$ at <$z$>=0.49 to $\Omega_{\mgii} \simeq$1.4$\times$10$^{-8}$ at <$z$>=2.1. \citet{BECKER2006, BECKER2011} also investigate the low ionisation ions \ion{O}{I}, \ion{Si}{II} and \ion{C}{II} in the redshift range 5.3<$z$<6.4. However, no current studies have explored the the comoving mass density of low ionisation ions in the redshift range 2<$z$<6. We explore this missing discovery space and present, for the first time, the evolution $\Omega_{\mgii}$ in the redshift range 2<$z$<5.45.}

 {The observations are described in Section \ref{sec:obs} and the \mgii candidate selection is described in \ref{sec:candidataselection}. We provide notes on the individual absorbers in Section \ref{sec:solnotes}. We discuss the impact of variable completeness and false positive contamination on our statistics in sections \ref{sec:variablecompleteness} and \ref{sec:falsepositive}. Our line statistics are presented in sections \ref{sec:linestats} and \ref{sec:d2N}. The $\Omega_{\mgii}$ calculations and values are in section \ref{sec:massdensity}. We discuss our results in Section \ref{sec:discussion}. Our findings are summarised in section \ref{sec:conclusion}. Throughout this paper we use a $\Lambda$CDM cosmology with $\Omega_{M}=0.308$ and $H_o=67.8$ kms$^{-1}$Mpc$^{-1}$ \citep{PLANCK2015}}.

\section{Observations and data reduction}
\label{sec:obs} 
Observations were conducted in service mode using the X-shooter spectrograph \citep{XSHOOTER} from July 2009 to October 2010 for the program 084.A-0390(A) under nominal conditions of seeing$\le$0.8" and any lunation. We utilised 0.7" and 0.6" slits to minimise the effect of OH sky lines, which deliver a nominal resolution of $R$=11,000 and {8,100 corresponding to a FWHM of 27 and 37 kms$^{-1}$} for the VIS and NIR arms, respectively. The final science exposure times on the 4 QSOs are given in Table \ref{tab:exposureTable}. The VIS and NIR spectra are collected simultaneously but not all data met the minimum quality control. Thus, total exposure times in the two arms are not equal for all QSOs. The data were reduced using customised IDL routines and the final spectra were binned to a resolution of 10 kms$^{-1}$pixel$^{-1}$. The QSO spectra were continuum fitted and normalised using the $UVES$\_$popler$ package \footnote{developed by \href{https://github.com/MTMurphy77}{Dr. Michael Murphy} \\ \href{https://github.com/MTMurphy77}{https://github.com/MTMurphy77}}.

\begin{table*}
 \caption{Total X-Shooter exposure times for the four QSOs. References for emission redshifts and magnitudes are given in the text descriptions of each QSO.}
 \label{tab:exposureTable}
 \begin{tabular}{|| c | c | c | c | c ||}
 \hline
 \hline
  &	 		& & Exposure Time & 		 Exposure Time 	 \\
  QSO &		$z_{\textrm{em}}$		& 		Magnitude & 		VIS		 	 & 		NIR	 \\
  &				& 		 & 		hours		 	 & 		hours	 \\
 \hline
\\
ULAS J0148+0600 & 5.98$\pm$0.01 & $^1y_{p1}$=18.77$\pm$0.10 & 12 & 11.5 \\
 \\
SDSS J0927+2001 & 5.79$\pm$0.02 & $J_{Vega}$=19.01$\pm$0.10 & 10 & 10 \\
 \\
 SDSS J1306+0356 & 5.99$\pm$0.03 & $J_{Vega}$=18.77$\pm$0.10 & 12 & 11.5 \\
\\
ULAS J1319+0959 & 6.1330$\pm$0.0007 & $J_{Vega}$=18.76$\pm$0.03 & 12 & 12 \\
 \hline
\hline
$^1$ \textrm{AB magnitude}
\end{tabular}
\end{table*}

\section{Candidate Selection}
\label{sec:candidataselection}

\subsection{Automatic search}
\label{sec:automaticdetection} 

{In order to identify possible \mgii absorbers we first create a candidate list using a detection algorithm which simply requires: }

$\bullet$ a minimum 3 consecutive pixel 3 or 5 $\sigma_i$ detection; 

\indent \hspace{2.1cm}$\sigma_i$=${(1-F_{\lambda}i)}/{E_{\lambda}}$

$\bullet$ $0.4 \le (W_{2796}+\sigma W_{2796}) / (W_{2803}+\sigma W_{2803})\le 3.5$ 

		\indent \hspace{3cm} or 
		
	 	 \indent  \hspace{1.5mm}   $0.4 \le (W_{2796}-\sigma W_{2796}) / (W_{2803}-\sigma W_{2803})\le 3.5$

$\bullet$ $W_{2796} / \sigma W_{2796} \ge 3$ or $W_{2803}/ \sigma W_{2803} \ge 3$

		\indent \hspace{3cm} or 

	 	 \indent  \hspace{1.5mm}   $W_{2796} / \sigma W_{2796} \ge 5$ or $W_{2803}/ \sigma W_{2803} \ge 5$

\noindent{{where $F_{\lambda}i$, $E_{\lambda}i$ are the flux error values associated with a pixel $i$ and $\sigma W_{2796}$, $\sigma W_{2803}$ are the rest equivalent width errors of the $\ion{Mg}{II}\lambda\lambda$2796 2803 features. This results in 103 candidates and we find no absorbers directly associated with the QSOs (within 3000 kms$^{-1}$).}}

\subsection{Visual Check}
\label{sec:visid} 

{The list of 103 candidates from the detection algorithm is then checked by eye by the lead author (AC). Absorbers with a large mis-match in the velocity profile of the $\lambda$2796 and $\lambda$2803 were rejected. These rejected systems generally occurred in regions of strong sky line residuals.}

{We also search for associated ions (\ion{C}{II}, \ion{Si}{II}, \ion{Mg}{I}, \ion{Al}{II}, \ion{Al}{III}, \ion{N}{V}, \ion{O}{I}, \ion{O}{VI}, \ion{Fe}{II} and \ion{Ca}{II}) with the identified doublets and they will be presented in a subsequent paper.  The visual check is also used to reject mis-identified \mgii that are in fact due to other doublets, such as \ion{C}{IV}, whose velocity profiles are well matched. However, given that this additional information is not available across the full redshift path of our survey or for all \mgii absorbers, we do not use associated transitions to strengthen the identification of individual \mgii absorbers. Following this visual inspection we are left with 52 \mgii absorbers.}

\subsection{Voigt Profiles $\&$ Equivalent Widths}
\label{sec:voigtprofiles}

We used VPFIT 10.0 to fit Voigt profiles to the absorption lines and, through a $\chi^2$ minimisation procedure determine the best fitting values of redshift (z), column density (N) and Doppler parameter (b). The Doppler parameter has contributions from thermal and turbulent motions; the latter were assumed to be the same for all the ions fitted, while for the former we fixed the gas temperature to be T=10000K in all cases. As the dominant source of uncertainty arises from the continuum fitting process, we adjust the continuum level by $\pm$5$\%$ and repeat the entire fitting procedure. This leads to the error bars associated with each set of Voigt profile parameters. 

\begin{figure}
	\includegraphics[width=8.5cm]{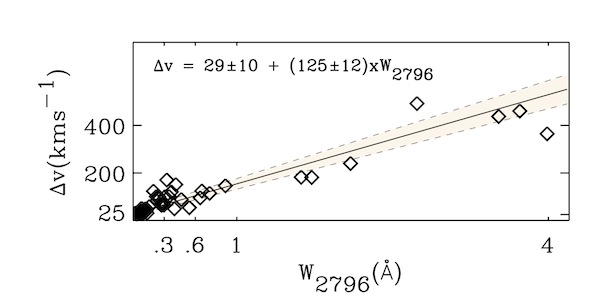}
 \caption{ $W_{2796}$ (\AA) vs velocity width ($\Delta v$; kms$^{-1}$) of identified systems. The linear fit $\Delta v$=29$\pm$10 kms$^{-1}$+(125$\pm$12)kms$^{-1}$\AA$^{-1}\times W_{2796}$ is plotted as a solid black line and the 1$\sigma$ boundaries are over-plotted with dashed lines and shaded in with light tan.}
 \label{fig:ewv}
\end{figure}

 {Individual components within 500 kms$^{-1}$ are considered a system and the equivalent width of the $\lambda$2796 transition ($W_{2796}$) is computed as the sum of each component's $W_{2796}$. The associated system equivalent width error ($\sigma W_{2796}$) is calculated by adding all individual component $W_{2796}$ errors in quadrature. The velocity width ($\Delta v$) is calculated by first identifying the boundary enclosing 90$\%$ of the optical depth ($\Delta \lambda$; \citealt{PROCHASKA2008}) which leads to $\Delta v$=$c\times\Delta \lambda$ $\div$ 2796.3542 where $c$ is the speed of light. The identified \mgii systems and their components, Voigt profile parameters, component $W_{2796}$, system $W_{2796}$, associated errors and recovery levels are presented below. The $W_{2796}$ vs velocity width distribution is plotted in Figure \ref{fig:ewv} with the best linear fit line overplotted as a solid black line. The 1$\sigma$ bounds are overplotted with black dash lines and filled in with light tan. The best linear fit is}

\begin{equation} 
 \Delta v=29\pm10 \text{kms}^{-1}+(125\pm12) (\text{kms}^{-1}  \text{\AA} ^{-1}) \times W_{2796} \\
	\label{eq:wdeltav}
\end{equation}

\begin{figure}
	\includegraphics[width=8.5cm]{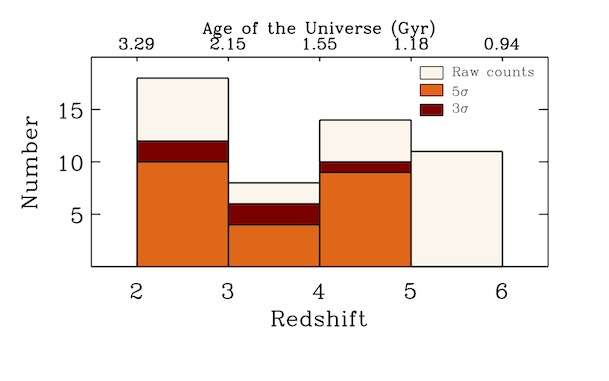}
	\includegraphics[width=8.5cm]{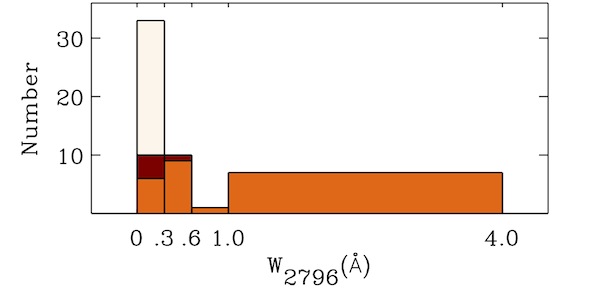}
 \caption{Histograms of $\mgii$ discovered systems. The raw counts, without any completeness or false positive considerations, are plotted in light tan. The 3 and 5$\sigma$ statistics are, respectively, plotted in orange and burgundy. The selection criteria are described in Section \ref{sec:completeness}. $Top$ $Panel:$ Redshift histograms of all identified systems. $Bottom$ $Panel:$ $W_{2796}$ histogram with boundaries denoting the distribution of identified weak ($W_{2796}$$\le$$0.3$\AA), intermediate ($0.3$$<$$W_{2796}$$\le$$0.6$\AA), medium ($0.6$$<$$W_{2796}$$\le$$1.0$\AA) and strong ($W_{2796}$$>$$1$ \AA) systems. }
 \label{fig:ewzhisto}
\end{figure}

\noindent {and the slope is very consistent with the results of MS12. However, we find a lower intercept value (29$\pm$10 kms$^{-1}$) than MS12 (80.2$\pm$8.6 kms$^{-1}$), reflecting the increased resolving power of X-Shooter VIS {($R$=11000)} vs. FIRE {(R=6000)}. The raw statistics as a function of $W_{2796}$ and redshift are given in Figure \ref{fig:ewzhisto}.}

 In order to be able to identify the entire family of \mgii absorbers we observed our targets for a minimum of $\sim$10h, more than doubling the average exposure time of MS12 and C16 for similar redshift objects. {While the increased exposure time will increase the likelihood of identifying possible weak absorbers, some weak systems may still exist with widths narrower than the instrumental resolution}. For example, both studies observe $SDSS$ $J1306+0356$ and $ULAS$ $J1319+0950$, but we increased the exposure times of the MS12 and C16 spectra ($\sim$4.35 and $\sim$5.35 hours respectively) to an average of $\sim$10 hours. Pairing these exposure times with the larger collection area of VLT vs Magellan (8m vs. 6.5m), we are able to confidently identify and classify weak \mgii absorption systems past z$\sim$5 at both 3 and 5$\sigma$ levels. 
 
 We investigate the impact of telluric contamination characteristic of NIR observations and find that there are portions of the spectra in which even the strongest systems can not be confidently identified. However, our selection criteria only reject weak visually identified systems. This gives us confidence that the selection criteria are robust and that the surviving systems are real. Our analysis only includes those systems which meet these  selection criteria and the calculation of the associated {recovery} level of each system is discussed in Section \ref{sec:completeness}.

 {In the following subsections we discuss the individual {lines} of sight and each of the discovered \mgii absorbers. The absorbers which do not meet our 5$\sigma$ selection criteria are marked with $^{\ast}$ and those which do not meet both 5$\sigma$ and 3$\sigma$ selection criteria are marked with $^{\ast\ast}$. {Individual components which are blended with other transitions are marked with $\Downarrow$ and individual components which are blended with poor residuals arising from sky lines are marked with $\Uparrow$. In these instances, the Voigt profile is used to obtain the associated $W_{2796}$ value.} The selection criteria are described in Section \ref{sec:completeness}. We visually identify 52 \mgii absorption systems and 23 pass our 5$\sigma$ selection criteria over a redshift path $\Delta$z=13.8. }

\subsection{Individual sight lines}
\label{sec:solnotes} 

\subsubsection{ULAS J0148+0600}
\label{sec:u0148} 

 {This QSO was first discovered in the UKIRT Infrared Deep Sky Survey (UKIDSS; \citealt{UKIDSS}) and was first presented in literature 	by \citealt{BANADOS2014} with $i_{p1}$=22.80$\pm$0.25, $z_{p1}$=19.46$\pm$0.02, $y_{p1}$=19.40$\pm$0.04 and {emission redshift}, z=5.96. The redshift was later refined by \citet{BECKER2015B} using X-Shooter on VLT to $z_{\textrm{em}}$=5.98$\pm$0.01. This sightline had not been previously investigated for the presence of metal line absorption systems but a $\sim$110 Mpc$h^{-1}$ Gunn-Peterson trough was discovered and discussed by \citealt{BECKER2015B}. It extends from z $\sim$5.5 to z $\sim$5.9 and we discovered no associated metal absorbers (\mgii, \ion{C}{IV} or \ion{Si}{IV}) in this redshift range. A caveat is that we are not more than 50$\%$ complete for \mgii over the full redshift range of interest (as indicated in Figure \ref{fig:completenesstest}). However, we have performed a comprehensive search for multiple ions and find no unidentified metal transitions which could be associated with unidentifed strong \mgii systems in that redshift range. }

We present 9 new \mgii systems. Systems 1, 6$^{\ast}$, 7, and 9$^{\ast\ast}$ are single component systems. System 2 has the most complex velocity structure with {five} components, {two} of which are saturated. Systems 3, 4 and 8 are {two} component systems and system 4 is fit with {three} components, one of which is saturated. Systems 6$^{\ast}$ only meets our 3$\sigma$ selection criteria described in section \ref{sec:stats}. System 9$^{\ast\ast}$ does not meet meet our 3 or 5$\sigma$ selection criteria described in section \ref{sec:stats}. 

\begin{figure*}
 	\includegraphics[width=17cm]{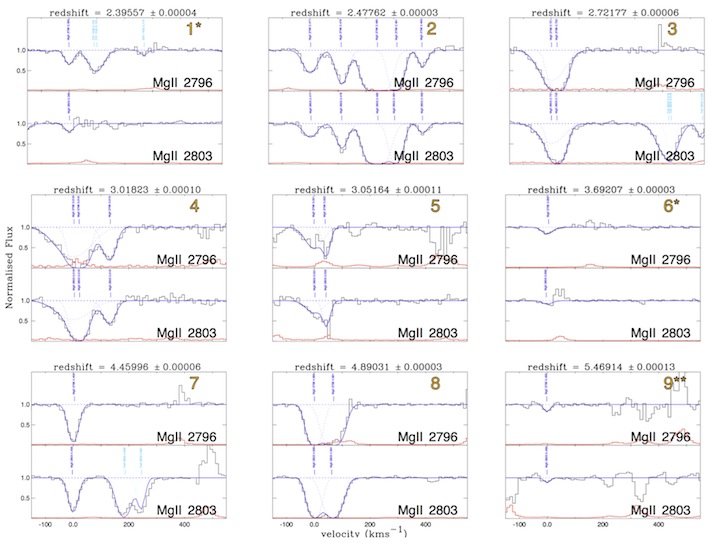} 

\begin{flushleft}

 $^{\ast}$ $\hspace{1.2mm}$ denotes system which does not meet our 5$\sigma$ selection criteria
 
 $^{\ast}$ $^{\ast}$ denotes system which does not meet our 5$\sigma$ and 3$\sigma$ selection criteria

\end{flushleft}	

\caption{All \mgii systems identified in the ULAS J0148+0600 sightline. The top panel of each system plot is the location of the $\lambda$2796 transition and the bottom panel is the associated $\lambda$2803 transition. In each panel, the vertical axis is the continuum normalised flux. The horizontal axis is the velocity separation (kms$^{-1}$) from the lowest redshift component of a system. The normalised spectrum is plotted in black and the associated error is in red. The solid blue line represents the full fit to the spectra and includes other ions besides \ion{Mg}{II}. Individual components are plotted with dashed lines and are identified by a vertical label. \mgii components are in solid blue and other transitions are in light blue.}	
	
 \label{fig:u0148systems}
\end{figure*}

The highest redshift absorber in this sightline that passes both of our 3 and 5$\sigma$ selection criteria is system 8 with $z_{\textrm{abs}}$=4.89031$\pm$4$\times$10$^{-5}$. All of the systems are plotted in Figure \ref{fig:u0148systems} and their {measured} component and system parameters can be seen in Table \ref{tab:u0148Table}.

\subsubsection{SDSS J0927+2001}
\label{sec:s0927} 

This QSO was first discovered as an SDSS $i$ band dropout with $i_{AB}$=22.12$\pm$0.17, $z_{AB}$=19.88$\pm$0.08 and a $J$-band magnitude of $J_{Vega}$=19.01$\pm$0.10 (2MASS; \citealt{2MASS}). It was spectroscopically confirmed by \citet{FAN2006} using the Multi-Aperture Red Spectrometer on the 4m Kitt Peak telescope with $z_{\textrm{em}}$=5.79$\pm$0.02. 

 {We present the first discovery of 22 \mgii systems in this sightline. System 1 is a {two} component system. System 3 is also a two component system and its $\lambda$2796 transition is blended with an \ion{Al}{II}$\lambda$1670 absorption feature associated with system 13 ($z_{\textrm{abs}}=4.60$). System 20$^{\ast\ast}$ is a {two} component system with the $\lambda$2803 transition residing in a wavelength range dominated by residuals from a sky line subtraction. The {remainder} of the systems are single component. Systems 6$^{\ast}$, 9$^{\ast}$, and 13$^{\ast}$ only meet our 3$\sigma$ selection criteria described in section \ref{sec:stats}. System 20 $^{\ast\ast}$ along with systems 2$^{\ast\ast}$, 4$^{\ast\ast}$, 5$^{\ast\ast}$, 8$^{\ast\ast}$, 10$^{\ast\ast}$, 11$^{\ast\ast}$, 14$^{\ast\ast}$, 15$^{\ast\ast}$, 16$^{\ast\ast}$, 17$^{\ast\ast}$, 18$^{\ast\ast}$, 19$^{\ast\ast}$, 21$^{\ast\ast}$ and 22$^{\ast\ast}$ do not meet meet our 3 or 5$\sigma$ selection criteria described in section \ref{sec:stats}. }

The highest redshift absorber in this sightline that passes both of our 3 and 5$\sigma$ selection criteria is system 7 with $z_{\rm{abs}}$=2.82038$\pm$4$\times$10$^{-5}$ and the highest redshift absorber in this sightline that passes only the 3{$\sigma$ recovery} selection criteria is system 13$^{\ast}$ with $z_{\textrm{abs}}$=4.60564$\pm$7$\times$10$^{-5}$. All of the systems are plotted in Figure \ref{fig:s0927systems} and their {measured} component and system parameters can be seen in Table \ref{tab:s0927Table}.

\subsubsection{SDSS J1306+0356}
\label{sec:s1306} 
{This QSO was first discovered as an SDSS $i$ band dropout with $i_{AB}$=22.58$\pm$0.26, $z_{AB}$=19.47$\pm$0.05 and a $J$-band magnitude of $J_{Vega}$=18.77$\pm$0.10 (2MASS). It was spectroscopically confirmed by \citealt{FAN2001} using the Echelle Spectrograph and Imager (ESI) on KeckII with $z_{\textrm{em}}$=5.99$\pm$0.03}. 

{MS12 {conducted follow-up observations} in the NIR with FIRE and first identified systems 3, 5, 7, 8 and 9 in Table \ref{tab:s1306Table}. They did not identify system 6 with $z_{\textrm{abs}}$=4.13 despite its} $W_{2796}$ being greater than that of system 5 with $z_{\textrm{abs}}=3.48$. {This is most likely due to the fact that our automatic detection algorithm (described in section \ref{sec:completeness}) only requires one of the doublet features to have a $5\sigma$ detection. This allows for the possibility that one transition of the $\ion{Mg}{II}\lambda\lambda$2796 2803 doublet falls within a wavelength} range affected by possible residuals from poor sky line removal, as is the case for the $\ion{Mg}{II}\lambda$2796 feature of system 6. We provide individual component Voigt profile information on the previously discovered systems and also present 3 other new systems.

System 1$^{\ast\ast}$ is a {two} component system where the $\ion{Mg}{II}\lambda$2803 feature is heavily blended with $\ion{C}{IV}\lambda$1548 associated with the $z_{\textrm{abs}}=4.86$ \mgii system. System 2 is a {two} component system. System 4$^{\ast\ast}$ is a newly identified system with a single component where its $\lambda$2796 component is heavily blended with the $\lambda$2803 components of system 3. Due to this blending, it is considered an isolated system and it does not meet our 3 or 5$\sigma$ selection criteria described in section \ref{sec:stats}. {We do not consider it in our incidence rate statistics or when computing $\Omega_{\mgii}$. }

{System 9 has been included in the incidence line statistics presented in MS12 but has been excluded by C16 from their study. We see no reason to exclude the system but we discuss the impact of this system upon our incidence line and $\Omega_{\mgii}$ statistics discussed in sections \ref{sec:linestats} and \ref{sec:massdensity}.} The highest redshift absorber in this sightline that passes both of our 3 and 5$\sigma$ selection criteria is system 9 with $z_{\textrm{abs}}$=4.87902$\pm$0.00011. All of the systems are plotted in Figure \ref{fig:s1306systems} and their {measured} component and system parameters can be seen in Table \ref{tab:s1306Table}.

\subsubsection{ULAS J1319+0950}
\label{sec:u1319} 

 {This QSO was first discovered as an SDSS $i$ band dropout with $i_{AB}$=22.83$\pm$0.32, $z_{AB}$=19.99$\pm$0.12 and $Y_{Vega}$=19.01$\pm$0.03, $J_{Vega}$=18.76$\pm$0.03 as measured in UKIDSS. It was spectroscopically confirmed by \citet{MORTLOCK2009} with $z_{\textrm{em}}$=6.127$\pm$0.004 and the emission redshift has been further refined by \citet{WANG2013} using the Atacama Large Millimetre/submillimiter Array (ALMA). They find $z_{em}$=6.1330$\pm$0.0007 and this is the emission redshift adopted in this work. It was further observed by MS12 and \citet{DODORICO2013} using FIRE on Magellan and X-Shooter on VLT, respectively. MS12 first identified system 8. \citet{DODORICO2013} first identified systems 2, 3, 7, 9 and confirmed system 8 in Table \ref{tab:u1319Table}. }

 {We present 7 new systems in this sightline. System 1$^{\ast}$ is a single component system with {its} $\lambda$2803 transition heavily blended with the \ion{Si}{IV}$\lambda$1393 feature of an absorption system ($z_{\textrm{abs}}=5.57$), which will be discussed in a {subsequent} paper. Systems 4, 5$^{\ast\ast}$, 6$^{\ast\ast}$, 10$^{\ast\ast}$ and 12$^{\ast\ast}$ are {two} component systems. System 12$^{\ast\ast}$ is heavily blended with sky lines. System 9 is a {three} component system and its $\lambda$2796 components are blended with a sky line. It is associated with a confirmed \ion{C}{IV}$\lambda\lambda$1548 1550 system \citep{DODORICO2013}. Systems 1$^{\ast}$ only meets our 3$\sigma$ selection criteria described in section \ref{sec:stats}. Systems 5$^{\ast\ast}$, 6$^{\ast\ast}$, 10$^{\ast\ast}$, 11$^{\ast\ast}$ and 12$^{\ast\ast}$ do not meet our 3 or 5$\sigma$ selection criteria described in section \ref{sec:stats}. }

The highest redshift absorber in this sightline that passes both of our 3 and 5$\sigma$ selection criteria is system 9 with $z_{\textrm{abs}}$=4.66297$\pm$6$\times$10$^{-5}$. All of the systems are plotted in Figure \ref{fig:u1319systems} and their {measured} component and system parameters can be seen in Table \ref{tab:u1319Table}.

\subsection{Survey Completeness}
\label{sec:completeness} 

{Our ability to retrieve absorption features as a function of redshift is impacted by the presence of OH sky emission lines and telluric absorption in the near-infrared sky which can blend with and obscure absorption features. Furthermore, the error computed from photon counting statistics may not be representative in wavelength areas heavily contaminated by sky lines. In order to quantify the impact of the variable S/N resulting from the contamination of sky lines, we performed a Monte Carlo simulation in which we inject {artificial systems in 100 \ang steps. We cover every Angstrom of our survey path by injecting $\sim$30 million artificial systems per sightline.} The injected systems are chosen to represent the observed $W_{2796}$ vs velocity distribution (see Figure \ref{fig:ewv}) and are uniformly distributed in the rest equivalent width range, $ W_{2796} \le 1\AA$. We apply the exact same philosophy to detect these inserted systems as we described in sections \ref{sec:automaticdetection} and \ref{sec:visid}.}

{The injected systems are first detected automatically by the same algorithm used to create the initial \mgii candidate list described in sec \ref{sec:automaticdetection}. The output of the detection algorithm is a Heaviside function, $H(W_{2796},z)$, where $W_{2796}$ is the rest frame equivalent width and $z$ is the redshift of the injected \mgii $\lambda$2796 feature. The values of the output are }

\begin{equation} 
 H(W_{2796},z)=\left\{
 \begin{array}{ll}
  0 &\textrm{, if the injected system is not detected} \\
  1 &\textrm{, if the injected system is detected}
 \end{array}
 \right.
	\label{eq:heaviside}
\end{equation}

\noindent {{Thus, the recovery fraction of \mgii is  }}

\begin{equation}
 L(dW_{2796}, dz)=\frac{1}{N} \sum\limits_{i=1}^{N} H(W_{2796},z)\
	\label{eq:lrecovery}
\end{equation}

\begin{figure*}
	\begin{center}
	\includegraphics[ width=8.5cm]{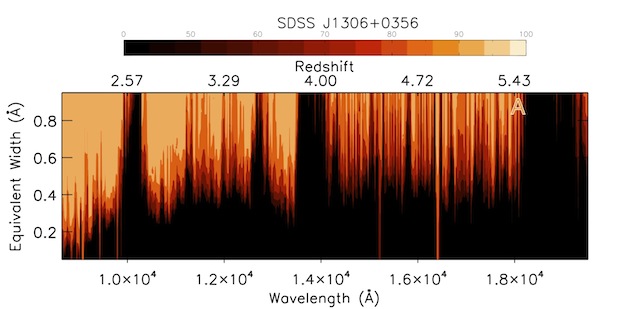}\includegraphics[ width=8.5cm]{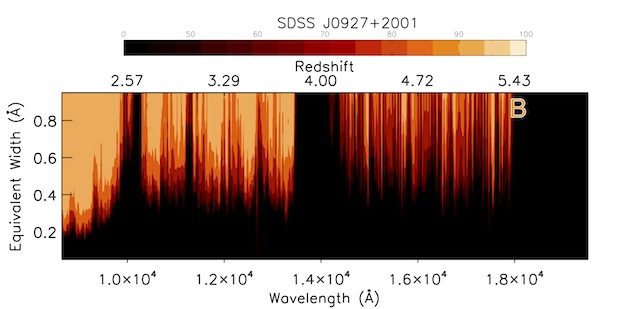}
	\includegraphics[ width=8.5cm]{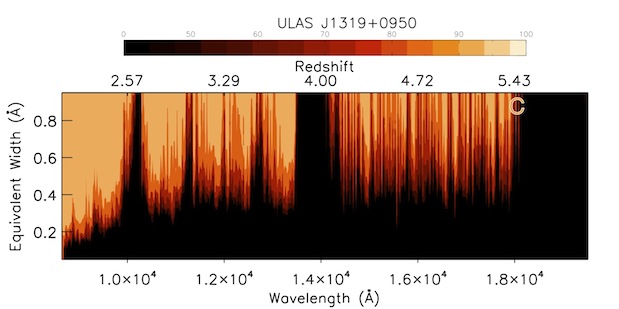}\includegraphics[ width=8.5cm]{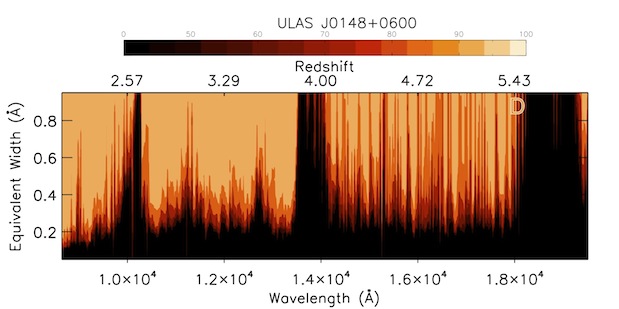}
 \caption{Completeness test results at the 5$\sigma$ level for each sightline. The $x$ and $y$ axis of each panel represent the wavelength (\AA) and $W_{2796}$ (\AA), respectively, of inserted systems. The top $x$ axis indicates the corresponding redshift of an inserted system. Their recovery rate, C($dW_{2796}$, $dz$), is denoted by the colour bar plotted in each panel. All recovery rates below 50$\%$ are shaded in black. Recovery rates for SDSS J1306+0356 are in panel $A$, SDSS J0927+2001 are in panel $B$, ULAS J1319+0950 are in panel $C$ and ULAS J0148+0600 are in panel $D$. The wavelength resolution of the recovery function, C($dW_{2796}$, $dz$), allows us to identify clean portions of the spectra in the near infrared as can be seen in panel $A$ at around $\sim$16660\AA$ $ where the recovery rate is above 50$\%$ for even the weakest of inserted systems. }
 \label{fig:completenesstest}
	\end{center}
\end{figure*}

\noindent{{where $dW_{2796}=0.01$ (rest equivalent width bin), $dz=0.01$ (redshift bin; corresponds to $\sim$28\AA$ $) and $N$ is the total number of elements of $H (W_{2796},z)$ in the respective $dW_{2796}$ and $dz$ bin. The 5$\sigma$ result is plotted in Figure \ref{fig:completenesstest}. Systems with $W_{2796}$$>$1\AA$ $ have been assigned the completeness level of a system with $W_{2796}$=1\ang as previous studies (MS12) found that systems with $W_{2796}$>0.95\ang have a similar completeness level as those with $W_{2796}$ $\simeq$0.95\AA.  However, given that the detection algorithm only provides an initial candidate list from which the user selects final candidates, we need to consider that the human selection of \mgii doublets (section \ref{sec:visid}) could be contaminated by false positives or that the user might miss true systems. }}

{In order to account for these biases, we follow the prescription put forth by MS12 and test the user on their ability to identify true absorbers and their likelihood of accepting false positives as true \mgii systems. We create a simple simulation which randomly chooses whether or not to insert a doublet and then prompts the user to vote on whether a \mgii doublet is present. If the simulation chooses to insert a doublet, it then randomly chooses whether a true or false positive system will be inserted. This randomisation ensures that the user has no a priori expectation in the voting process. The artificial \mgii doublet has double the separation of the true \mgii doublet with transitions at $\lambda$2796.3542 and $\lambda$2810.7087 and will be referred to as \ion{Mg}{II'} from here on. We run the voting process for close to $\sim$10000 iterations and bin our results as a function of signal to noise defined as a boxcar S/N of the inserted feature, SNR $\equiv W_{2796}/\sigma W_{2796}$. }

\begin{figure}
	\begin{center}
		\includegraphics[ width=8.5cm]{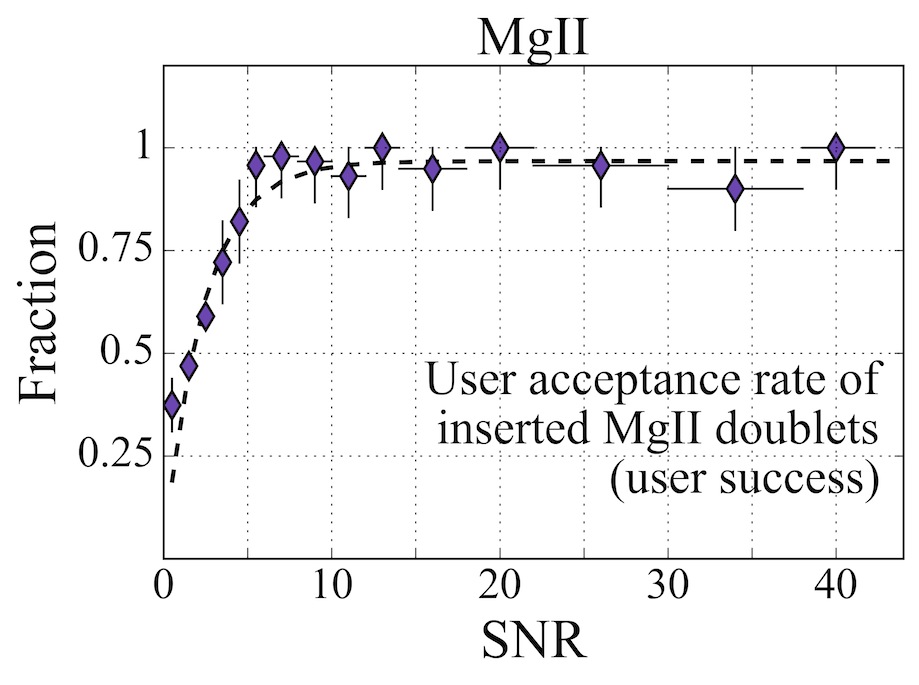} 
		
		\includegraphics[ width=8.5cm]{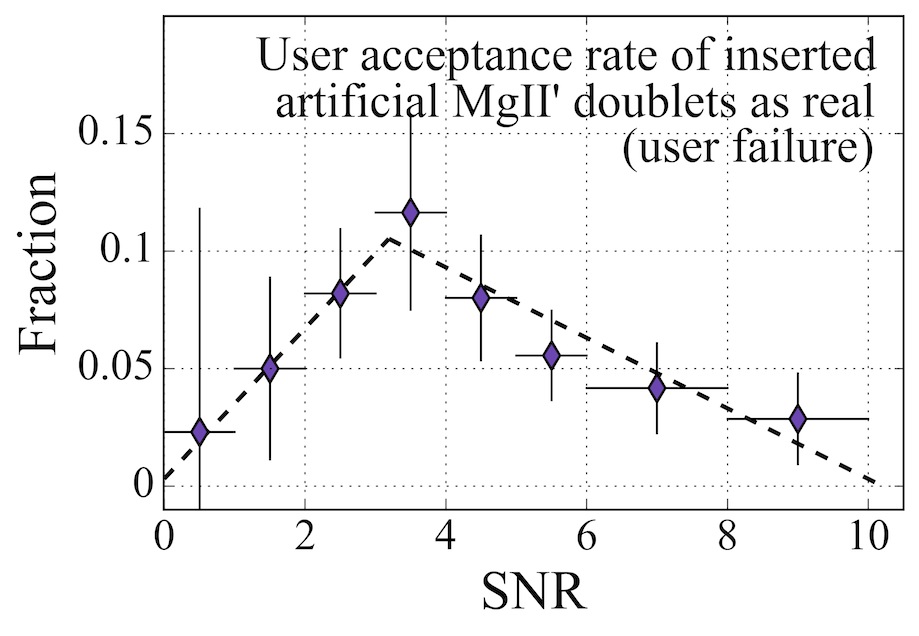}

	 	\caption{Plotted with purple diamonds are the binned user success (top panel) and user failure (bottom panel). Plotted with dashed lines are the respective best fits (eqs. \ref{eq:pmg2} and \ref{eq:pfmg2}). The horizontal bounds denote the S/N bin considered and the vertical error bars correspond to the associated 95$\%$ Wilson confidence interval.}
 		\label{fig:acceptancestats}
	\end{center}
\end{figure}

{We then calculate the user success rates as the ratio between the number of inserted systems to the number of identified  \mgii  systems. This fraction should be close to 1 if the user can accurately identify \mgii doublets (see top panel of Figure \ref{fig:acceptancestats}). We also calculate the user failure as the ratio between the number of artificial (\ion{Mg}{II}') inserted systems to the number of \ion{Mg}{II}' identified as true absorbers (\ion{Mg}{II}). This fraction should be close to 0 if the user can accurately distinguish between true and artificial \mgii systems (see bottom panel of Figure \ref{fig:acceptancestats}). For both binomial distributions we calculate error bars corresponding to the 95$\%$ Wilson confidence interval.}

{We fit the user success distribution (using a $\chi^{2}$ minimisation technique) with an exponential function (similar to MS12 and C16) of the form}

\begin{equation}
 P^{MgII}(SNR)=P_{\infty} (1-e^{S/SNR})
	\label{eq:pmg2}
\end{equation}

\noindent{{where P$_{\infty}$ (best fit value: 0.967) is the probability that the user will accept a true \mgii system and S (best fit value: 2.36) is an SNR exponential scale factor. Similarly to the findings of MS12 and C16, we find that even for the best S/N regions, the user acceptance rate is not 100$\%$ as the SN profile sharply decreases in regions polluted by narrow sky lines or telluric absorption.  }}

{In order to fit the user failure distribution, we use a triangle function  }

\begin{equation}
 P^{FP}(SNR)=\left\{
 \begin{array}{ll}
   P^{FP}_{max}(SNR/s_{p})&, $ $ SNR \le s_{p}\\
   P^{FP}_{max}(\frac{SNR-s_{f}}{s_{p}-s_{f}})&, $ $ SNR>s_{p}
 \end{array}
 \right.
	\label{eq:pfmg2}
\end{equation}

\noindent{{where $P^{FP}_{max}$ is the maximum contamination rate (best fit value: 0.10) which arises at $s_{p}$=3.19. We find that the user acceptance of injected false doublets as real approaches 0 as the SNR reaches 10.2. }}

{In order to fold the user success and failure into our completeness calculations, we turn the functional fits (eqs. \ref{eq:pmg2} and \ref{eq:pfmg2}) into grids binned in the same manner as the recovery fraction grids (see eq. \ref{eq:recovery}). The resulting grid for user acceptance is denoted as $A^{\ion{Mg}{II}}(dW_{2796}, dz)$ and the resulting grid for user failure is denoted as $A^{FP}(dW_{2796}, dz)$. We combine the user success grids with the recovery grids (for each sightline) and define the completeness as the product}

\begin{equation}
 C(dW_{2796}, dz)=L(dW_{2796}, dz) \times A^{\ion{Mg}{II}}(dW_{2796}, dz).\
	\label{eq:recovery}
\end{equation}

\section{MgII line statistics}
\label{sec:stats}
When considering absorption systems within an equivalent width bin  {$(dW_{i}$, $W_{2796} \in dW_{i}$)} and redshift bin {($dz_{j}$, $z_1 \leq [z \in dz_j] \leq z_2$)}, the number of systems is described by the population densities 

\begin{equation}
 	\Bigg(\frac{d^2 N}{dzdW}\Bigg)_{ij}=\frac{N}{\Delta W_i \Delta z_j} \\
	\label{eq:d2ndzdw}
\end{equation}

\begin{equation}
 	\Bigg(\frac{dN}{dz}\Bigg)_j=\frac{N}{\Delta z_j}\\
	\label{eq:dndz}
\end{equation}
\begin{equation}
 	\Bigg(\frac{dN}{dw}\Bigg)_i=\frac{N}{\Delta W_i}\\
	\label{eq:dndw}
\end{equation}

\noindent where $N$ is the true number of systems, $\Delta W$ is the full range of the equivalent width bin and $\Delta z$ is the total redshift path in the bin. The main focus of our investigation into incidence rate statistics is to create a bridge between statistics from previous studies and our final measurements of the comoving mass density of \ion{Mg}{II}, $\Omega_{\mgii}$. As such, we first focus on the incidence rate of \mgii absorbers ($dN/dz$; eq. \ref{eq:dndz}). This focus is motivated by the need to first identify the true redshift path ($dz$) over which a \mgii system (with $W_{2796} \in dW_i$ and $z_1 \leq [z \in dz_j] \leq z_2$) and the false positive contamination rate, both critical steps in calculating $\Omega_{\mgii}$ as will be described in section \ref{sec:massdensity}.

\subsection{Adjusting for variable completeness}
\label{sec:variablecompleteness} 

Given the fine resolution of the $C(dW_i, dz_j)$ function we must account for varying completeness across a larger bin of interest. In order to account for this variability, we first define a visibility function $R(W_{2796},z)$ which identifies regions probed by our survey where a \mgii feature with $W_{2796}$ and redshift $z$ could be identified with a recovery rate of $\ge$50$\%$. Previous studies (MS12, C16) have excluded wavelength regions contaminated by poor telluric subtractions but completeness also varies as a function of $W_{2796}$ (Figure \ref{fig:completenesstest}) in the \textit{J, H, K} bands. In order to account for this variation, we {first define a simple step function $R(dz_j)$ (dz=0.01, dW=0.01\AA$ $) which accounts for the \mgii path of each QSO. It is defined as 1 in the redshift range from 1000 kms$^{\textrm{-1}}$ redwards of the $Ly\alpha$ emission peak to the end of the $H$ band corresponding to an absorption redshift $\sim$5.45 for \ion{Mg}{ii}. Everywhere else, the value of $R(dz)$ is 0. In order to calculate the redshift path density \citep{LANZETTA1987, STEIDEL1992} of our survey across a redshift bin (dz) we then combine the visibility function ($R(dz_j)$) with the recovery grids (eq. \ref{eq:recovery})}

\begin{equation} 
 G(dW_i, dz_j)=\left\{
 \begin{array}{ll}
  0 &\textrm{, if $C(dW_i, dz_j)$<0.50} \\
  1 &\textrm{, if $C(dW_i, dz_j)$$\ge$0.50 $\&$ $R(dz_j)$=1}
 \end{array}
 \right.
	\label{eq:rwz}
\end{equation}

{The completeness adjusted path of our survey is then,}

\begin{equation}
 	g(dW_i, dz_j) \equiv \sum_{s} G(dW_i, dz_j) \\
	\label{eq:gwz}
\end{equation}

\noindent {where \textit{s} represents the number of sightlines, each with $i$ and \textit{j} elements of \textit{C($dW_i$, $dz_j)$} such that $W\in dW_i$ and $z_1$ $\leq$ $dz_j$ $\leq$ $z_2$. This formulation is slightly different than the usual description of the redshift path density. If completeness depends mostly on the strength of the absorber then the redshift path density can be used to identify a completeness level through a sharp drop-off point which corresponds to a minimum $W_{2796}$ that can be identified. However, given the structure in the completeness results (see Figure \ref{fig:completenesstest}) we implement a minimum completeness level (50$\%$; see eq. \ref{eq:rwz}) cutoff rather than a minimum $W_{2796}$ cutoff. {Thus, eq. \ref{eq:gwz} accounts for all redshift bins in which a \mgii system with $W_{2796}$ is identified at least 50$\%$ of the time by the detection algorithm and human interaction step. The total cumulative path of our survey across a redshift bin is then}}

\begin{equation}
 	z(dW_i)=\int_{z1}^{z2} g(dW_i, dz_j)\mathrm{d}z\\
	\label{eq:zpath}
\end{equation}

\begin{figure}
	\includegraphics[width=8.5cm]{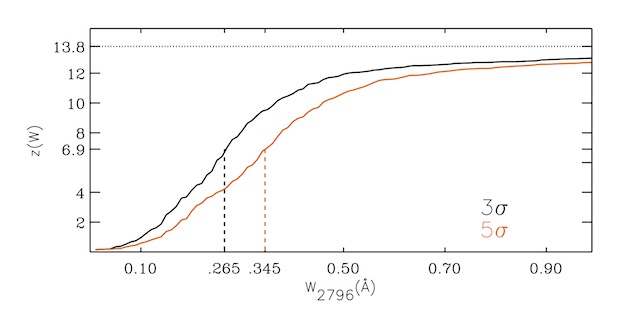}
 \caption{Integrated redshift path for the survey, z(W). Plotted with a solid black line is z(W) calculated using only a 3$\sigma$ selection on the redshift path density equation ( g(W, z); see eq. \ref{eq:gwz}). Plotted with a solid orange line is z(W) calculated using a 5$\sigma$ selection. {The vertical dashed lines highlight the $W_{2796}$ value for which we are 50$\%$ complete. The horizontal dot-dot line represents the full redshift path of our sample.}}
 \label{fig:zw}
\end{figure}

\noindent {and the total cumulative path of the survey can be seen in Figure \ref{fig:zw}. The total redshift path of our survey is 13.8 and is plotted as a horizontal dot-dot line in Figure \ref{fig:zw}. We are 50$\%$ complete down to an equivalent width $W_{2796}$=0.265\AA$ $ when we consider a 3$\sigma$ recovery criteria and we are 50$\%$ complete down to an equivalent width $W_{2796}$=0.345\AA$ $ when we consider a 5$\sigma$ recovery criteria (see the vertical dashed lines in Figure \ref{fig:zw}).}

{If completeness does not vary drastically between redshift and equivalent width bins we can then compute the average completeness, $\bar{C}$. Following the lead of MS12, we use the completeness adjusted path (eq. \ref{eq:rwz}) along with the visibility function and define} 

\begin{equation}
 	\bar{C}(dW_i, dz_j)=\frac{\int \int G(dW_i, dz_j) \frac{d^2 N}{dzdW} dz dW}{\int \int R(dz_j) \frac{d^2 N}{dzdW} dz dW}\\
	\label{eq:barc}
\end{equation}

\noindent {If the number density of absorbers also does not vary across the same redshift bin, we can use $\bar{C}$ along with the total number of systems in the same $dW_i$ and $dz_j$ bin ($\ddot{N}$) and find the true number of absorbers}

\begin{equation}
 	N(dW_i, dz_j) \equiv \frac{\ddot{N}(dW_i, dz_j)} {\bar{C}(dW_i, dz_j)} \\
	\label{eq:N}
\end{equation}

{However, while the expression for the true numbers of absorbers takes into account the probability that the user might miss some \mgii absorbers (user success quantified in eq. \ref{eq:pmg2}) it does not take into account the probability that the user might misidentify a non-\mgii as a \mgii absorber (user failure quantified in eq. \ref{eq:pfmg2}).  In order to incorporate this into our statistics we compute the average recovery rate ($\bar{L}$)  and average failure rate ($\bar{A}^{FP}$)  }

\begin{equation}
 	\bar{L}(dW_i, dz_j)=\frac{\int \int R(dz_j) \times L(dW_i, dz_j) \frac{d^2 N}{dzdW} dz dW}{\int \int R(dz_j) \frac{d^2 N}{dzdW} dz dW} \\
	\label{eq:barl}
\end{equation}

\begin{equation}
 	\bar{A}^{FP}(dW_i, dz_j)=\frac{\int \int  R(dz_j) \times {A}^{FP}(dW_i, dz_j) \frac{d^2 N}{dzdW} dz dW}{\int \int R(dz_j)  \frac{d^2 N}{dzdW} dz dW}\\
	\label{eq:barafp}
\end{equation}

\noindent {and rewrite eq. \ref{eq:N} }

\begin{equation}
 	N(dW_i, dz_j) \equiv \frac{\ddot{N}(dW_i, dz_j)} {\bar{C}(dW_i, dz_j)-\bar{L}(dW_i, dz_j) \times \bar{A}^{FP}(dW_i, dz_j)} \\
	\label{eq:finalNcomp}
\end{equation}

While we have adjusted our discovered statistics for the impact of user failure and variable completeness resulting from sky line pollution we still have not addressed the contamination by false positive identifications of doublet features. We discuss the implementation of such considerations in the follow-up section.

\subsection{Adjusting for false positives}
\label{sec:falsepositive} 
In order to account for the likely contamination by false positive detections, we doubled the rest separation of the $\ion{Mg}{II}\lambda\lambda$2796 2803 doublet and searched for an artificial $\ion{Mg'}{II}\lambda\lambda$2796 2810 doublet {in the continuum normalised spectra}. We used the same parameters for the doublet identification algorithm as described in section \ref{sec:completeness} {and obtained an initial list of 273 $\ion{Mg'}{II}$ candidates. From this list, the lead author identified 26 plausible $\ion{Mg'}{II}$ candidates. We emphasise here that no fake lines are inserted at this stage, the search algorithm is simply tuned to the separation of the artificial doublet.}

{Next, we compute the probability that the lead author can accurately identify an inserted $\ion{Mg'}{II}$ system ($P^{\ion{Mg'}{II}}$) and the probability that a $\ion{Mg'}{II}$ system is misidentified ($P'^{FP}$) as functions of S/N in the same fashion as eqs. \ref{eq:pmg2} and \ref{eq:pfmg2}. The results for user success and user failure for this artificial doublet ($\ion{Mg'}{II}$) are plotted in Figure \ref{fig:falseacceptancestats}. Following this, we compute the recovery function for this artificial doublet ($L^{'}(dW'_i, dz_j)$) with the same $W^{'}_{2796}$ and redshift binning ($dW^{'}_i$, $dz_j$) as eq. \ref{eq:recovery}. We follow the exact steps described in section \ref{sec:completeness}. It is only in the computation of the recovery and user success/failure functions that fake artificial systems are inserted and searched for.}

\begin{figure}
	\includegraphics[width=8.5cm]{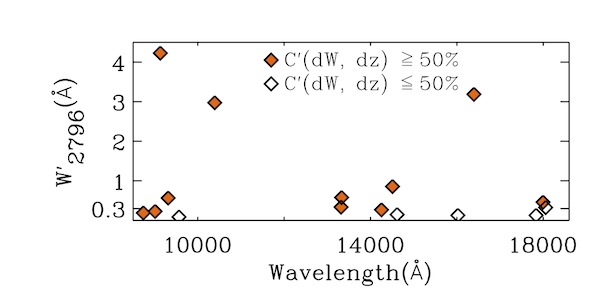}
 \caption{$W'_{2796}$ (\AA) vs wavelength (\AA) of identified false positive systems. Empty diamonds represent those systems with a recovery rate below 50$\%$ at a 5$\sigma$ level. Those systems with a recovery rate$\ge$50$\%$ when we consider a 5$\sigma$ selection criteria are plotted with filled orange diamonds. }
 \label{fig:false_ewz}
\end{figure}

\begin{figure}
	\includegraphics[width=8.5cm]{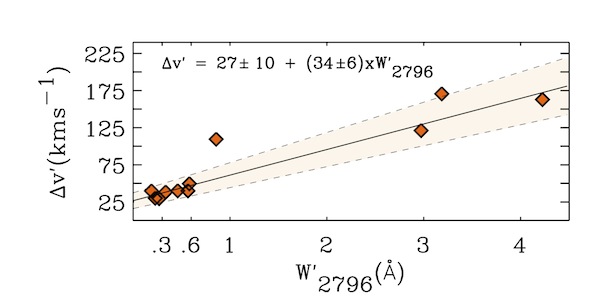}
 \caption{ $W'_{2796}$ (\AA) vs velocity width $\Delta v'$ (kms$^{-1}$) of identified false positive systems with a recovery rate$\ge$50 $\%$. The linear fit $\Delta v'$=24$\pm$11 $\textrm{kms}^{-1}$+(29$\pm$5)\textrm{(kms$^{-1}$\AA$^{-1}$)}$\times W'_{2796}$ is plotted as a solid black line and the 1$\sigma$ boundaries are over-plotted with dashed lines and shaded in with light tan.} 
 \label{fig:false_ewv}
\end{figure}

{Next, we combine the recovery function with the user success grids to compute the completeness ($C^{'}(dW'_i, dz_j)$) of artificial \ion{Mg'}{II} systems in the same fashion and with the same granularity of eq. \ref{eq:recovery}. The resulting 5$\sigma$ completeness grids for each sightline can be seen in Figure \ref{fig:artificialcompletenesstest}.}  The wavelength vs. $W_{2796}$ distribution of $\ion{Mg'}{II}\lambda\lambda$2796 2810 can be seen in Figure \ref{fig:false_ewz} and their $W'_{2796}$ vs $\Delta v'$ distribution can be seen Figure \ref{fig:false_ewv}. {Only 12 systems have a $C^{'}(dW'_i, dz_j)$ value$\ge$50$\%$.}

{In order to identify the true number of \ion{Mg'}{II} absorbers ($N'$) we then compute the average completeness ($\bar{C'}$), average recovery rate ($\bar{L'}$) and user failure ($\bar{A'}^{FP}$) of this population in exactly the same manner as described in eqs. \ref{eq:barc}, \ref{eq:barl} and \ref{eq:barafp}. We can now express the true number of artificial \ion{Mg'}{II} absorbers as, }

\begin{equation}
 	N^{'}(dW'_i, dz_j) \equiv \frac{\ddot{N'}(dW'_i, dz_j)} {\bar{C'}(dW'_i, dz_j)-\bar{L'}(dW'_i, dz_j) \times \bar{A'}^{FP}(dW'_i, dz_j)} \\ \\
	\label{eq:totalfalse}
\end{equation}

\noindent where $\ddot{N}^{'}$ is the number of artificial $\mgii'\lambda\lambda$2796 2810 detections with $W_{2796} \in dW_i$ and $z_1 \leq [z \in dz_j] \leq z_2$. The false positive {correction factor} of the same bin is then

\begin{equation}
 	F(dW_i, dz_j)=1-\frac{N^{'}(dW_i, dz_j)} {N(dW_i, dz_j)}\\
	\label{eq:fprate}
\end{equation}

\noindent where $N^{'}$ is the true number of artificial \mgii systems and $N$ is the true number of real \mgii systems in a $dW_i$, $dz_j$ bin as calculated in eqs. \ref{eq:finalN} and \ref{eq:totalfalse}. {We then combine the variable completeness and false positive corrections applied to the incidence line statistics and define a final correction factor A($dW_i$, $dz_j$)}

\begin{equation}
 A(dW_i, dz_j)=\frac{\bar{F}(dW_i, dz_j)}{ \bar{C}(dW_i, dz_j)-\bar{L}(dW_i, dz_j) \times \bar{A}^{FP}(dW_i, dz_j)} \
	\label{eq:abc}
\end{equation}

\noindent {Thus, the true number of absorbers adjusted for completeness and false positive contamination is}

\begin{equation}
 	N(dW_i, dz_j) =  A(dW_i, dz_j) \times \ddot{N}(dW_i, dz_j) \\
	\label{eq:finalN}
\end{equation}

\noindent {with associated Poisson error}

\begin{equation}
 	\sigma N(dW_i, dz_j) \equiv A(dW_i, dz_j) \times \sqrt{\ddot{N}(dW_i, dz_j)}   \\
	\label{eq:finalNerror}
\end{equation}

\noindent {for an equivalent width and redshift bin ($dW_i, dz_j$) such that $W_{2796} \in dW_i$ and $dz_j\in[z_1, z_2]$.}

\subsection{$dN/dz$ and $dN/dX$}
\label{sec:linestats} 

{We can now calculate our incidence rates as eq. \ref{eq:finalN} provides us with the total number of absorbers in any redshift and equivalent width bin. The error associated with this true number of absorbers is given in eq. \ref{eq:finalNerror}.}

\begin{figure}
	\includegraphics[width=8.5cm]{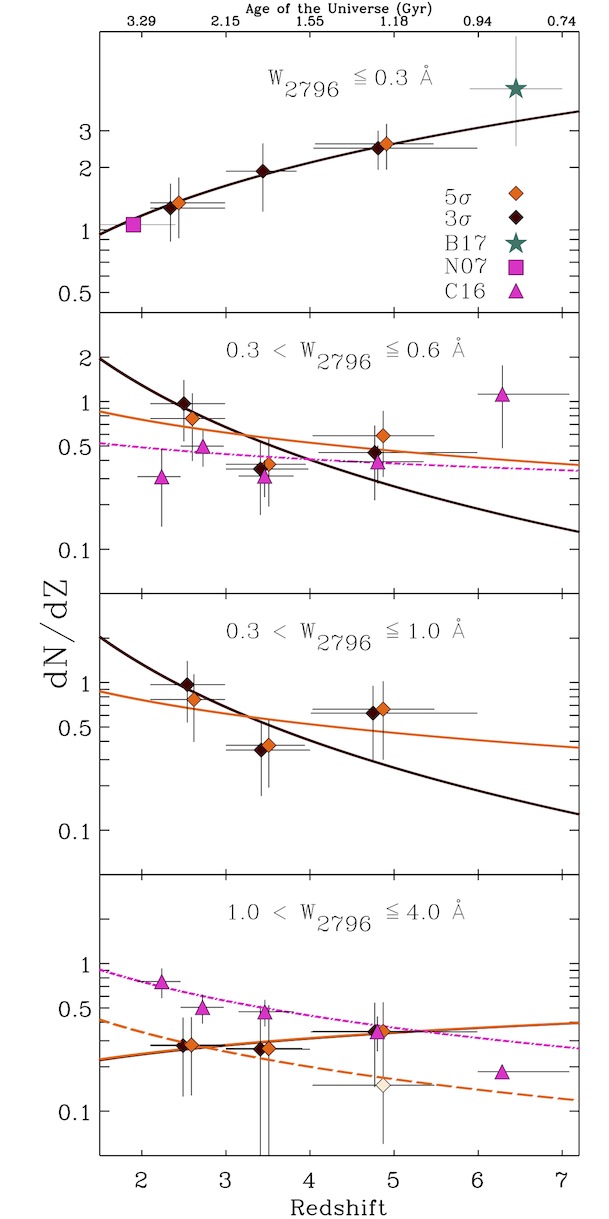}
 \caption{ {Incidence rate of \mgii absorbers, $dN/dz$, separated in four $W_{2796}$ bins selected to highlight the evolution of weak ($W_{2796}$$\le$0.3\AA), intermediate (0.3<$W_{2796}$$\le$0.6\AA), intermediate/medium (0.3<$W_{2796}$$\le$1.0\AA) and strong ($W_{2796}$>1\AA$ $) systems. The 5$\sigma$ values are offset by +0.1 in redshift for clarity. The green star in the top panel is from \citet{BOSMAN2017}.  The pink square in the top panel is from \citet{NARAYANAN2007} and the pink triangles in the second from the top and bottom panels are from C16. Plotted with filled diamonds are the values from this study with orange and burgundy identifying values calculated when considering 3 or 5$\sigma$ selection criteria, respectively. {The light coloured diamond represents the incidence rate calculated by excluding system 9 in the SDSS1306+0356 sightline as discussed in the text}. The values for all data points are in Table \ref{tab:fptable}. {The best fit lines for this work, the adjusted rate of strong absorbers and C16 are plotted with solid, long dash-dash and a dot-dash line respectively. The best fit lines computed in this work are plotted with solid lines. The colours of the best fit lines match the associated data points given in Table \ref{tab:parameterTable}. }}}
 \label{fig:dndz}
\end{figure}

 {In order to compare with previous results, we separate our total sample in three redshift bins (2<$z$$\le$3, 3<$z$$\le$4 and 4<$z$$\le$6) and four equivalent width bins: weak ($W_{2796}$$\le$0.3\AA), intermediate (0.3<$W_{2796}$$\le$0.6\AA), intermediate/medium (0.3<$W_{2796}$$\le$1.0\AA) and strong ($W_{2796}$>1\AA) systems. Our results can be seen in Table \ref{tab:fptable} and Figure \ref{fig:dndz}. {We follow previous studies in fitting the redshift evolution of the line density with a power law of the form} $dN/dz=N_{*}(1+z)^{\beta}$ and the best fit parameters with 1$\sigma$ bounds are presented in Table \ref{tab:parameterTable}. {Given the small sample size, we use a $\chi^{2}$ minimisation technique to the binned data to perform the fits (see Table \ref{tab:parameterTable} for all fit parameters).}}

\begin{table*}

 \caption{\mgii absorber line density ($dN/dz$), comoving absorber line density ($dN/dX$), false positive rates ($F(dW, dz)$), number of discovered systems ($\ddot{N}$), true number of absorbers ($N$), redshift bin ($\Delta z$), median redshift values (<$z$>) and the average of the binned $dN/dX$ values ( <$dN/dX$>) for 3 and 5$\sigma$ detection limits and $W_{2796}$ bins. A dash denotes no detections in the bin at the specified $\sigma$ level. Note that median redshift values are different for individual W$_{2796}$ bins reflecting the varying completeness levels of those bins (see Figures \ref{fig:completenesstest} and \ref{fig:completenesstest3sigma}).}
 \label{tab:fptable}

 \begin{tabular}{|| p{0.15 \textwidth} | p{0.09\textwidth} | p{0.095\textwidth} | p{0.07\textwidth} | p{0.11\textwidth} | p{0.16 \textwidth} | p{0.11\textwidth} | p{0.085 \textwidth} ||}
 \hline
 \hline
 \hspace{4.5mm}$<$z$>$ &    $\Delta z$     & $F(dW, dz)$ &   $\ddot{N}$ &    $N$ &   $dN/dz$ &   $dN/dX$ &  <$dN/dX$>  \\

 \hline
 \end{tabular}

  \begin{tabular}{|| p{0.03\textwidth} | p{0.05\textwidth} | p{0.06\textwidth}  | p{0.06\textwidth}  | p{0.02\textwidth}   | p{0.025\textwidth}  | p{0.015\textwidth} | p{0.02\textwidth}  | p{0.03\textwidth} | p{0.05\textwidth} | p{0.07\textwidth} | p{0.07\textwidth} | p{0.07\textwidth} | p{0.05\textwidth} | p{0.03\textwidth} | p{0.03\textwidth}  ||}

 3$\sigma$ & 5$\sigma$ & 3$\sigma$ & 5$\sigma$& 3$\sigma$ & 5$\sigma$ & 3$\sigma$ & 5$\sigma$ & 3$\sigma$ & 5$\sigma$ & 3$\sigma$ &  5$\sigma$ &  3$\sigma$ &  5$\sigma$ & 3$\sigma$ &  5$\sigma$ \\

 \hline
\hline
\end{tabular}

\begin{centering}
 $ W_{2796} \le 0.3$\AA $ $ \\
 \end{centering}

 \begin{tabular}{|| c | c | c | c | c | c | c | c | c | c | c | c | c | c | c | c |c  ||}
 \hline

 2.41 & 2.34 & 2.00-2.99 & 2.00-2.99 & 0.49 & 0.57 & $ $ 5$ $ & $ $4        &   10.3      &     9.48      & 1.26$\pm$0.56 & 1.35$\pm$0.58 & 0.38$\pm$0.17 			& 0.41$\pm$0.18 & 0.49 & 0.51 \\
 3.41 &-      & 3.00-3.84 & 	-             & 1.00 &-      &$ $ 2$ $ &$ $-      &    7.67      &     -           &   1.92$\pm$0.69 &-& 0.51$\pm$0.18 			                 &-&  &\\
 4.78 & 4.81 & 4.04-5.45 & 4.06-5.45 & 0.65 & 0.93    &$ $ 3$ $ & $ $2        &  21.9     &     16.1      &  2.45$\pm$0.65 & 2.58$\pm$0.67 & 0.57$\pm$0.15 			& 0.60$\pm$0.16 &  &\\
 \hline
 \hline
  \end{tabular}
 
 \begin{centering}
  $0.3<W_{2796} \le 0.6$\AA $ $ \\
\end{centering}

  \begin{tabular}{|| c | c | c | c | c | c | c | c | c | c | c | c | c | c | c | c |c  ||}
 \hline
 2.53 & 2.50 & 2.00-2.99 & 2.00-2.99 & 0.77 & 0.72 &$ $ 4$ $ &$ $ 3        &    5.03     &    4.27      & 0.97$\pm$0.49 & 0.77$\pm$0.44 & 0.30$\pm$0.15 				& 0.23$\pm$0.13 & 0.16 & 0.16\\
 3.41 & 3.41 & 3.00-3.65 & 3.00-3.95 & 0.37 & 0.35 &$ $ 3$ $ &$ $ 3         &    3.81     &    4.29      & 0.35$\pm$0.30 & 0.38$\pm$0.31 & 0.09$\pm$0.08 				& 0.10$\pm$0.08  &  &\\
 4.76 & 4.77 & 4.01-5.45 & 4.03-5.45 & 0.72 & 0.77 &$ $ 3$ $ &$ $ 3         &     3.63    &    4.43      & 0.45$\pm$0.28 & 0.59$\pm$0.32 & 0.11$\pm$0.07 				& 0.14$\pm$0.07 &  &\\
\hline
\hline
\end{tabular}

\begin{centering}
 $0.3<W_{2796} \le 1.0$\AA $ $ \\
 \end{centering}

 \begin{tabular}{|| c | c | c | c | c | c | c | c | c | c | c | c | c | c | c | c | c  ||}
 \hline
2.54 & 2.52 & 2.00-2.99 & 2.00-2.99 & 0.77 & 0.72 &$ $ 4$ $ &$ $ 3              &    5.03     &   4.27        & 0.97$\pm$0.49 & 0.77$\pm$0.44 & 0.30$\pm$0.15 				& 0.23$\pm$0.13  & 0.18 & 0.17 \\

3.42 & 3.41 & 3.00-4.00 & 3.00-3.94 & 0.37 & 0.35 &$ $ 3$ $ &$ $ 3              &    3.81     &   4.29        & 0.35$\pm$0.30 & 0.38$\pm$0.31 & 0.09$\pm$0.08 				& 0.10$\pm$0.08  &  &\\

4.75 & 4.77 & 4.01-5.45 & 4.03-5.45 & 0.77 & 0.81 &$ $ 4$ $ &$ $ 4              &    4.66     &   5.46        & 0.62$\pm$0.33 & 0.66$\pm$0.36 & 0.15$\pm$0.08 				& 0.18$\pm$0.08  &  &\\

\hline
\hline
\end{tabular}

\begin{centering}
 $1.0< W_{2796} \le 4.0$\AA $ $ \\
\end{centering}

 \begin{tabular}{|| c | c | c | c | c | c | c | c | c | c | c | c | c | c | c | c | c  ||}
 \hline
2.49 & 2.49 & 2.00-2.99 & 2.00-2.99 & 0.33 & 0.33 & $ $ 3 $ $ & $ $ 3                 &     3.32    &      3.37     & 0.27$\pm$0.26 & 0.28$\pm$0.15 & 0.09$\pm$0.05 			 	& 0.08$\pm$0.05 & 0.08 & 0.08 \\
3.43 & 3.41 & 3.00-4.00 & 3.00-3.91 & 1.00 & 1.00 & $ $ 1 $ $ & $ $ 1                   &      1.05   &      1.06     & 0.26$\pm$0.25 & 0.27$\pm$0.26 & 0.07$\pm$0.06 				& 0.07$\pm$0.06  &  & \\
4.75 & 4.77 & 4.01-5.45 & 4.03-5.45 & 0.66 & 0.67 & $ $ 3 $ $ & $ $ 3                   &       3.00   &      3.02     & 0.34$\pm$0.24 & 0.35$\pm$0.25 & 0.08$\pm$0.06 				& 0.08$\pm$0.06   &  & \\

\hline
\hline
 \end{tabular}

\begin{centering}
 $ W_{2796} \le 1.0$\AA $ $ \\
 \end{centering}

 \begin{tabular}{|| c | c | c | c | c | c | c | c | c | c | c | c | c | c | c | c | c  ||}
 \hline
2.49 & 2.51 & 2.00-2.99 & 2.00-2.99 & 0.58 & 0.62 &$ $ 9$ $ &$ $ 7              &    15.3     &   13.7        & 2.23$\pm$0.75 & 2.12$\pm$0.73 & 0.68$\pm$0.23 				& 0.65$\pm$0.22  & 0.67 & 0.51 \\
3.42 & 3.41 & 3.00-4.00 & 3.00-3.94 & 0.79 & 0.35 &$ $ 5$ $ &$ $ 3              &    11.5     &    4.3        & 2.27$\pm$0.75 & 0.38$\pm$0.31 & 0.60$\pm$0.20 				& 0.10$\pm$0.08  &  &\\
4.78 & 4.81 & 4.01-5.45 & 4.03-5.45 & 0.67 & 0.90 &$ $ 7$ $ &$ $ 6              &    26.6     &   21.6        & 3.08$\pm$0.73 & 3.34$\pm$0.76 & 0.72$\pm$0.17				& 0.78$\pm$0.18   &  &\\
\hline
\hline
\end{tabular}

\begin{centering}
 $ W_{2796} \le 4.0$\AA $ $ \\
 \end{centering}

 \begin{tabular}{|| c | c | c | c | c | c | c | c | c | c | c | c | c | c | c | c | c  ||}
 \hline
2.49 & 2.49 & 2.00-2.99 & 2.00-2.99 & 0.54 & 0.56 & 12    & 10                  &      18.6   &      17.1     & 2.50$\pm$0.79 & 2.40$\pm$0.77 & 0.76$\pm$0.24 			 	& 0.73$\pm$0.24 & 0.74 & 0.59 \\
3.43 & 3.41 & 3.00-4.00 & 3.00-3.91 & 0.81 & 0.48 & 6$ $ & 4$ $               &      12.5   &      5.35     & 2.53$\pm$0.80 & 0.64$\pm$0.40 & 0.67$\pm$0.21 				& 0.17$\pm$0.11  &  & \\
4.75 & 4.77 & 4.01-5.45 & 4.03-5.45 & 0.67 & 0.87 & 10    & 9$ $               &      29.6   &      24.6     & 3.42$\pm$0.77 & 3.69$\pm$0.80 & 0.80$\pm$0.18 				& 0.86$\pm$0.19   &  & \\

\hline
\hline

 \end{tabular}
\end{table*}

\begin{table}
 \caption{Best fit parameters with 1$\sigma$ errors to the \mgii absorber line density $dN/dz$=$N^*$(1+z)$^\beta$ }
 \label{tab:parameterTable}
 \begin{tabular}{|| c | c | c | c | c ||}
 \hline
 \hline
$\Delta W_{2796}$ (\AA) & $\Delta z$ & $\sigma$ & $N^*$ 		  & $\beta$ \\
 \hline
 \hline
0.0-0.3    & 2.00-5.45 & 3  & 0.33$\pm$0.08 & 1.15$\pm$0.15 \\

\hline
0.3-0.6    & 2.10-5.45 & 3  & 15.70$\pm$5.71 & -2.27$\pm$0.27 \\
0.3-0.6    & 2.10-5.45 & 5  &   1.63$\pm$0.32 & -0.70$\pm$0.14 \\

\hline
0.3-1.0    & 2.10-5.45 & 3  & 5.05$\pm$1.67 & -1.40$\pm$0.23 \\
0.3-1.0    & 2.10-5.45 & 5  &   1.05$\pm$0.37 & -0.37$\pm$0.35 \\

\hline
1.0-4.0    & 2.10-5.45 & 3  & 0.14$\pm$0.09 &  0.48$\pm$0.20 \\
1.0-4.0    & 2.10-5.45 & 5  & 0.14$\pm$0.09 &  0.48$\pm$0.20 \\

\hline
0.3-0.6$^a$  	& 1.9-6.3 &	& 0.728$\pm$0.688 & -0.362$\pm$0.624 \\
0.6-1.0$^a$   	& 1.9-6.3 &	& 0.092$\pm$0.071 & 0.803$\pm$0.503 \\
1.0+$^a$   	& 1.9-6.3 &	& 2.344$\pm$1.589 & -1.034$\pm$0.474 \\
 \hline
\hline

 \end{tabular}
 
$^a$ Parameter fits from \citep{CHEN2016} with 1$\sigma$ errors. 
\end{table}

 {We present, for the first time, the incidence rate of weak \mgii systems with $z$$>$2.5 in the top panel of Figure \ref{fig:dndz} with a data point from \citet{NARAYANAN2007} overplotted as a pink square. When we consider 5$\sigma$ recovery selected systems, we measure an incidence rate {$dN/dz$=1.35$\pm$0.58} at <$z$>=2.34 and find that by <$z$>=4.81 it almost {doubles} with {$dN/dz$=2.58$\pm$0.67}. We do not discover any weak \mgii systems {satisfying the} 5$\sigma$ selection criteria in the redshift range $z\in$ [3, 4]. }

{We do not isolate medium systems (0.6<$W_{2796}$$\le$1.0\AA$ $) as none were found below redshift 4 and only 1 was found in total. We combine this system with the intermediate systems (0.3<$W_{2796}$$\le$0.6\AA$ $) and find that the combined incidence rate of intermediate/medium systems (0.3<$W_{2796}$$\le$1.0\AA$ $) decreases with redshift, $\beta$=-0.37$\pm$0.35 and $N^*$=1.05$\pm$0.37. However, in our 3$\sigma$ selected sample, which includes one extra system in the lowest redshift bin, we find that their combined incidence rate also decreases with redshift, $\beta$=-1.40$\pm$0.23 and $N^*$=5.05$\pm$1.67. The discrepancy between the $\beta$ and $N^*$ best fit values and confidence intervals highlights the large impact a single system can have upon our statistics given the small sample size.}

{We find that the incidence rates of strong \mgii systems (bottom panel of Figure \ref{fig:dndz}) {increases} with redshift, a trend that is in disagreement with the MS12 and C16 results. Using our full sample of strong absorbers we find $\beta$=0.48$\pm$0.20 while C16 compute $\beta$=-1.034$\pm$0.474 for a similar redshift and equivalent width bin. However, C16 have 87 strong \mgii absorbers in their sample while we only have 7.  Thus, we attribute the difference to our smaller sample size.}

{As we discussed in subsection \ref{sec:s1306} which describes the absorbers in sightline $SDSS$ $J1306$+$0356$, C16 do not consider system 9 in their statistics while we and MS12 do. The false positive corrections described in section \ref{sec:falsepositive} should in principle account for such contaminations but, given that a single system can have a large impact on the statistical properties of our sample, we remove system 9 from our sample and recalculate the false positive correction ($F(dw, dz)$; see eq. \ref{eq:fprate}) and incidence rate of strong absorbers for our highest redshift bin. We find $F(dw, dz)$=0.44 and $dN/dz$=0.15$\pm$0.09 (light coloured diamond in bottom panel of Figure \ref{fig:dndz}).}

{Next, we recompute the best fits to the incidence rates of strong absorbers using the above $dN/dz$ value for our highest redshift bin and find $N^*$=1.10$\pm$0.43 and $\beta$=-1.06$\pm$0.27 (orange dash-dash line in bottom panel of Figure \ref{fig:dndz}). The newly computed $\beta$ value is in agreement with the C16 results ($\beta$=-1.034$\pm$0.474). Despite this agreement, we do not have any evidence to exclude system 9 from our sample of strong \mgii absorbers. Another point of difference between our sample and that of C16 is the relative number of absorbers found below z=4.345. C16 identified 87 strong absorbers  over the redshift range 1.947$\le$z$\le$5.350 and 72 of them ($\sim$83$\%$) are found below z=4.345. We identify a total of 7 strong absorbers with only 4 ($\sim$57$\%$) of them found below z=4.345. }

{Given these discrepancies, we see that the differences between the best fit parameters $N^*$ and $\beta$ for strong \mgii absorbers between our sample and that of C16 is most likely caused by the fact that we only investigate 4 lines of sight. C16 have 100 total lines of sight with 32 of them having an emission redshift $z_{em}\ge$ 5.79. Still, despite the limitations of our sample, the error bars of our incidence rates overlap with the error bars of C16 (see bottom panel of Figure \ref{fig:dndz}). }

 {We also investigate the evolution of the comoving incidence rate of \mgii system, $dN/dX$ where $dX$ is the absorption path.  In the $\Lambda$CDM cosmology adopted here, the absorption distance for a given redshift $z$ is defined as }

\begin{equation}
X(z)=\frac{2}{3\Omega_{M}} [\Omega_{M}(1+z)^{3}+\Omega_{\Lambda} ]^{1/2} \
	\label{eq:deltax}
\end{equation}

\noindent {thus for a redshift bin [$z_1$, $z_2$] with $z_2$>$z_1$, the absorption path between $z_2$ and $z_1$ is }

\begin{equation}
dX_{(z_1, z_2)}=X(z2)- X(z1)\
	\label{eq:deltaxdistance}
\end{equation}

\begin{figure}
	\includegraphics[width=8.5cm]{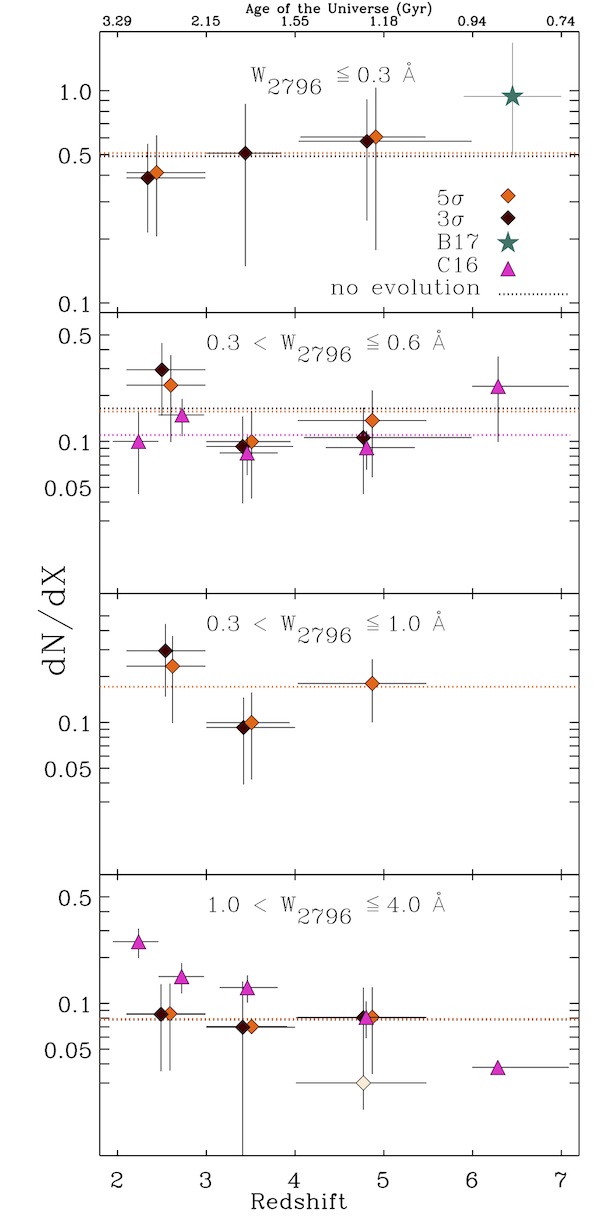}
 \caption{ {Comoving incidence rate of \mgii absorbers, $dN/dX$, separated in four $W_{2796}$ bins selected to highlight the evolution of weak, intermediate, intermediate/medium and strong systems. Symbols are as described in Figure \ref{fig:dndz}. {The mean values are shown to test the 'no evolution' hypothesis and are plotted with a dotted line with the same colours adopted for the data points}. The data points and associated mean values are in Table \ref{tab:fptable}.}}
 
 \label{fig:dndx}
\end{figure}

\noindent {and our results can be seen in Figure \ref{fig:dndx} and the numerical values are given in Table \ref{tab:fptable}. }

A flat behaviour in the evolution in $dN/dX$ represent no comoving evolution. In order to test this hypothesis, we compare the measured $dN/dX$ values to their mean, <$dN/dX$>, which is overplotted as a dashed line in each panel. We find a flat evolution of $dN/dX$ for {all} systems. {However, as we have previously discussed, the redshift evolution of strong systems is subject to limitations arising from the small sample size of our survey. For example, if we remove system 9 in sightline $SDSS$ $J1306$+$0356$ from our sample for the same reasons discussed above, we measure  $dN/dX$=0.03$\pm$0.01 (light coloured diamond in bottom panel of Figure \ref{fig:dndx}).  With this new value, we find a similar evolution in $dN/dX$ as  C16. Interestingly, systems with $W_{2796} \le$0.3\AA $ $ exhibit a flat evolution with $dN/dX$=0.41$\pm$0.18 at <$z$>=2.34 and  $dN/dX$=0.60$\pm$0.16 at <$z$>=4.81.}{ In the latter stages of preparing this manuscript, we became aware of \citet{BOSMAN2017} (B17 from here on) which reports on the search for \mgii absorbers in the redshift range 5.9$<$z$<$7. Their incidence rates\footnote{We have adjusted the B17 $dN/dX$ value to reflect the Plank cosmology used in this paper.} can be seen in the top panels of figures \ref{fig:dndz} and \ref{fig:dndx}. In particular, their incidence rate ($dN/dz$) is in very good agreement to the power law fit which best describes our 3$\sigma$ selected sample.}

\subsection{$d^{2}N/dzdW$}
\label{sec:d2N} 
 
 {We also investigate the equivalent width distribution}

\begin{equation}
 	 \frac{d^2 N}{dzdW}=\frac{N}{\Delta z \Delta W}\
	\label{eq:d2ndzdwequation}
\end{equation}

\noindent{{with associated Poisson error}}

 \begin{equation}
 	 \frac{d^2 N}{dzdW}=\frac{\sqrt{N}}{\Delta z \Delta W}\
	\label{eq:d2ndzdwerror}
\end{equation}

\noindent {Following previous studies we fit the functional form, }

 \begin{equation}
 	 \frac{d^2 N}{dzdW}=\frac{N^{\ast}}{W^{\ast}} $ $e^{(-W/W^{\ast}})\
	\label{eq:d2ndzdwequationfit}
\end{equation}

\begin{figure}
	\includegraphics[width=8.5cm]{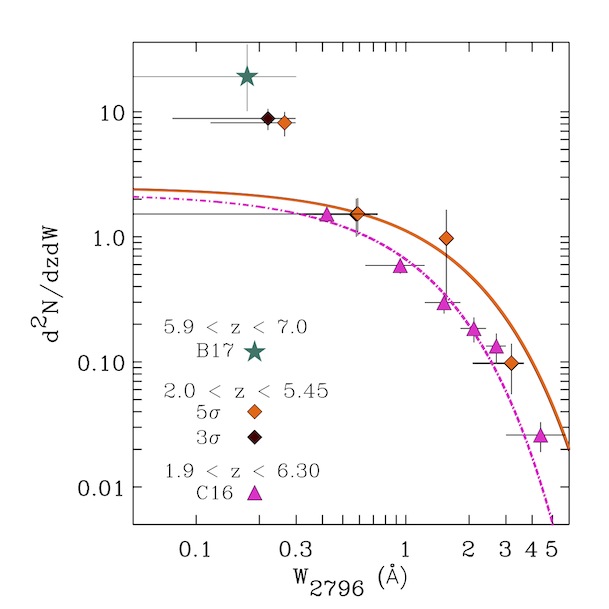}
 \caption{ {Incidence rate of \mgii absorbers, $d^{2}N/dzdW$, per unit redshift per unit equivalent width as measured in our study and those of C16 and B17. Symbols and line styles are as described in Figure \ref{fig:dndz}. The data points and best fit values are given in Table \ref{tab:d2ndzdwTable}.}}
 
 \label{fig:d2ndzdw}
\end{figure}

\noindent {{using a $\chi^2$ minimisation technique and the best fit parameters, along with the binned data points,} are presented in Table \ref{tab:d2ndzdwTable} and plotted in Figure \ref{fig:d2ndzdw}.}

\begin{table*}
 \caption{Incidence rate of \mgii absorbers, $d^{2}N/dzdW$, per unit redshift per unit equivalent width for our entire data set over the redshift range, 2<$z$<6.  <$W_{2796}$> is the median of the $\Delta W_{2796}$ equivalent width bin and $N^{\ast}$ and $W^{\ast}$ are the corresponding best fit parameter to the function $d^{2}N/dzdW$=($N^{\ast}/W^{\ast})*e^{(-W/W^{\ast})}$}
 \label{tab:d2ndzdwTable}
 \noindent\adjustbox{max width=\textwidth}{
 \begin{tabular}{|| c | c | c | c | c | c | c | c | c | c ||}
 \hline
 \hline
$<W_{2796}>$ & 		&$ \Delta$ $W_{2796}$& & $d^{2}N/dzdW$ & & $N^{\ast}$  &		& $W^{\ast}$ & \\ 

 \hline
 3$\sigma$ & 5$\sigma$ 	& 3$\sigma$ & 5$\sigma$	 & 3$\sigma$ & 5$\sigma$ & 3$\sigma$ & 5$\sigma$ 	& 3$\sigma$ & 5$\sigma$ 	 \\
\hline
\hline
0.22 	       &	0.26 		& 	0.077-0.298	& 	0.117-0.298	& 8.84$\pm$1.70	& 8.16$\pm$1.81       & - &	-  & - & -\\
0.58		&	0.59		& 	0.309-0.735	& 	0.308-0.735	& 1.51$\pm$0.51	& 1.53$\pm$0.51	         & 3.10$\pm$0.63	&	 3.10$\pm$0.64 & 1.24$\pm$0.48 & 1.25$\pm$0.49 \\
1.56		&	1.56 		& 	1.482-1.637	& 	1.482-1.637 	& 0.97$\pm$0.67	& 0.98$\pm$0.67	& 	&	 & & \\
3.18		&	3.18 		& 	2.082-3.655	& 	2.082-3.655 	& 0.10$\pm$0.04	& 0.10$\pm$0.04	& 	&	 & & \\

\hline
\hline
 \end{tabular}
 }
 \end{table*}


 {In contrast to the results of C16 (plotted in pink in Figure \ref{fig:d2ndzdw}) we are unable to fit the {full} $d^{2}N/dzdW$ distribution with a single exponential function. Instead, we find that we must limit our fitting range to systems with $W_{2796}$>0.3\AA$ $. Using our 5$\sigma$ selected sample we measure $N^{\ast}$=3.10$\pm$0.64 and $W^{\ast}$=1.25$\pm$0.49. This leads to an expectation of $\sim$2.01$_{-0.16}^{+0.29}$ absorbers at <$W_{2796}$>$\sim$0.26\AA$ $. We measure $d^2 N/dzdW$=8.16$\pm$1.81 at <$W_{2796}$>=0.26\AA$ $.}

 {An excess of absorbers with $W_{2796}$$\le$0.3\AA$ $ is also observed by  \citet{CHURCHILL1999} and \citet{NARAYANAN2007} in the redshift range 0.4$\le$z$\le$1.4 and 0.4$\le$z$\le$2.4 respectively. However, we can only highlight this excess of weak absorbers over the full redshift path of our sample as we can not further subdivide it into smaller redshift and equivalent width bins and maintain significant densities of absorbers.}

 {Alternatively, we could subdivide the equivalent width bin $W_{2796}$$\le$0.3\AA$ $ and investigate if a second function (exponential or power-law) is necessary to describe the distribution as found by \citet{CHURCHILL1999} and \citet{NESTOR2005}. However, we are unable to do so as we have only 6 absorbers in the narrow equivalent width range $\Delta W_{2796}$=[0.117-0.298] when we consider our 5$\sigma$ selected sample. }

 {Given these limitations, we can not comment on the evolution of the best fit parameters $N^{\ast}$ and $W^{\ast}$ as a function of redshift as done by MS12 and C16 nor can we investigate the best fit functional form of a secondary function required to explain the high number of  \mgii absorbers with $W_{2796} \le$0.3\AA}. Further observations of independent lines of sight with similar S/N and resolution as those presented here are required to investigate this population of weak \mgii absorbers. {B17 also discover an excess of weak \mgii absorbers when considering expectations from a functional fit to the distribution of \mgii absorbers with $W_{2796}$>0.3\AA.  They measure $d^2N/dzdW$=19.1$^{+15.5}_{-9.0}$ over the equivalent width range 0.05<$W_{2796}$<0.3\AA{} in the redshift range 5.9<$z$<7.0. }

\section{The comoving mass density of MgII, $\Omega_{\textrm{MgII}}$}
\label{sec:massdensity} 

The comoving mass density is defined as the {first moment} of the column density distribution function (CDDF) normalised to the critical density today and, for \mgii, it can be written as

\begin{equation}
 \Omega_{\mgii}=\frac{H_o m_{\mgii}} {c\rho_{crit}} \int Nf(N)dN \
	\label{eq:omegamg2integral}
\end{equation}

\noindent where $m_{\mgii}$ is the mass of a \mgii ion and $\rho_{crit}$=1.89$\times$10$^{-29}$h$^2$gcm$^{-3}$ and $f(N)$ is the CDDF. In practice, it is approximated as 

\begin{equation}
 \Omega_{\mgii}=\frac{H_o m_{\mgii}} {c\rho_{crit}}  \frac{1}{dX} \sum_{s} \sum_{\ddot{N}}\sum_{k} N(\mgii)\
	\label{eq:omegamg2}
\end{equation}

\noindent {where $s$ represents the number of sightlines with $\ddot{N}$ discovered systems, each with $k$ components with respective column density $N(\mgii)$. By combining eq. \ref{eq:omegamg2} with eq. \ref{eq:abc} we can calculate $\Omega_{\mgii}$ for the same equivalent width and redshift bin ($dW_i, dz_j$) }
\begin{equation}
 \Omega_{\mgii}(dW_i, dz_j)=\frac{A(dW_i, dz_j)}{dX_j} \frac{H_o m_{\mgii}}{c\rho_{crit}} \sum_{s} \sum_{\ddot{N}}\sum_{k} N(\mgii)\
	\label{eq:omegamg2final}
\end{equation}

\noindent {and account for the variable completeness and false positive contamination described in sections \ref{sec:variablecompleteness} and \ref{sec:falsepositive} respectively.}

\begin{figure}
	\includegraphics[width=8.5cm]{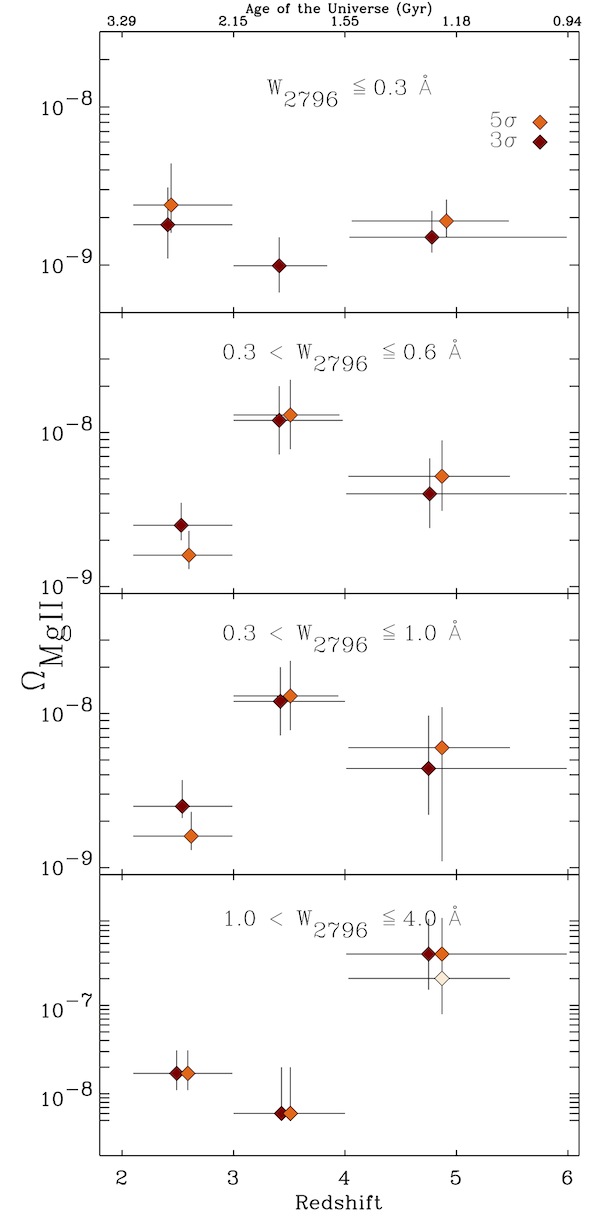}
 \caption{Evolution of the comoving mass density of $\mgii$, $\Omega_{\mgii}$, separated in five $W_{2796}$ bins. Symbols are as described in Figure \ref{fig:dndz}. The horizontal bars denote the $dz$ boundaries in Table \ref{tab:fptable}. {The light coloured diamond represents the computed value of $\Omega_{\mgii}$ as calculated by excluding system 9 in the SDSS J1306+0356 sightline as discussed in the text.}The values and associated errors for all data points are in Table \ref{tab:omegatable}. Note that the vertical axis ranges are different in each panel.}
  \label{fig:omega}
\end{figure}

 {In order to calculate the uncertainty of each $\Omega_{\mgii}$ measurement we bootstrap across all discovered systems. During this process, a discovered $\ddot{N}$ system can be selected multiple times or not selected at all. We iterate 1000 times across each {($dW_i, dz_j$)} bin. {The reported uncertainties denote the 66$\%$ confidence limits.} Since we discover only one strong system in the redshift window {$z$$\in$[3, 4] we simply consider the lower and upper bounds of the column density measurements of each associated component (system 4 in Table \ref{tab:u0148Table})}. The $\Omega_{\mgii}$ values and the error bars can be seen in Table \ref{tab:omegatable} and Figure \ref{fig:omega}. }
 
\begin{table}
 \caption{Comoving mass density of $\mgii$ ($\Omega_{\mgii}$) with 1$\sigma$ bootstrap errors and median redshift values (<$z$>) for 3 and 5$\sigma$ detection limits and $W_{2796}$ bins. A dash denotes no detections in the bin at the specified $\sigma$ level. Note that median redshift values are different for individual W$_{2796}$ bins reflecting the varying completeness levels of those bins (see Figures \ref{fig:completenesstest} and \ref{fig:completenesstest3sigma}).}
 \label{tab:omegatable}
 \noindent\adjustbox{max width=\textwidth}{
 \begin{tabular}{|| c | c | c | c ||}
 \hline
 \hline
 $<$z$>$ & 		 & $\Omega_{\mgii}$ &    \\

 \hline
 3$\sigma$ & 5$\sigma$ & 3$\sigma$ & 5$\sigma$ \\

 \hline
\hline
  &	 		&$ W_{2796} \le 0.3$\AA & 		 		 \\
 \hline
 
 2.41 	& 2.34 		 	  &	 $1.9^{+1.2}_{-0.8}\times10^{-9}$ &           $2.4^{+2.0}_{-0.8}\times10^{-9}$ \\
\\
 3.41 	&-	       	 	  	  &     $9.9^{+5.1}_{-3.2}\times10^{-10}$ &          -\\
\\
 4.78 	& 4.81 	 	 	  &     $1.5^{+0.7}_{-0.2}\times10^{-9}$ &             $1.9^{+7.0}_{-0.4}\times10^{-9}$ \\
\\
\hline
 \hline
 
 &	 		&$0.3<W_{2796} \le 0.6$\AA & 		 		 \\
 \hline
 2.53 & 2.50 	 	 	 & $2.5^{+1.2}_{-0.4}\times10^{-9}$       &           $1.6^{+0.7}_{-0.3}\times10^{-9}$ \\
\\ 
 3.41 & 3.41 	 	 	 & $1.2^{+0.8}_{-0.5}\times10^{-8}$       &            $1.3^{+0.9}_{-0.5}\times10^{-8}$ \\
\\
 4.76 & 4.77 	 	 	 & $4.0^{+2.8}_{-1.6}\times10^{-9}$       &            $5.2^{+3.7}_{-3.0}\times10^{-9}$ \\
\\
\hline
\hline

  &	 		&$0.3<W_{2796} \le 1.0$\AA & 		 		 \\
 \hline
 2.53 & 2.50 	 	 	 & $2.5^{+1.2}_{-0.4}\times10^{-9}$       &           $1.6^{+0.7}_{-0.3}\times10^{-9}$ \\
\\ 
 3.41 & 3.41 	 	 	 & $1.2^{+0.8}_{-0.5}\times10^{-8}$       &            $1.3^{+0.9}_{-0.5}\times10^{-8}$ \\
\\
 4.75 & 4.77 	 	 	 & $4.4^{+5.3}_{-2.2}\times10^{-9}$       &            $5.9^{+5.6}_{-4.8}\times10^{-9}$ \\
\\

 \hline
\hline 

  &	 		&$ W_{2796} \le 1$\AA & 		 		 \\
  \hline
  
 2.49 	& 2.51 	 & $4.4^{+1.7}_{-1.0}\times10^{-9}$ & $4.0^{+1.9}_{-0.7}\times10^{-9}$ \\
\\ 
 3.42 	& 3.41 	 & $1.3^{+1.2}_{-0.6}\times10^{-8}$ & $1.3^{+3.3}_{-0.4}\times10^{-8}$ \\
\\
 4.78 	& 4.81   & $6.0^{+4.4}_{-2.3}\times10^{-9}$ & $7.8^{+4.3}_{-6.0}\times10^{-9}$ \\
\\
\hline
\hline

 &	 		&$1.0< W_{2796} \le 4.0$\AA & 		 	 \\
 \hline
 
 2.49 & 2.49 	 	 	 & 	 $1.7^{+1.4}_{-0.6}\times10^{-8}$ &             $1.7^{+1.4}_{-0.6}\times10^{-8}$ \\
\\

 3.43 & 3.41 	 	 	 & 	 $6.0^{+14.4}_{-1.0}\times10^{-9}$ &            $6.0^{+14.4}_{-0.9}\times10^{-9}$ \\
\\

 4.75 & 4.77 	 	 	 & 	 $3.8^{+5.8}_{-1.3}\times10^{-7}$ &            $3.8^{+5.9}_{-1.3}\times10^{-7}$ \\
\\

 \hline
 \hline

 &	 		&$ W_{2796} \le 4.0$\AA & 		 	 \\
 \hline
 2.49 & 2.48 	 	 	 & 	 $2.1^{+3.1}_{-0.9}\times10^{-8}$ &             $2.1^{+6.3}_{-0.6}\times10^{-8}$ \\
\\

 3.42 & 3.41 	 	 	 & 	 $1.9^{+1.1}_{-0.6}\times10^{-8}$ &             $1.9^{+2.9}_{-0.2}\times10^{-8}$ \\
\\

 4.76 & 4.77 	 	 	 & 	 $3.8^{+6.2}_{-2.3}\times10^{-7}$ &             $3.9^{+7.1}_{-2.4}\times10^{-7}$ \\
\\

\hline
\hline

 \end{tabular}
}
\end{table}

\begin{figure}
	\includegraphics[width=8.5cm]{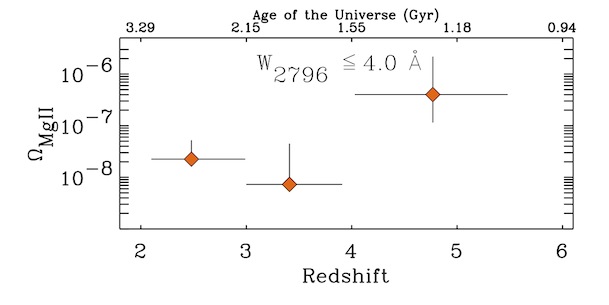}
 \caption{{Evolution of the comoving mass density of $\mgii$, $\Omega_{\mgii}$, as calculated when applying the 5$\sigma$ selection criteria for all systems. The horizontal bars denote the $dz$ boundaries in the last entry of  Table \ref{tab:fptable}. The values and associated errors can be seen in the last entry of Table \ref{tab:omegatable}.}}
 \label{fig:omega1ang}
\end{figure}

 {We observe a flat evolution in $\Omega_{\mgii}$ when we consider systems with $W_{2796} \le$1\AA$ $ however, when we consider strong systems, we observe an increase of over an order of magnitude from $\Omega_{\mgii}$=2.1$^{+6.3}_{-0.6}\times10^{-8}$ at <$z$>=2.49 to $\Omega_{\mgii}$=3.9$^{+7.1}_{-2.4}\times10^{-7}$ at <$z$>=4.77. Given that these strong systems contribute the largest fraction to the total $\Omega_{\mgii}$ budget, this trend continues when we consider all systems (see Figure \ref{fig:omega1ang} and last entry in Table \ref{tab:omegatable}). In order to investigate the nature of the absorbers, we discuss the comoving incidence rates ($dN/dX$) and the comoving mass density $\Omega_{\mgii}$ values in section \ref{sec:discussion}.}

\section{Discussion}
\label{sec:discussion}

We are able to measure, for the first time, $\Omega_{\mgii}$ in the redshift range 2<$z$<5.45. We implement a recovery rate selection {and} adjust for the impact of user doublet selection, variable completeness and false positive contamination as a function of wavelength. This is especially important in the NIR where telluric absorption and OH sky line emission lines have a significant impact on the signal to noise profile. The fine binning of the completeness function (see eq. \ref{eq:recovery} and Figures \ref{fig:completenesstest}, \ref{fig:completenesstest3sigma}) allows for the implementation of a minimum recovery rate cutoff (50$\%$; see eq. \ref{eq:rwz}) which isolates all possible wavelength locations where absorbers can be found and the total redshift path over which absorbers can be identified.

The computation of high resolution recovery maps for each sightline ($dW_{2796}$=0.01\AA, $dz$=0.01) is critical in calculating the incidence rates of \mgii absorbers ($dN/dz$ and $dN/dX$), their equivalent width distribution in a redshift bin ($d^2N/dzdW$) and finally, their comoving mass density ($\Omega_{\mgii}$).

Our study is the first to measure $\Omega_{\mgii}$ beyond redshift 2 and our lowest redshift bin value (2.1$^{+3.1}_{-0.6} \times$10$^{-8}$ at <$z$>=2.48) is in good agreement with the results of \citealt{MATHES2017} (1.4$\pm$0.2$\times$10$^{-8}$ at <$z$>=2.1). The incidence rates and nature of \mgii absorbers ($W_{2796}$>0.3\AA) in the redshift range 2<$z$<7.08 have also been studied by MS12, MS13, and C16  {and we find that our calculated incidence rates overlap with those measured by these previous studies. We also present, for the first time, the incidence rate of weak  \mgii absorbers ($W_{2796} \le$0.3\AA) in the redshift range 2<$z$<5.45. Our highest redshift absorber is system 8 in sightline $SDSS$ $U0148+0600$ with {z=4.89031$\pm$4$\times$10$^{-5}$}. }

\subsection{The evolution of $\Omega_{\textrm{MgII}}$}
\label{sec:omegadiscussion} 

{Intervening \mgii provides a powerful and independent probe of metal enriched gas across a large range of redshift space.} The large oscillator strength and doublet ratio allows for a straightforward identification technique which is well understood. The large rest frame separation ($\Delta\lambda$ $\sim$1580\AA) between $Ly\alpha$ and \mgii emission of the QSO allows for a very large redshift path to be considered in each sightline {unimpeded by the $Ly\alpha$ forest}. \mgii is also an important {probe of low ionisation gas}. It has an ionisation energy of 1.1 Ryd and has been established as a tracer of \ion{H}{i} in the redshift range 0.5$\le$z$\le$1.4 \citep{MENARDCHELOUCH2009}.

Furthermore, both MS13 and \citet{BERG2016} find that the fraction of \mgii absorbers associated with DLAs increases with redshift. MS13 find that at the mean redshift of their sample (<$z$>$\simeq$3.402) the ratio of \mgii absorbers that are DLAs is $\sim$40.7$\%$ which is up from 16.7$\%$ at <$z$>=0.927. \citet{BERG2016} also find that the ratio of DLAs associated with \mgii systems with $W_{2796}$~$\le$0.6\ang increases by a factor of 5 from z $\sim$1 to z $\sim$4. These findings suggest that, with increasing redshift, \mgii systems with $W_{2796}$~$\le$0.6\ang become even more important in tracking cool neutral gas.

 We first compare the evolution of weak systems ($W_{2796} \le$0.3\AA) with that of intermediate/medium systems (0.3<$W_{2796} \le$1\AA). We find that weak systems {contribute a significant fraction to} the \mgii budget for systems with $W_{2796} \le$1\AA$ $. {When we consider a 5$\sigma$ selection criteria, they contribute $\sim$60$\%$ at <$z$>=2.34 but their contribution decreases to $\sim$24$\%$ by <$z$>=4.81.}

  {Of particular interest is the increase in $\Omega_{\mgii}$ towards z$\sim$5 {present} in $W_{2796}$ bins with $W_{2796}$>0.3\AA $ $ (see Figure \ref{fig:omega}). In order to quantify this evolution we test two hypotheses. We first test the hypothesis that systems do not evolve with redshift by {calculating the $\chi^2$ between {$\Omega_{\mgii}$ redshift bin values} and their mean. Secondly, we test the hypothesis that {$\Omega_{\mgii}$ bin values} are best described by a linearly evolving redshift dependent function using a $\chi^2$ minimisation technique. }}

 {We find that the 3$\sigma$ weak systems are described {well by their mean ($\chi^2$=1.6) while intermediate/medium systems are not described well by either hypothesis}. Given the ambiguity in interpreting the evolution of $\Omega_{\mgii}$ from systems with $W_{2796} \le$1\AA$ $ and that weak systems account for a significant fraction of $\ion{Mg}{II}$, we {next consider the comoving incidence rates,} $dN/dX$, and $\Omega_{\mgii}$ for all systems with $W_{2796} \le$1.0\ang and test the above hypotheses. {We find that both hypotheses are rejected}. When considering 5$\sigma$ completeness selected systems with $W_{2796} \le$1\AA, we find that $\Omega_{\mgii}$ increases from 4.0$^{+1.9}_{-0.7}\times$10$^{-9}$ at <$z$>=2.51 to 7.8$^{+4.3}_{-6.0}\times$10$^{-9}$ at <$z$>=4.81 while their comoving incidence rates ($dN/dX$) also increases from 0.61$\pm$0.23 at <$z$>=2.49 to 0.78$\pm$0.18 at <$z$>=4.81. In order to reject a hypothesis, we require a p-value$\le$0.01 which corresponds to $\chi^2$ $\sim$9.21 for $N_{DOF}$=2.}

 {For strong systems, we again find {that their $\Omega_{\mgii}$ binned values} are not described well by either hypothesis with the {middle} redshift bin (<$z$>=3.41) contributing to the discrepancy. It is in this redshift bin where we discover only one strong and no medium systems. {Another} important concern for the $\Omega_{\mgii}$ measurements associated with these strong systems is that measuring the column density of saturated components {may introduce} a large uncertainty which can not be properly accounted for if only fitting the individual \mgii systems. However, we have performed full spectra fits which account for other associated metal transitions (\ion{Mg}{I}, \ion{Fe}{II}, \ion{Ca}{II}, \ion{Si}{II}, \ion{Al}{II} and \ion{C}{IV}) and have confidence that our bootstrap errors reflect the true uncertainty.}

{Finally, we consider all of the identified \mgii systems in each redshift bin (see last entry in Table \ref{tab:omegatable}) and find that both hypotheses are rejected. However, we again find that $\Omega_{\mgii}$ increases from 2.1$^{+6.3}_{-0.6}\times$10$^{-8}$ at <$z$>=2.48 to 3.9$^{+7.1}_{-2.4}\times$10$^{-7}$ at <$z$>=4.77 (see Figure \ref{fig:omega1ang}) when we consider the 5$\sigma$ selected systems. This order of magnitude increase is in disagreement to the comoving incidence rate ($dN/dX$) evolution which is best described  by their mean. The reverse trends between the evolution of the comoving incidence rate of all \mgii absorbers and their associated $\Omega_{\mgii}$ values suggest an increasing metalicity of \mgii absorbers with redshift and/or that these same absorbers track different galaxy populations at different redshifts. In order to further investigate this trend, we next consider expected statistics from correlating \mgii absorbers with known galaxy populations.}

\subsection{Expectations from known galaxy populations}
\label{sec:expectations} 

 {If \mgii absorption traces enriched CGM gas around galaxies, as is generally thought, we can use a redshift dependent galaxy luminosity function paired with gas halo size and covering fraction to build an expectation for the comoving incidence rate $dN/dX$. MS12 explore two scenarios to account for their $dN/dX$ statistics when considering their full sample or strong systems only. They explore if $dN/dX$ statistics are best fit when considering galaxies whose halo size track a fixed luminosity with redshift or a luminosity which evolves with redshift. They find that a scenario where galaxies with fixed halo size track a fixed luminosity best fits the data but can not reject the second hypothesis. We focus on whether the highest redshift incidence rate values observed {(including the $weak$ systems) can be explained through a one to one association with galaxy halos at redshift $\sim$5.} }

As galaxies grow with time, we consider the case that a galaxy halo scales with a redshift dependent characteristic luminosity.  {For the same halo, we assume that the covering fraction of \mgii does not evolve with redshift. In order to estimate the gas halo size we use a Holmberg-like scaling}

\begin{equation}
   R(L_{Bmin}, W_{2796})=R_{W_{2796}}^{\ast}\bigg( \frac{L_{Bmin}}{L_{\ast}}\bigg)^{\beta_{W_{2796}}}\
	\label{eq:rlb}
\end{equation}

\noindent {and consider the results of \citet{NIELSEN2013b} (their Table 3; K-band luminosity scaled halo absorption radii) for the $R_{W_{2796}}^{\ast}$ and $\beta_{W_{2796}}$ parameters with the caveat that the gas halo sizes and \mgii covering fractions ($f_{R}(L)$) presented in their "\mgii Absorber-Galaxy Catalog" (MA\ion{G}{ii}CAT) are based on 182 intermediate redshift (0.072$\le$z$\le$1.w120) galaxy-absorber pairs. The MA\ion{G}{ii}CAT catalog has the added benefit that the scalings are separated by the equivalent width of the \mgii absorber, $W_{2796}$. This allows us to calculate the cross section presented by a \mgii gas halo associated with a galaxy based on the strength of the $W_{2796}$ feature {and the limiting luminosity of the galaxy population considered ($L_{Bmin}$)},}
 
\begin{equation}
  \sigma_{\mgii}=f_{R}(L) \pi  R(L_{Bmin}, W_{2796})^2 \  
   	\label{eq:crosssection}
\end{equation}
 
{We then consider the total number of galaxies per Mpc$^{3}$ expected at a redshift $z$,}

\begin{equation}
   N(L_{Bmin}, z)=  \int_{L_{Bmin}}^\infty \phi(L_{\ast}, z) \  
   	\label{eq:nlbz}
\end{equation}

\noindent {where $\phi(L_B, z)$ is the B-band luminosity function at redshift $z$ and $L_{Bmin}$ is the minimum luminosity considered. We then combine the expected cross section of a \mgii gas halo associated with a galaxy of minimum luminosity ratio ($L_{Bmin}$/$L_{\ast}$) with the expected total number of galaxies per Mpc$^3$ with the same minimum luminosity ratio and calculate the expected comoving incidence rate in a similar fashion as \citet{CHURCHILL1999} and MS12}

\begin{equation}
  \frac{dN}{dX}=  \frac{c\pi}{H_0} f_{R}(L) R^{\ast2}_{W_{2796}}  \phi^{\ast}(z) \Gamma(\alpha+2\beta_{W_{2796}}+1, \frac{L_{Bmin}}{L_{\ast}})\     
  	\label{eq:dndxfromg}
\end{equation}

\begin{figure}

	\includegraphics[width=8.5cm]{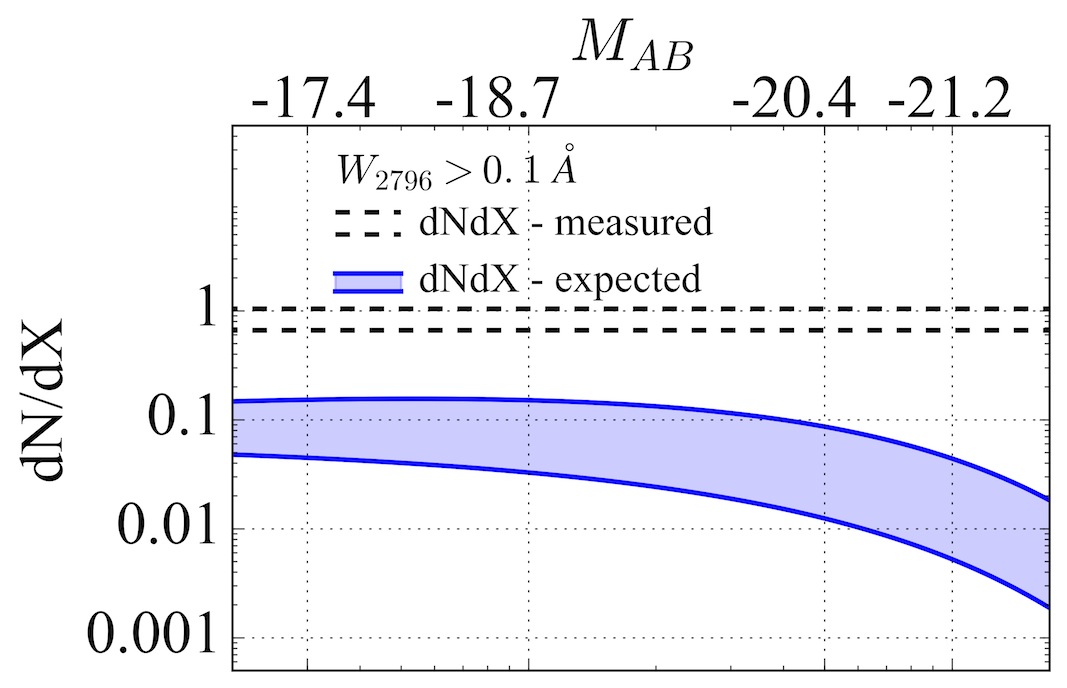}
	\includegraphics[width=8.5cm]{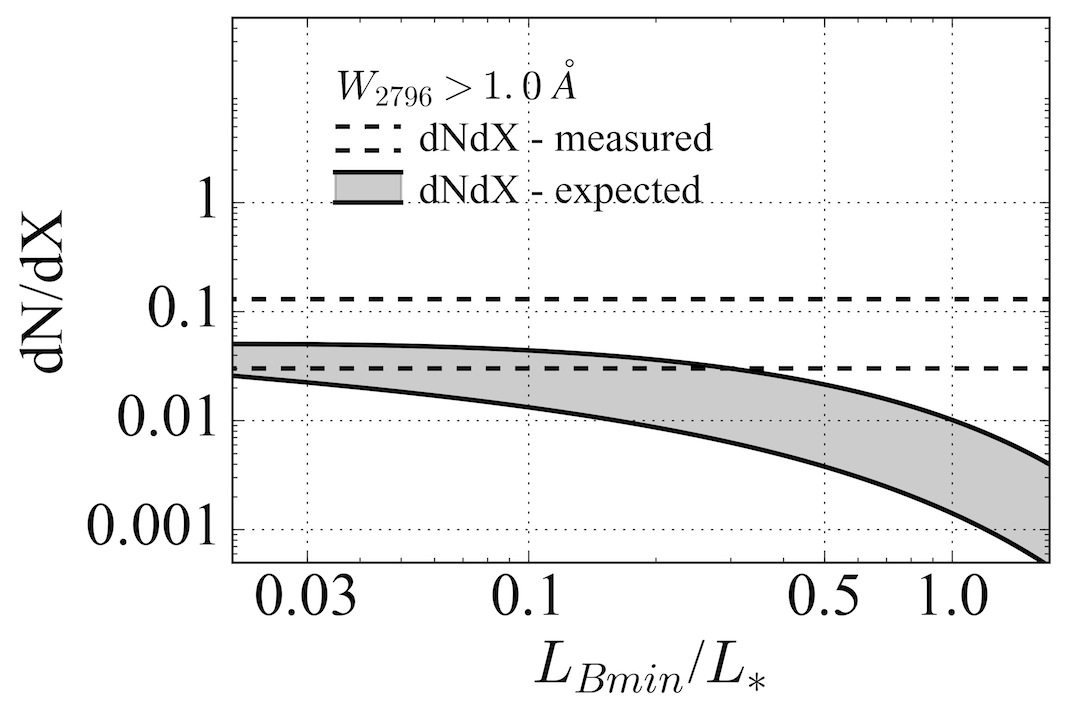}
    \caption{{Expectations of incidence rates as calculated using the luminosity function provided by \citet{MASON2016} for z $\sim$5 ($\alpha$=-1.75$\pm$0.13; $M^{\ast}$=-21.2$\pm$0.2; log($\phi^{\ast}$)=-3.12$^{-0.15}_{+0.24}$[mag$^{-1}$Mpc$^{-3}$]) with the luminosity scaled absorption radii provided by \citet{NIELSEN2013b} ($W_{cut}$=0.1\AA, $R^{\ast}_{W_{2796}}$=75$^{+40}_{-6}$ kpc, $\beta$=0.23$^{+0.01}_{-0.01}$ and $f_{R}(L)$=0.84$^{+0.04}_{-0.04}$; $W_{cut}$=1.0\AA, $R^{\ast}_{W_{2796}}$=62$^{+21}_{-1}$ kpc, $\beta$=0.20$^{+0.01}_{-0.03}$ and $f_{R}(L)$=0.34$^{+0.06}_{-0.05}$). The solid lines denote the boundaries of the $expected$ incidence rates as calculated using eq. \ref{eq:dndxfromg} while the dashed lines represent the $measured$ incidence rate (Table \ref{tab:fptable}).}}

\label{fig:dndxexpected}
\end{figure}

\noindent {where $c$ is the speed of light, $f_{R}(L)$ is the covering fraction, $\Gamma(a, b)$ is the upper incomplete gamma function and $\phi^{\ast}(z)$ along with $\alpha$ are the \citet{SCHECHTER1976} function parameters for the respective redshift dependent luminosity function. We evaluate eq. \ref{eq:dndxfromg} with the luminosity function provided by \citet{MASON2016} for z $\sim$5 and compare to our highest redshift bin $dN/dX$ 5$\sigma$ values (<$z$>=4.77 for systems with $W_{2796}$>0.1\ang and $W_{2796}$>1\AA$ $). The results are presented in Figure \ref{fig:dndxexpected}.}

{When we consider all systems ($W_{2796}$>0.1\AA; top panel in Figure \ref{fig:dndxexpected}), we find that we can not reproduce the measured incidence rate ($dN/dX$=0.86$\pm$0.19 at <$z$>=4.77; Table \ref{tab:fptable}). If we consider all galaxies down to the limiting magnitude over which the luminosity function is defined ($M_{AB} \simeq$-17.5), we calculate $dN/dX_{expected}$=0.05$^{+0.10}_{-0.01}$. Thus, we measure a comoving incidence rate for \mgii absorbers $\simeq$7 to 17 times higher than the expected value if a single absorber is associated with a single galaxy with $M_{AB} \le$-17.5.}

{The expected incidence rate (eq. \ref{eq:dndxfromg}) is a function of covering fraction ($f_{R}(L)$), gas halo size (eq. \ref{eq:rlb}) and the total number of galaxies considered (eq. \ref{eq:nlbz}) which itself depends on the luminosity function \citep{MASON2016}. When considering all systems with $W_{2796}$>0.1\AA{}, \citet{NIELSEN2013b} compute a covering fraction $f_{R}(L)$=0.85$^{+0.03}_{-0.04}$ thus, an increase in covering fraction only (which is already close to unity) can not connect the measured $dN/dX$ to $dN/dX_{expected}$. Given this, the measured incidence rate of \mgii absorbers with $W_{2796}$>0.1\AA{} could be explained by either an increase in the characteristic gas halo size ($R^{\ast}_{W_{2796}}$), a sharper increase in the gas halo size (ie. larger $\beta_{W_{2796}}$), a steeper slope for the luminosity function ($\alpha$), lower limit of integration for the luminosity function or a lower $L_{\ast}$ value (ie. the luminosity associated with a fixed halo size evolves with redshift) or a combination of all of the above. Identifying the $absorption$ $halo$ properties of \mgii absorbers is outside the scope of this paper, thus we can only highlight that an evolving covering fraction can not $alone$ bridge the gap between the measured incidence rates of systems with $W_{2796}$>0.1\AA{} and $dN/dX_{expected}$.}

{When we consider strong systems ($W_{2796}$>1.0\AA; bottom panel in Figure \ref{fig:dndxexpected}) we find that, if the physical properties of strong absorber haloes do not evolve with redshift, then we must integrate down to $L_{Bmin}/L_{\ast} \simeq$0.29 in order for the computed $dN/dX_{expected}$ values to match the boundary of the measured comoving incidence rate of strong \mgii absorbers ($dN/dX$=0.08$\pm$0.06; Table \ref{tab:fptable}). However, if strong \mgii absorbers are associated with galaxies with $L \ge 0.5L_{\ast}$ as suggested by \citet{SEYFFERT2013} we can then calculate their physical cross section as}

\begin{equation}
  \sigma_{phys}=\frac{H_0}{c}  \frac{1}{N(L_{Bmin, z})} \frac{dN}{dX} \bigg|_{measured} \  
  \label{eq:sphys}
\end{equation}

\noindent {We calculate $\sigma_{phys} \simeq$0.041 $Mpc^2$ at <$z$>=4.77.}

{One point of difference between the strong systems used by \citep{SEYFFERT2013} and our sample is the equivalent width considered. They do not include strong systems with $W_{2796}$>3\AA{} as they could be biased tracers of intragroup gas rather than be associated with the enriched halo of a single galaxy.  With these consideration, we find that $\sigma_{phys}$ of strong absorbers increases from 0.015 $Mpc^2$ at $z$=2.3 \citep{SEYFFERT2013} to 0.041 $Mpc^2$ at <$z$>=4.77 if they continue to be associated with galaxies with $L$$\ge$0.5$L_{\ast}$. However, an increasing covering fraction or association with lower luminosity galaxies ($L$<0.5$L_{\ast}$) can also explain the incidence rates of strong \mgii systems.}

\subsection{The evolution of high vs. low ionisation absorption systems}
\label{sec:lowvshigh} 

Absorption line systems in the spectra of QSOs offer a luminosity independent view of the universe and are generally separated into $low$ (e.g. \ion{C}{ii}, \ion{Mg}{ii}) and $high$ (e.g. \ion{C}{iv}, \ion{Si}{iv}) ionisation systems which probe the ionisation and enrichment of the universe in different temperature and density environments. In order to build a full picture of the evolution of metals it is then useful to compare and contrast the evolution of these systems across cosmic time.

Observations of $high$ ionisation systems \citep{RYANWEBER2009, BECKER2009, SIMCOE2011, DODORICO2013} report a drop in $\Omega_{\ion{C}{iv}}$ {by factors between} 2 to 4 beyond redshift 5. In order to investigate if all of their identified \ion{C}{iv} systems are tracing high temperature low density gas associated with the IGM, \citet{DODORICO2013} also search their sight lines for the presence of \ion{C}{ii} and \ion{Si}{iv}. They compare their observed z $\sim$5 ratios of \ion{Si}{iv}/\ion{C}{iv} and \ion{C}{ii}/\ion{C}{iv} with the z $\sim$3 results of \citet{BOKSENBERG2015} and conclude that they are drawn from different parent populations. By extending their comparison to a set of \textit{CLOUDY} ionisation models, \citet{DODORICO2013} also conclude that some of their \ion{C}{iv} absorbers trace a very dense and neutral environment where the ionising photons are provided by local sources rather than the global UV background.

As we move towards the epoch of reionisation, the universe becomes more neutral. Thus, $low$ $ionisation$ systems which probe a similar ionisation energy ( $\sim$1Ryd), are critical in probing the reionisation history of hydrogen. \citet{FINLATOR2015, FINLATOR2016} and \citet{GARCIA2017} compare their simulations with the observations of \citet{BECKER2006, BECKER2011} which suggest that \ion{C}{II} is more abundant than \ion{C}{IV} by z $\sim$6. {They find that this can be explained by either an increase in the proper density of metal enriched regions, their associated \ion{C}{II} mass fraction or a decrease in the CGM metallicity}. {Also, as \citet{BECKER2006} noted, the increase in the cross section of $low$ $ionisation$ can be driven by the increase in the radius out to which haloes can become self-shielded as the global UVB decreases towards redshift $\sim$6.} We investigate if our observations of \mgii at z $\sim$4.77 can provide further insight, especially as \mgii has already been connected to \ion{H}{I} \citep{RAO2006} at lower redshifts.

Absorption systems in the spectra of high redshift QSOs are important tracers of cool neutral gas especially as current 21-cm surveys which directly track \ion{H}{i} can only probe up to z $\sim$0.25 (\citealt{CATINELLA2015}). The latest measurements of $\Omega_{\ion{H}{i}}$ as traced by DLAs identified in the spectra of hundreds of QSOs by \citet{SANCHEZRAMIREZ2016} and \citet{CRIGHTON2015} extend the evolution of $\Omega_{\ion{H}{i}}$ to redshift z $\sim$5. {Remarkably, they find a fairly constant value for $\Omega_{\ion{H}{i}}$ from redshift 5 to 3}. However, even such studies are unable to comment on the evolution of $\Omega_{\ion{H}{i}}$ past redshift 5 as the effective optical depth of \ion{H}{i} leads to the presence of large absorption features known as Gunn-Peterson troughs \citep{GUNNPETERSON1965}.

{In order to compare $\Omega_{\mgii}$ with $\Omega_{\ion{H}{I}}$ we calculate the expected $\Omega_{\mgii}$ assuming that \mgii traces cold, neutral gas as it does at lower redshifts \citet{RAO2006}. We combine the mean metallicity at z $\simeq$4.85 (<$Z$>=-2.03$^{+0.09}_{-0.11}$) as measured by \citet{RAFELSKI2014} with the comoving mass density of \ion{H}{I} at <$z$>=4.9 ($\Omega_{\ion{H}{I}}$=0.98$^{+0.20}_{-0.18}\times$ 10$^{-3}$) as measured by \citet{CRIGHTON2015} and calculate }

\begin{equation}
 \Omega_{\mgii}^{expected}=\Omega_{\ion{H}{I}} \times \frac{m_{\mgii}} {m_{\ion{H}{I}}} \times[Mg/H]_{\sun} \times10^{<Z>} \
 	\label{eq:omegaexpectedfromh1}
\end{equation}

\noindent {where $[Mg/H]_{\sun}$ is the solar abundance ratio of Mg to \ion{H}{I} (10$^{7.6}$/10$^{12}$). We obtain $\Omega_{\mgii}^{expected}$=8.85$^{+2.25}_{-3.25}\times$10$^{-9}$ while {we measure $\Omega_{\mgii}=3.9_{-2.4}^{+7.1}\times10^{-7}$} at <$z$>=4.77. {There is between {$\sim$44 times} more \mgii gas} than expected from the comoving \ion{H}{I} mass density and the global DLA metallicity. At face value, this suggests that $\Omega_{\mgii}$ (which is dominated by strong systems) is associated with both ionised and neutral gas perhaps indicating an increased link between \mgii absorbers and Lyman-limit systems at higher redshifts. Alternatively, these strong \mgii absorbers have a higher metallicity than typical DLAs at this redshift as suggested by MS13. We will investigate the metallicity of strong vs. weak systems in an upcoming paper. }

\citet{BECKER2006, BECKER2011} also provide observations of \ion{O}{I}, \ion{Si}{II} and \ion{C}{II} {in the redshift range 5.3<$z$<6.4}. They find a relatively high incidence rate for their $low$ $ionisation$ absorbers ($dN/dX$ $\simeq$$0.25^{+0.21}_{-0.13}$) which they point out is comparable to the total number density of the combined DLA and sub-DLA populations {at z $\sim$3}. This suggests that the incidence rate of $low$ $ionisation$ systems does not evolve in the redshift range 3<$z$<6, similarly to the evolution of weak \mgii absorbers as measured in this study. Our highest redshift bin covers this gap in discovery space. If we consider all of our \mgii systems, we find $dN/dX$=0.38$\pm$0.13 at <$z$>=4.77.

 {Given that \mgii systems probe a larger range of environments (ie. not all \mgii absorbers will have \ion{O}{I}, \ion{C}{II}, etc.) suggests that the number density of low ionisation systems is not evolving rapidly over the redshift range 3<$z$<6 with the caveat {that the cross-section of \mgii absorbers is most likely larger than that of \ion{O}{I} and \ion{C}{II} absorbers}. However, given that over the same redshift range $\Omega_{\ion{Mg}{II}}$ increases by a factor of $\sim$10, we see this as evidence that the evolution in $\Omega_{\ion{C}{IV}}$ is driven by a change in ionisation state rather than a rapid increase in the enrichment of the IGM.}
 
{Further studies could confirm if this is due to a bias in our sample or if indeed we are observing an increase in $\Omega_{\ion{Mg}{II}}$ towards redshift 5 and beyond. It is important to note that investigations of \ion{C}{IV} in this redshift range are in the optical range (below 1$\micron$) while observations of \mgii are pushed into the H and K bands for redshifts greater than 4. Thus, directly comparing the evolution of \ion{C}{IV} and \ion{Mg}{II} is complicated by systematics arising from the presence of telluric contamination in the NIR. Future space based NIR observations with the James Webb Space Telescope could remove such complications but, unfortunately, the instrumentation lacks the resolution necessary to identify the full family of \mgii absorbers.}

{Tracking the ionisation state and abundance of metals has proven itself a difficult task, in both observations and simulations. It must account for the complex interaction between the strength of the global UV background, cross section of absorbers and relative strengths of different ions which characterise the reionisation of the universe. In order to investigate possible systematics associated with different cosmological scale simulations and hydro-solvers, \citet{KEATING2016} investigate the evolution of both low and high ionisation systems and compare their findings with observations. It is encouraging that they conclude that at least \ion{O}{i} and \ion{C}{ii} systems are not influenced by choice of feedback schemes. Unfortunately, current cosmological simulations struggle to reproduce the incidence rates of \mgii at z $\sim$6 (C16). }

{This current lack of {spatial} resolution {in hydrodynamic simulations} is unfortunate as these $low$ $ionisation$ systems can be associated with the low-mass haloes needed to reionise the universe (M$_{lim}$>-17; \citealt{MASON2016}). These same systems contribute a significant amount of metals to the total metal budget {and are most affected by resolution effects in simulations}. Our findings suggest that within the first Gyr of the universe's history}, \mgii is already well established in the haloes of galaxies and that low mass haloes contribute an increasing amount to the total $\Omega_{\mgii}$ budget with increasing redshift. However, in order to establish this relationship, \mgii absorbers have to be investigated beyond z $\sim$6 and more sightlines have to be observed in the redshift range explored by this study.

\section{Conclusions}
\label{sec:conclusion}

We use VLT X-Shooter (VIS and NIR) to investigate four QSOs (z $\sim$6) for the presence of intervening absorption systems and we present our analysis of identified \mgii systems in the redshift range z $\simeq$ 2-6. We first {create a candidate list using an automated detection algorithm and then} visually {confirm} transitions and associated ions across the entire wavelength range using the $plotspec$ package\footnote{developed by \href{https://github.com/nhmc/plotspec/}{Dr. Neil Crighton} \\ \href{https://github.com/nhmc/plotspec/}{https://github.com/nhmc/plotspec/}} (see section \ref{sec:visid}). {We} fit each feature using $VpFIT$ 10.0. This allows us to measure the column density and doppler parameter of each component. In order to compare to previous works, we also measure the equivalent width of each \mgii $\lambda$2796 feature ($W_{2796}$) and associated velocity width ($\Delta v$) {(see figure \ref{fig:ewv})}.

In order to investigate the impact of sky lines on our ability to identify \mgii systems, we inject $\sim$30 million artificial systems and search for them using {the same automated detection algorithm} and consider the human bias introduced in the visual confirmation step (see section \ref{sec:completeness}). In order to identify the false positive rates, we relax the rest frame separation of the \mgii doublet $\lambda \lambda$2796 2803 and search for an artificial \ion{Mg'}{ii} doublet $\lambda \lambda$2796 2810 (see section \ref{sec:falsepositive}). We combine the results of the false positive analysis (see Table \ref{tab:fptable}) with high resolution completeness maps ($\Delta z$=0.01 $\Delta W_{2796}$=0.01\AA; see Figures \ref{fig:completenesstest} and \ref{fig:completenesstest3sigma}) in order to investigate the line statistics of the observed \mgii absorbers (see sections \ref{sec:linestats} and \ref{sec:d2N}) and to measure the comoving mass density of \mgii ($\Omega_{\mgii}$; see section \ref{sec:massdensity}). Our main findings are:

\begin{enumerate} 

\item {We visually identify 52 \mgii systems with {28} passing our 3$\sigma$ selection criteria and {24} passing our 5$\sigma$ selection criteria. The visually selected systems range in equivalent width range 0.021$\le$$W_{2796}$$\le$3.655\AA{} and redshift range z=2.00-5.92. {The highest redshift absorber which meets the 3$\sigma$ and 5$\sigma$ selection criteria is system 8 in sightline $SDSS$ $U0148+0600$ with {z=4.89031$\pm$4$\times$10$^{-5}$}. }}\\

\item {We identify {10} weak systems ($W_{2796}$$\le$0.3\AA) when we consider 3$\sigma$ selection criteria. We identify {6} weak systems when we consider 5$\sigma$ selection criteria with the weakest of these \mgii absorbers having W$_{2796}$=0.117$\pm$0.006. {The rejected systems (for both 3 and 5$\sigma$ considerations) are weak \mgii systems except for system 6 in sightline $SDSS J0927+2001$ which has $W_{2796}$=0.500$\pm$0.031.}} \\

\item {We measure the incidence rate $dN/dz$ in four equivalent width bins in order to highlight the evolution of weak ($W_{2796}$$\le$0.3\AA), intermediate (0.3<$W_{2796}$$\le$0.6\AA), intermediate/medium (0.3<$W_{2796}$$\le$1.0\AA ) and strong ($W_{2796}$>1\AA) systems. Our results can be seen in Figure \ref{fig:dndz} and Table \ref{tab:fptable}. {For weak systems we measure an {incidence rate $dN/dz$=1.35$\pm$0.58 at <$z$>=2.34 and find that it almost doubles with $dN/dz$=2.58$\pm$0.67} by <$z$>=4.81}. We fit each distribution with the functional form $dN/dz$=$N^*$(1+z)$^\beta$ and the parameter fits can be seen in Table \ref{tab:parameterTable}. At a 3$\sigma$ selection criteria, we find that the incidence rate of weak systems increases towards redshift $\sim$5 (ie. $\beta$>0). For {all} systems ($W_{2796}$<4\AA) we find that their incidence rate also increases towards redshift $\sim$5 {from 2.40$\pm$0.77 at <$z$>=2.49 to 3.69$\pm$0.80 at <$z$>=4.77.}} \\

\item {We measure the comoving incidence rate $dN/dX$ in the same four equivalent width bins and our results can be seen in Figure \ref{fig:dndx} and Table \ref{tab:fptable}. {We find that the incidence rates of systems with $W_{2796}$$\le$0.3\ang does not evolve with redshift with a mean $dN/dX$=0.51. For all systems ($W_{2796}$<4\AA), we find that their evolution is best described  by their mean. While the incidence rate values overlap with those presented in C16, we acknowledge that our survey path for strong systems is smaller.}} \\

\item {We find that the comoving incidence rate of all \mgii absorbers at <$z$>=4.77 ($dN/dX$=0.86$\pm$0.19) is comparable to incidence rate of the $low$ $ionisation$ absorbers \ion{O}{I}, \ion{Si}{II} and \ion{C}{II} presented by \citet{BECKER2011} with $z_{\textrm{abs}}$$>$5.7 ($dN/dX\simeq$0.22$^{+0.21}_{-0.13}$). As \citet{BECKER2011} have pointed out that their incidence rate is comparable to the total number density of the combined DLA and sub-DLA populations at z $\sim$3, we find that the number density of low ionisation systems is not evolving rapidly over the redshift range 3<$z$<6 and note the connection with point 9 below. } \\

\item {{We identify an excess of weak absorbers ($W_{2796}$$\le$0.3\AA) when comparing to an expectation from an exponential fit to the equivalent width distribution of \mgii absorbers with $W_{2796}$>0.3\AA}. Our bin values and parameter fits can be seen in Table \ref{tab:d2ndzdwTable} and section \ref{sec:d2N}. This is similar to the findings of \citet{CHURCHILL1999} and \citet{NARAYANAN2007} in the redshift ranges of 0.4$\le$$z$$\le$~1.4 and 0.4$\le$$z$$\le$2.4 respectively. {In the latter stages of preparing this manuscript, we became aware of \citet{BOSMAN2017} which also reports an excess of weak \mgii absorbers in the redshift range 5.9<$z$<7.0.}} \\

\item {{We measure $\Omega_{\mgii}$=2.1$^{+6.3}_{-0.6}\times10^{-8}$, 1.9$^{+2.9}_{-0.2} \times10^{-8}$, 3.9$^{+7.1}_{-2.4}\times10^{-7}$ at <$z$>=2.48, 3.41, 4.77 respectively when we consider a 5$\sigma$ selection criteria. We also measure the comoving mass density $\Omega_{\mgii}$ of weak, intermediate/medium and strong systems. Our results can be seen in Figure \ref{fig:omega}, the values are presented in Table \ref{tab:omegatable} and we discuss these results in section \ref{sec:omegadiscussion}. We find that weak systems provide a significant amount of \mgii when considering systems with $W_{2796}$<1\AA. We find that $\Omega_{\mgii}$ from strong systems increases by an order of magnitude from $\Omega_{\mgii}$=1.7$^{+1.4}_{-0.6}\times10^{-8}$ at <$z$>=2.49 to $\Omega_{\mgii}$=3.8$^{+5.9}_{-1.3}\times10^{-7}$ at <$z$>=4.77. For $\Omega_{\mgii}$, as traced by weak and intermediate/medium systems, we find a similar result.}} \\

\item {{We calculate expected incidence rates (see eq. \ref{eq:dndxfromg}) by pairing \mgii absorption halo properties in the redshift range 0.072$\le$z$\le$1.120 presented by \citet{NIELSEN2013b} with the z $\sim$5 luminosity function presented by \citet{MASON2016}. The results can be seen in Figure \ref{fig:dndxexpected}. We find that we can not account for the incidence rates of \mgii systems with $W_{2796}$>0.1\AA{} if the absorption halo properties of \mgii do not evolve from z $\simeq$1.12 to z $\simeq$5 even if we integrate the luminosity function down to the lowest magnitude for which it is defined. This discrepancy can not be explained by an evolution in covering fraction alone and we discuss this in section \ref{sec:expectations}. When we consider strong systems, we find that in order to explain their incidence rate we must consider all galaxies with $L$$\ge$0.29$L_{\ast}$. If instead, strong systems are only associated with galaxies with $L\ge$0.50$L_{\ast}$ \citep{SEYFFERT2013} then their physical cross-section ($\sigma_{phys}$) increases to 0.041 Mpc$^2$ at <$z$>=4.77 from 0.015 Mpc$^2$ at $z$=2.3. }}\\

\item {We compare our highest redshift bin value for $\Omega_{\mgii}$ at z $\sim$4.77 with an expected value calculated using eq. \ref{eq:omegaexpectedfromh1} which incorporates the global DLA metallicity ($<$Z$>$=-2.03$^{+0.09}_{-0.11}$; \citealt{RAFELSKI2014}) and the comoving mass density of neutral hydrogen present in DLAs as measured by \citet{CRIGHTON2015} ($\Omega^{}_{\ion{H}{I}}$=0.98$^{+0.20}_{-0.18} \times$10$^{-3}$). We compute an expected value $\Omega_{\mgii}^{expected}$=8.85$^{+2.25}_{-3.25}$ $\times$10$^{-9}$ while we measure {$\Omega_{\mgii}$=3.9$^{+7.1}_{-2.4}\times10^{-7}$}. Thus, we conclude that although \mgii traces \ion{H}{I} at lower redshift, our results show that current scaling relations can not account for all of the $\Omega_{\ion{H}{I}}$ at z $\sim$4.5 to 5.5. This suggests that $\Omega_{\mgii}$ (which is dominated by strong systems) could be associated with both ionised and neutral gas or that these strong systems have a higher metallicity than typical DLAs at this redshift as suggested by MS13. }

\end{enumerate}

Our findings suggest that \mgii absorbers are well established after the first Gyr of evolution of the universe with the comoving incidence rate of weak \mgii systems showing little to no evolution over the following $\sim$2~Gyr, down to z $\sim$2. While these systems and the nature of the associated galaxies has been investigated up to z $\sim$2.4 by \citet{CHURCHILL1999} and \citet{NARAYANAN2007}, no such work has been undertaken in the redshift range of our survey. The evolution of the comoving mass density of these same systems (top panel of Figure \ref{fig:omega}) shows little to no evolution while the total fraction of \mgii in these systems increases towards redshift $\sim$5. This suggests that tracking these weak systems becomes critical when attempting to trace the total budget of metals in the early universe.

 {In order to account for the impact of sky lines, we implement wavelength dependent and recovery rate based completeness and false positive corrections and find that even weak \mgii systems can be identified past $z$ $\sim$5 at both 5 and 3$\sigma$ considerations (see Figures \ref{fig:completenesstest} and \ref{fig:completenesstest3sigma} respectively). This is encouraging as more than a hundred QSOs have been identified past $z$ $\sim$5 \citep{JIANG2016, BANADOS2016} which can be followed up with current spectroscopic instruments. Integral field spectrographs also provide an exciting opportunity to simultaneously investigate intervening absorbers and the associated galaxies with a single observation.}

\section*{Acknowledgements}

 {We thank Glenn Kacprzak, Michael Murphy and {Nikki Nielsen} for useful discussions and Manodeep Singha for his expertise with code optimisation. {We also thank the anonymous referee for their comments and attention to detail which substantially improved the manuscript}. This work is based on observations made with the ESO telescopes at the La Silla Paranal Observatory under programme ID 084.A-0390(A). Parts of this research have made use of the Matplotlib library \citep{HUNTER2007}. ERW and AC acknowledge the \textit{Australian Research Council} for \textit{Discovery Project} grant DP1095600 which supported this work. AC is also supported by a Swinburne University Postgraduate Research Award (SUPRA) scholarship. Parts of this research were conducted by the Australian Research Council Centre of Excellence for All-sky Astrophysics (CAASTRO), through project number CE110001020.}




\bibliographystyle{mnras}
\bibliography{paper1_5_biblio} 





\appendix

\section{System tables and plots}
\begin{figure*}
	\includegraphics[trim=4.5cm 8.0cm 4.5cm 7.5cm, clip=true, width=18cm]{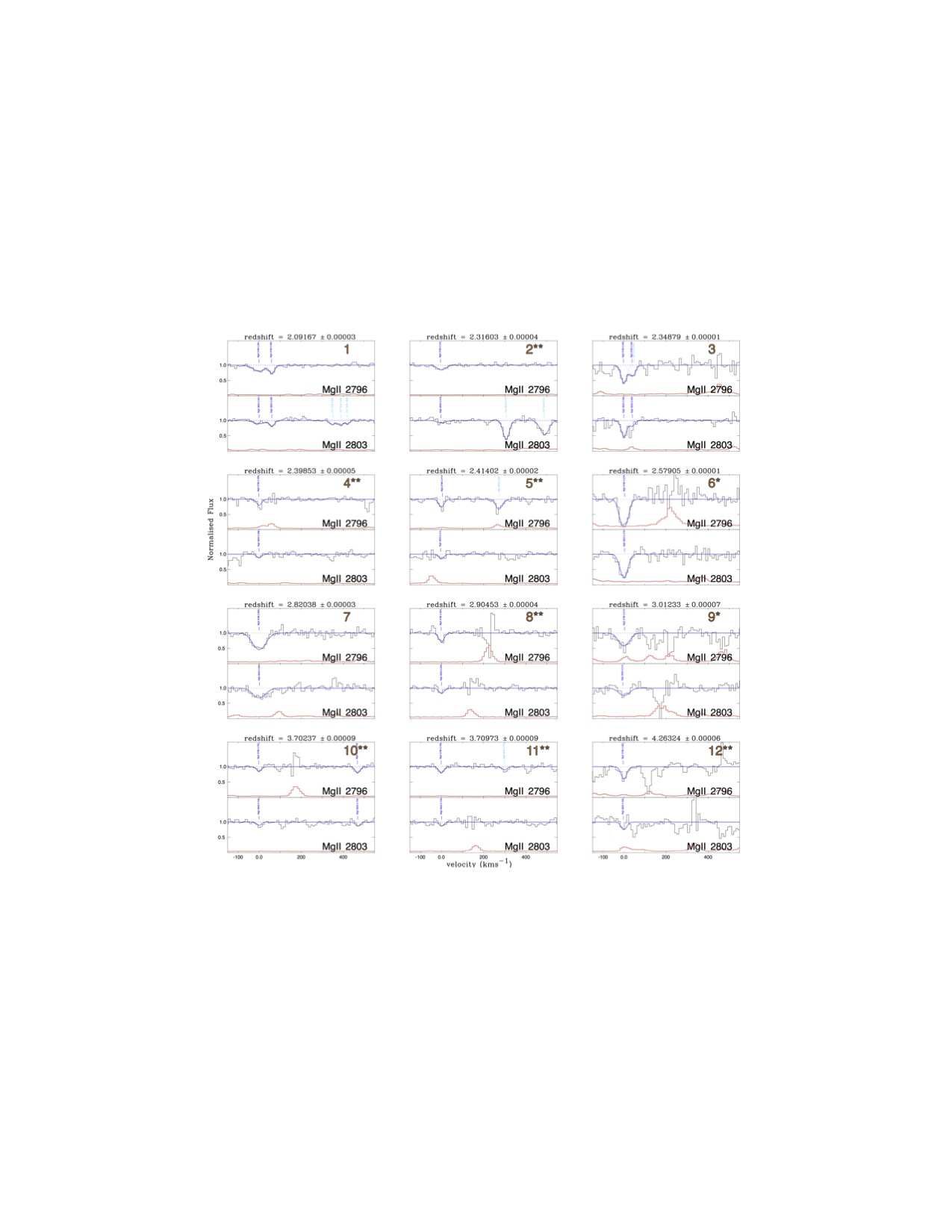} 

\begin{flushleft}
 $^{\ast}$ $\hspace{1.2mm}$ denotes system which does not meet our 5$\sigma$ selection criteria 
 
 $^{\ast}$ $^{\ast}$ denotes system which does not meet our 5$\sigma$ and 3$\sigma$ selection criteria

\end{flushleft}

\caption{All \mgii systems identified in the SDSS J0927+2001 sightline. The top panel of each system plot is the location of the $\lambda$2796 transition and the bottom panel is the associated $\lambda$2803 transition. In each panel, the vertical axis is the continuum normalised flux. The horizontal axis is the velocity separation (kms$^{-1}$) from the lowest redshift component of a system. The normalised spectrum is plotted in black and the associated error is in red. The solid blue line represents the full fit to the spectra and includes other ions besides \ion{Mg}{II}. Individual components are plotted with dashed lines and are identified by a vertical label. \mgii components are in solid blue and other transitions are in light blue.}

 \label{fig:s0927systems}
\end{figure*}

\begin{figure*}
	\includegraphics[trim=0.1cm 0.1cm 0.1cm 0.001cm, clip=true, width=18cm]{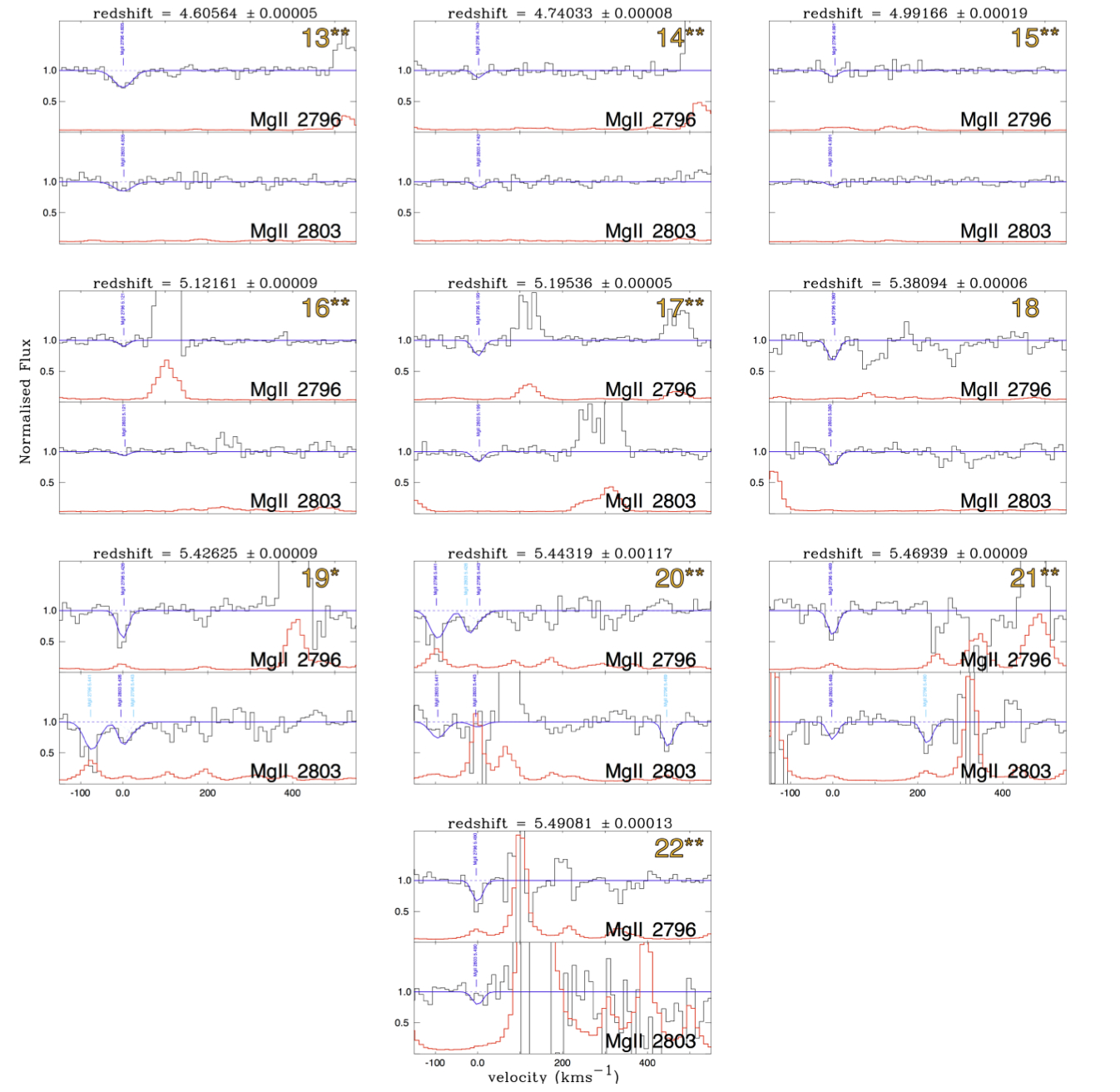} 

\begin{flushleft}
 $^{\ast}$ $\hspace{1.2mm}$ denotes system which does not meet our 5$\sigma$ selection criteria 
 
 $^{\ast}$ $^{\ast}$ denotes system which does not meet our 5$\sigma$ and 3$\sigma$ selection criteria

\end{flushleft}

\caption{All \mgii systems identified in the SDSS J0927+2001 sightline. The top panel of each system plot is the location of the $\lambda$2796 transition and the bottom panel is the associated $\lambda$2803 transition. In each panel, the vertical axis is the continuum normalised flux. The horizontal axis is the velocity separation (kms$^{-1}$) from the lowest redshift component of a system. The normalised spectrum is plotted in black and the associated error is in red. The solid blue line represents the full fit to the spectra and includes other ions besides \ion{Mg}{II}. Individual components are plotted with dashed lines and are identified by a vertical label. \mgii components are in solid blue and other transitions are in light blue.}

 \label{fig:s0927systems_b}
\end{figure*}

\begin{figure*}
 	\includegraphics[trim=0.1cm 0.1cm 0.1cm 0.001cm, clip=true, width=18cm]{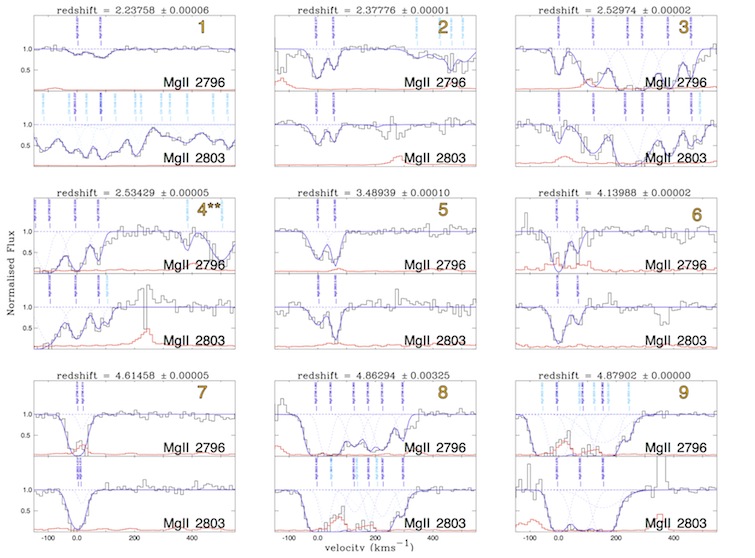} 

\begin{flushleft}

 $^{\ast}$ $\hspace{1.2mm}$ denotes system which does not meet our 5$\sigma$ selection criteria

 $^{\ast}$ $^{\ast}$ denotes system which does not meet our 5$\sigma$ and 3$\sigma$ selection criteria

\end{flushleft}

\caption{All \mgii systems identified in the SDSS J1306+0356 sightline. The top panel of each system plot is the location of the $\lambda$2796 transition and the bottom panel is the associated $\lambda$2803 transition. In each panel, the vertical axis is the continuum normalised flux. The horizontal axis is the velocity separation (kms$^{-1}$) from the lowest redshift component of a system. The normalised spectrum is plotted in black and the associated error is in red. The solid blue line represents the full fit to the spectra and includes other ions besides \ion{Mg}{II}. Individual components are plotted with dashed lines and are identified by a vertical label. \mgii components are in solid blue and other transitions are in light blue.}
 \label{fig:s1306systems}
\end{figure*}

\begin{figure*}
	 \includegraphics[trim=0.1cm 0.1cm 0.1cm 0.001cm, clip=true, width=18cm]{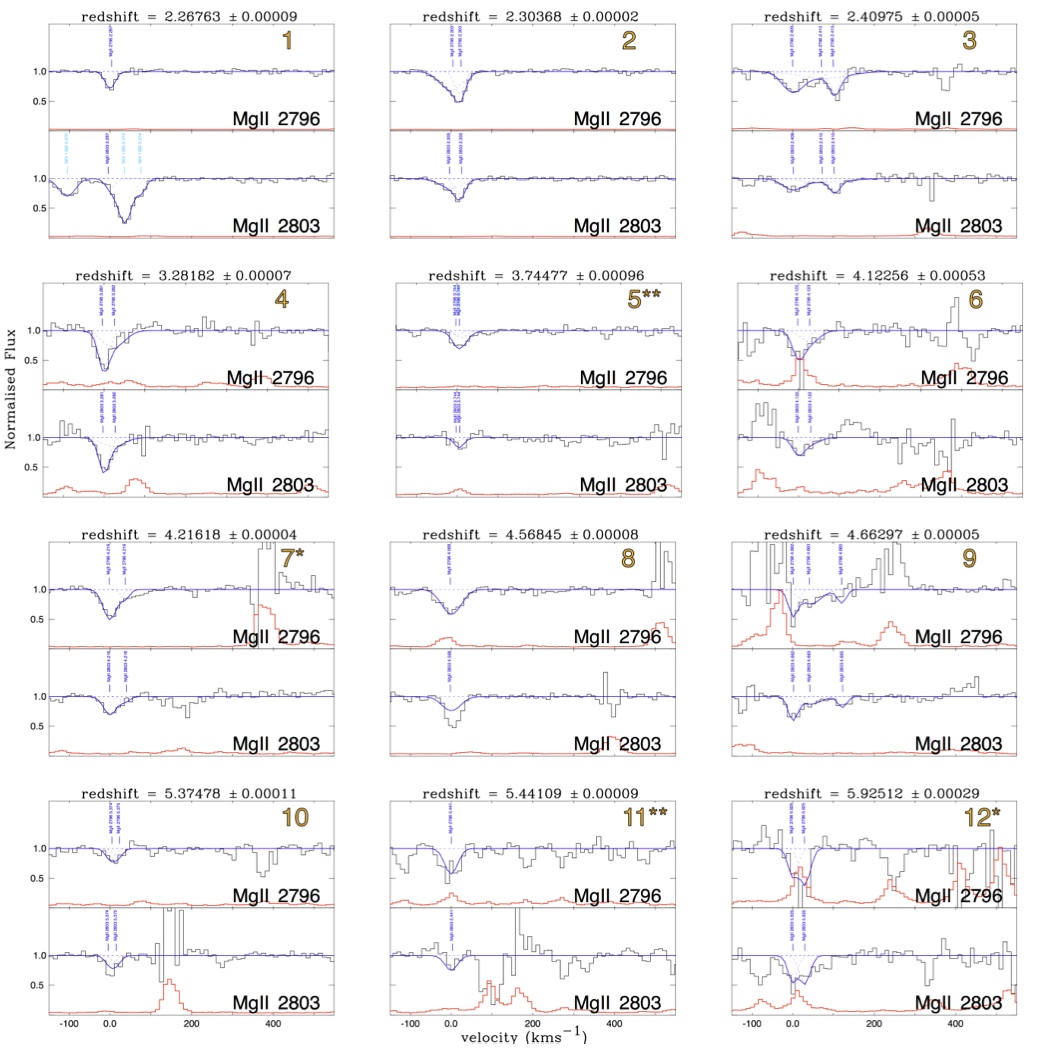} 

\begin{flushleft}
 $^{\ast}$ $\hspace{1.2mm}$ denotes system which does not meet our 5$\sigma$ selection criteria 
 
 $^{\ast}$ $^{\ast}$ denotes system which does not meet our 5$\sigma$ and 3$\sigma$ selection criteria

\end{flushleft}

\caption{All \mgii systems identified in the ULAS J1319+0959 sightline. The top panel of each system plot is the location of the $\lambda$2796 transition and the bottom panel is the associated $\lambda$2803 transition. In each panel, the vertical axis is the continuum normalised flux. The horizontal axis is the velocity separation (kms$^{-1}$) from the lowest redshift component of a system. The normalised spectrum is plotted in black and the associated error is in red. The solid blue line represents the full fit to the spectra and includes other ions besides \ion{Mg}{II}. Individual components are plotted with dashed lines and are identified by a vertical label. \mgii components are in solid blue and other transitions are in light blue.}
 
 \label{fig:u1319systems}
\end{figure*}

Here we present all associated components of identified \mgii absorption systems. The numerical values of all \mgii components identified the in sightline $ULAS$ $J0148+0600$ can be seen in Table \ref{tab:u0148Table}. The numerical values of all \mgii components identified the in sightline $SDSS$ $J0927+2001$ are given in Table \ref{tab:s0927Table} and all systems can be seen in Figures \ref{fig:s0927systems} and \ref{fig:s0927systems_b}. The numerical values of all \mgii components identified the in sightline $SDSS$ $J1306+0356$ are given in Table \ref{tab:s1306Table} and all systems can be seen in Figure \ref{fig:s1306systems}. The numerical values of all \mgii components identified the in sightline $ULAS$ $J1319+0950$ are given in Table \ref{tab:u1319Table} and all systems can be seen in Figure \ref{fig:u1319systems}.

\begin{table*}
	\centering
	\caption{\mgii systems identified in ULAS J0148+0600 sightline. A system is defined as all components within 800 kms$^{-1}$ of the lowest redshift component. The Table lists $z$, $log$($N$) and $b$ which are the redshift, column density and doppler parameter for each component Voigt profile fit. The 3 and 5$\sigma$ selection criteria are defined in eq. \ref{eq:recovery}. No lower bound is presented for systems with $b$=1 kms$^{-1}$ as the minimum doppler parameter we allow for a Voigt Profile is 1 kms$^{-1}$.}

\label{tab:u0148Table}
\noindent\adjustbox{max width=\textwidth}{
\begin{tabular}{c c c c c c c c c}
   &    &      & 	  \Large{ULAS J0148+0600}  & 	    &  	   & 	   &   \\[3pt]
\hline
\hline
   &    &      & 	   & 	    &  	 comp  & 	 sys   & sys 5$\sigma$ & sys 3$\sigma$  \\[1pt]
 System & Component & z     & log(N)    & b     & EW$_{2796}$ & EW$_{2796}$ & recovery rate & recovery rate\\[1pt]
   &    &      & cm$^{-2}$    & kms$^{-1}$    &\AA   &\AA   & $\%$ & $\%$ \\[1pt]
\hline
 & & & & & & & \\[1pt]
 1 & a & 2.39557$\pm$4$\times10^{-5}$ & $12.57_{-0.11}^{+0.07}$ & $13.9_{-4.65}^{+2.43}$ & 0.117$\pm$0.005 & 0.117$\pm$0.005 & 62.1 & 77.7 \\[2pt]
 & & & & & & & \\[1pt]
 2 & a & 2.47762$\pm$3$\times10^{-5}$ & $13.01_{-0.08}^{+0.09}$ & $28.1_{-2.41}^{+3.73}$ & 0.303$\pm$0.007 & 2.556$\pm$0.015 & 96.7 & 96.7 \\[2pt]
  & b & 2.47894$\pm$2$\times10^{-5}$ & $13.29_{-0.06}^{+0.07}$ & $20.6_{-1.87}^{+2.11}$ & 0.392$\pm$0.005 & & & \\[2pt]
  & c & 2.48039$\pm$0.00034 & $14.43_{-0.05}^{+0.84}$ & $34.7_{-0.87}^{+0.20}$ & 1.112$\pm$0.006 & & & \\[2pt]
  & d & 2.48119$\pm$0.00034 & $14.11_{-0.07}^{+0.43}$ & $18.8_{-0.91}^{+7.06}$ & 0.602$\pm$0.005 & & & \\[2pt]
  & e & 2.48231$\pm$4$\times10^{-5}$ & $12.66_{-0.11}^{+0.12}$ & $14.2_{-4.66}^{+4.62}$ & 0.144$\pm$0.008 & & & \\[2pt]
 & & & & & & & \\[1pt]
 3 & a & 2.72177$\pm$6$\times10^{-5}$ & $13.45_{-0.05}^{+0.03}$ & $53.1_{-5.66}^{+11.3}$ & 0.840$\pm$0.015 & 1.636$\pm$0.019 & 96.7 & 96.7 \\[2pt]
  & b & 2.72203$\pm$2$\times10^{-5}$ & $14.19_{-0.11}^{+0.56}$ & $22.0_{-1.14}^{+2.40}$ & 0.796$\pm$0.011 & & & \\[2pt]
 & & & & & & & \\[1pt]
 4 & a$\Uparrow$ & 3.01823$\pm$0.00010 & $13.56_{-0.05}^{+0.07}$ & $73.0_{-5.72}^{+14.6}$ & 1.005$\pm$0.041 & 2.081$\pm$0.052 & 96.7 & 96.7 \\[2pt]
  & b$\Uparrow$ & 3.01850$\pm$2$\times10^{-5}$ & $14.16_{-0.10}^{+0.97}$ & $24.4_{-1.47}^{+5.70}$ & 0.629$\pm$0.029 & & & \\[2pt]
  & c & 3.01997$\pm$2$\times10^{-5}$ & $13.27_{-0.06}^{+0.02}$ & $23.1_{-1.61}^{+1.66}$ & 0.446$\pm$0.011 & & & \\[2pt]
 & & & & & & & \\[1pt]
 5 & a$\Uparrow$ & 3.05164$\pm$0.00011 & $13.05_{-0.06}^{+0.13}$ & $37.7_{-2.75}^{+8.26}$ & 0.279$\pm$0.024 & 0.433$\pm$0.034 & 92.8 & 92.8 \\[2pt]
  & b$\Uparrow$ & 3.05223$\pm$3$\times10^{-5}$ & $14.50_{-0.59}^{+0.04}$ & $4.34_{-0.46}^{+0.91}$ & 0.154$\pm$0.023 & & & \\[2pt]
 & & & & & & & \\[1pt]
 6$^{\ast}$ & a$\Uparrow$ & 3.69207$\pm$4$\times10^{-5}$ & $12.26_{-0.21}^{+0.31}$ & $20.5_{-6.65}^{+22.5}$ & 0.077$\pm$0.005 & 0.077$\pm$0.005 & 37.9 & 55.1 \\[2pt]
 & & & & & & & \\[1pt]
 7 & a & 4.45996$\pm$6$\times10^{-5}$ & $13.65_{-0.08}^{+0.03}$ & $13.5_{-1.21}^{+1.77}$ & 0.389$\pm$0.006 & 0.389$\pm$0.006 & 96.7 & 96.7 \\[2pt]
 & & & & & & & \\[1pt]
 8 & a & 4.89031$\pm$4$\times10^{-5}$ & $14.21_{-0.12}^{+0.23}$ & $19.0_{-1.13}^{+0.33}$ & 0.699$\pm$0.004 & 1.481$\pm$0.019 & 93.3 & 93.3 \\[2pt]
  & b & 4.89161$\pm$4$\times10^{-5}$ & $14.07_{-0.09}^{+0.15}$ & $30.3_{-1.42}^{+1.65}$ & 0.782$\pm$0.019 & & & \\[2pt]
 & & & & & & & \\[1pt]
 9$^{\ast\ast}$ & a & 5.46914$\pm$0.00013 & $12.28_{-0.51}^{+0.39}$ & $2.59_{-1.58}^{+0.17}$ & 0.046$\pm$0.007 & 0.046$\pm$0.007 & 30.7 & 48.5 \\[2pt]
 & & & & & & & \\[1pt]
 \hline
 \hline
\end{tabular}

}

\begin{flushleft}

$\Downarrow$ $\hspace{0.8mm}$ denotes a component with a blended feature

$\Uparrow$ $\hspace{0.8mm}$ denotes a component polluted by a sky line or poor subtraction residual

 $^{\ast}$ $\hspace{1.2mm}$ denotes system which does not meet our 5$\sigma$ selection criteria

 $^{\ast}$ $^{\ast}$ denotes system which does not meet our 5$\sigma$ and 3$\sigma$ selection criteria

\end{flushleft}
\end{table*}

\begin{table*}
	\centering
	\caption{\mgii systems identified in SDSS J0927+2001 sightline. A system is defined as all components within 800 kms$^{-1}$ of the lowest redshift component. The Table lists $z$, $log$($N$) and $b$ which are the redshift, column density and doppler parameter for each component Voigt profile fit. The 3 and 5$\sigma$ selection criteria are defined in eq. \ref{eq:recovery}. No lower bound is presented for systems with $b$=1 kms$^{-1}$ as the minimum doppler parameter we allow for a Voigt Profile is 1 kms$^{-1}$.}

	\label{tab:s0927Table}
\noindent\adjustbox{max width=\textwidth}{

\begin{tabular}{c c c c c c c c c}
   &    &      & 	  \Large{SDSS J0927+2001}  & 	    &  	   & 	   &   \\[3pt]
\hline
\hline
   &    &      & 	   & 	    &  	 comp  & 	 sys   & sys 5$\sigma$ & sys 3$\sigma$  \\[1pt]
 System & Component & z     & log(N)    & b     & EW$_{2796}$ & EW$_{2796}$ & recovery rate & recovery rate\\[1pt]
   &    &      & cm$^{-2}$    & kms$^{-1}$    &\AA   &\AA   & $\%$ & $\%$ \\[1pt]
\hline
 & & & & & & & \\[1pt]
 1 & a & 2.09167$\pm$4$\times10^{-5}$ & $12.52_{-0.24}^{+0.21}$ & $33.8_{-8.81}^{+14.0}$ & 0.118$\pm$0.009 & 0.193$\pm$0.011 & 89.7 & 89.7 \\[2pt]
  & b & 2.09230$\pm$2$\times10^{-5}$ & $12.49_{-0.07}^{+0.09}$ & $4.38_{-1.39}^{+0.07}$ & 0.074$\pm$0.006 & & & \\[2pt]
 & & & & & & & \\[1pt]
 2$^{\ast\ast}$ & a & 2.31603$\pm$4$\times10^{-5}$ & $12.27_{-0.34}^{+0.27}$ & $23.7_{-10.5}^{+13.4}$ & 0.067$\pm$0.008 & 0.067$\pm$0.008 & 18.4 & 40.2 \\[2pt]
 & & & & & & & \\[1pt]
 3 & a & 2.34879$\pm$1$\times10^{-5}$ & $13.46_{-0.07}^{+0.14}$ & $5.66_{-0.47}^{+0.35}$ & 0.173$\pm$0.011 & 0.269$\pm$0.014 & 58.7 & 89.7 \\[2pt]
  & b$\Uparrow$ & 2.34922$\pm$5$\times10^{-5}$ & $12.50_{-0.04}^{+0.13}$ & $9.18_{-4.31}^{+3.19}$ & 0.095$\pm$0.009 & & & \\[2pt]
 & & & & & & & \\[1pt]
 4$^{\ast\ast}$ & a$\Uparrow$ & 2.39853$\pm$5$\times10^{-5}$ & $12.24_{-0.32}^{+0.20}$ & $5.07_{-2.72}^{+11.2}$ & 0.055$\pm$0.020 & 0.055$\pm$0.020 & 11.6 & 27.9 \\[2pt]
 & & & & & & & \\[1pt]
 5$^{\ast\ast}$ & a & 2.41402$\pm$3$\times10^{-5}$ & $12.43_{-0.15}^{+0.31}$ & $5.05_{-1.47}^{+0.17}$ & 0.071$\pm$0.012 & 0.071$\pm$0.012 & 22.0 & 43.9 \\[2pt]
 & & & & & & & \\[1pt]
 6$^{\ast}$ & a & 2.57905$\pm$2$\times10^{-5}$ & $13.49_{-0.07}^{+0.07}$ & $16.8_{-1.12}^{+1.15}$ & 0.500$\pm$0.031 & 0.500$\pm$0.031 & 31.0 & 75.9 \\[2pt]
 & & & & & & & \\[1pt]
 7 & a & 2.82038$\pm$4$\times10^{-5}$ & $13.08_{-0.09}^{+0.06}$ & $37.0_{-2.54}^{+1.91}$ & 0.310$\pm$0.016 & 0.310$\pm$0.016 & 79.4 & 93.2 \\[2pt]
 & & & & & & & \\[1pt]
 8$^{\ast\ast}$ & a & 2.90453$\pm$5$\times10^{-5}$ & $12.44_{-0.13}^{+0.10}$ & $6.63_{-5.63}^{+4.85}$ & 0.092$\pm$0.013 & 0.092$\pm$0.013 & 16.4 & 19.7 \\[2pt]
 & & & & & & & \\[1pt]
 9$^{\ast}$ & a$\Uparrow$ & 3.01233$\pm$7$\times10^{-5}$ & $12.86_{-0.13}^{+0.10}$ & $30.1_{-5.09}^{+3.82}$ & 0.252$\pm$0.046 & 0.252$\pm$0.046 & 17.1 & 63.8 \\[2pt]
 & & & & & & & \\[1pt]
 10$^{\ast\ast}$ & a & 3.70237$\pm$9$\times10^{-5}$ & $12.22_{-0.15}^{+0.33}$ & $2.64_{-1.64}^{+1.55}$ & 0.048$\pm$0.011 & 0.048$\pm$0.011 & 2.8 & 6.2 \\[2pt]
 & & & & & & & \\[1pt]
 11$^{\ast\ast}$ & a & 3.70973$\pm$9$\times10^{-5}$ & $12.32_{-0.24}^{+0.18}$ & $3.43_{-2.43}^{+2.24}$ & 0.059$\pm$0.011 & 0.059$\pm$0.011 & 0.10 & 0.42 \\[2pt]
 & & & & & & & \\[1pt]
 12$^{\ast\ast}$ & a$\Uparrow$ & 4.26324$\pm$7$\times10^{-5}$ & $12.65_{-0.09}^{+0.07}$ & $9.50_{-7.56}^{+5.67}$ & 0.121$\pm$0.017 & 0.121$\pm$0.017 & 16.3 & 21.2 \\[2pt]
 & & & & & & & \\[1pt]
 13$^{\ast}$ & a & 4.60564$\pm$6$\times10^{-5}$ & $12.57_{-0.15}^{+0.11}$ & $25.7_{-1.84}^{+1.38}$ & 0.129$\pm$0.009 & 0.129$\pm$0.009 & 48.2 & 55.1 \\[2pt]
 & & & & & & & \\[1pt]
 14$^{\ast\ast}$ & a & 4.74033$\pm$9$\times10^{-5}$ & $12.32_{-0.61}^{+0.54}$ & $1.16_{-0.16}^{+0.58}$ & 0.026$\pm$0.008 & 0.026$\pm$0.008 & 0.00 & 12.5 \\[2pt]
 & & & & & & & \\[1pt]
 15$^{\ast\ast}$ & a & 4.99166$\pm$0.00019 & $11.94_{-0.48}^{+0.33}$ & $3.50_{-2.50}^{+11.4}$ & 0.029$\pm$0.017 & 0.029$\pm$0.017 & 0.00 & 0.00 \\[2pt]
 & & & & & & & \\[1pt]
 16$^{\ast\ast}$ & a & 5.12161$\pm$0.00010 & $12.10_{-0.57}^{+0.47}$ & $1.07_{-0.07}^{+0.89}$ & 0.021$\pm$0.008 & 0.021$\pm$0.008 & 0.00 & 1.17 \\[2pt]
 & & & & & & & \\[1pt]
 17$^{\ast\ast}$ & a & 5.19536$\pm$5$\times10^{-5}$ & $12.50_{-0.23}^{+0.17}$ & $3.23_{-1.58}^{+1.19}$ & 0.070$\pm$0.009 & 0.070$\pm$0.009 & 9.86 & 9.86 \\[2pt]
 & & & & & & & \\[1pt]
 18$^{\ast\ast}$ & a & 5.38094$\pm$7$\times10^{-5}$ & $12.58_{-0.15}^{+0.11}$ & $5.18_{-2.61}^{+1.94}$ & 0.089$\pm$0.013 & 0.089$\pm$0.013 & 29.3 & 32.6 \\[2pt]
 & & & & & & & \\[1pt]
 19$^{\ast\ast}$ & a & 5.42625$\pm$0.00010 & $12.83_{-0.14}^{+0.10}$ & $5.55_{-1.86}^{+1.49}$ & 0.121$\pm$0.022 & 0.121$\pm$0.022 & 34.1 & 37.2 \\[2pt]
 & & & & & & & \\[1pt]
 20$^{\ast\ast}$ & a$\Uparrow$ & 5.44117$\pm$0.00015 & $12.76_{-0.14}^{+0.10}$ & $18.7_{-8.60}^{+6.48}$ & 0.180$\pm$0.063 & 0.236$\pm$0.068 & 0.00 & 0.08 \\[2pt]
  & b & 5.44319$\pm$0.00117 & $12.17_{-0.19}^{+0.13}$ & $18.0_{-7.76}^{+5.23}$ & 0.056$\pm$0.024 & & & \\[2pt]
 & & & & & & & \\[1pt]
 21$^{\ast\ast}$ & a$\Uparrow$ & 5.46939$\pm$0.00010 & $12.78_{-0.17}^{+0.13}$ & $4.83_{-1.75}^{+1.29}$ & 0.106$\pm$0.023 & 0.106$\pm$0.023 & 11.8 & 42.7 \\[2pt]
 & & & & & & & \\[1pt]
 22$^{\ast\ast}$ & a$\Uparrow$ & 5.49081$\pm$0.00013 & $12.53_{-0.13}^{+0.09}$ & $6.82_{-5.82}^{+4.97}$ & 0.093$\pm$0.035 & 0.093$\pm$0.035 & 0.00 & 23.3 \\[2pt]
 & & & & & & & \\[1pt]
 \hline
 \hline
\end{tabular}

}

\begin{flushleft}

$\Downarrow$ $\hspace{0.8mm}$ denotes a component with a blended feature

$\Uparrow$ $\hspace{0.8mm}$ denotes a component polluted by a sky line or poor subtraction residual

 $^{\ast}$ $\hspace{1.2mm}$ denotes system which does not meet our 5$\sigma$ selection criteria

 $^{\ast}$ $^{\ast}$ denotes system which does not meet our 5$\sigma$ and 3$\sigma$ selection criteria

\end{flushleft}

\end{table*}

\begin{table*}
	\centering
	\caption{\mgii systems identified in SDSS J1306+0356 sightline. A system is defined as all components within 800 kms$^{-1}$ of the lowest redshift component. The Table lists $z$, $log$($N$) and $b$ which are the redshift, column density and doppler parameter for each component Voigt profile fit. The 3 and 5$\sigma$ selection criteria are defined in eq. \ref{eq:recovery}. No lower bound is presented for systems with $b$=1 kms$^{-1}$ as the minimum doppler parameter we allow for a Voigt Profile is 1 kms$^{-1}$.}
	\label{tab:s1306Table}

\noindent\adjustbox{max width=\textwidth}{

\begin{tabular}{c c c c c c c c c}
   &    &      & 	  \Large{SDSS J1306+0356}  & 	    &  	   & 	   &   \\[3pt]
\hline
\hline
   &    &      & 	   & 	    &  	 comp  & 	 sys   & sys 5$\sigma$ & sys 3$\sigma$  \\[1pt]
 System & Component & z     & log(N)    & b     & EW$_{2796}$ & EW$_{2796}$ & recovery rate & recovery rate\\[1pt]
   &    &      & cm$^{-2}$    & kms$^{-1}$    &\AA   &\AA   & $\%$ & $\%$ \\[1pt]
\hline
 & & & & & & & \\[1pt]
 1$^{\ast\ast}$ & a$\Downarrow$ & 2.23758$\pm$7$\times10^{-5}$ & $12.21_{-0.23}^{+0.14}$ & $23.4_{-3.53}^{+2.25}$ & 0.059$\pm$0.006 & 0.183$\pm$0.009 & 41.4 & 43.2 \\[2pt]
  & b$\Downarrow$ & 2.23848$\pm$5$\times10^{-5}$ & $12.54_{-0.15}^{+0.10}$ & $33.8_{-3.90}^{+2.49}$ & 0.123$\pm$0.006 & & & \\[2pt]
 & & & & & & & \\[1pt]
 2 & a & 2.37776$\pm$2$\times10^{-5}$ & $13.10_{-0.07}^{+0.04}$ & $19.0_{-2.41}^{+1.52}$ & 0.306$\pm$0.010 & 0.446$\pm$0.013 & 96.7 & 96.7 \\[2pt]
  & b & 2.37840$\pm$2$\times10^{-5}$ & $13.13_{-0.14}^{+0.09}$ & $4.97_{-0.71}^{+0.45}$ & 0.140$\pm$0.008 & & & \\[2pt]
 & & & & & & & \\[1pt]
 3 & a & 2.52974$\pm$2$\times10^{-5}$ & $13.04_{-0.08}^{+0.05}$ & $8.33_{-1.39}^{+0.88}$ & 0.184$\pm$0.011 & 3.507$\pm$0.062 & 96.7 & 96.7 \\[2pt]
  & b$\Uparrow$ & 2.53114$\pm$4$\times10^{-5}$ & $13.58_{-0.06}^{+0.03}$ & $39.0_{-2.65}^{+0.58}$ & 0.766$\pm$0.049 & & & \\[2pt]
  & c & 2.53254$\pm$0.00041 & $14.80_{-0.12}^{+0.04}$ & $24.3_{-1.11}^{+0.08}$ & 0.908$\pm$0.017 & & & \\[2pt]
  & d & 2.53316$\pm$0.00142 & $13.75_{-0.05}^{+0.04}$ & $36.7_{-0.44}^{+0.27}$ & 0.864$\pm$0.020 & & & \\[2pt]
  & e & 2.53429$\pm$6$\times10^{-5}$ & $13.54_{-0.06}^{+0.04}$ & $25.3_{-0.88}^{+0.78}$ & 0.558$\pm$0.018 & & & \\[2pt]
  & f$\Downarrow$ & 2.53515$\pm$3$\times10^{-5}$ & $13.13_{-0.07}^{+0.04}$ & $10.5_{-1.40}^{+1.00}$ & 0.223$\pm$0.015 & & & \\[2pt]
 & & & & & & & \\[1pt]
 4 $^{\ast\ast}$ & a$\Downarrow$ & 2.54463$\pm$6$\times10^{-5}$ & $14.35_{-0.51}^{+0.24}$ & $2.13_{-0.24}^{+0.19}$ & 0.100$\pm$0.018 & 0.100$\pm$0.018 & 18.5 & 18.5 \\[2pt]
 & & & & & & & \\[1pt]
 5 & a & 3.48939$\pm$0.00010 & $13.09_{-0.10}^{+0.09}$ & $23.0_{-4.12}^{+4.12}$ & 0.320$\pm$0.013 & 0.563$\pm$0.023 & 96.7 & 96.7 \\[2pt]
  & b & 3.49031$\pm$8$\times10^{-5}$ & $14.71_{-0.21}^{+0.27}$ & $6.22_{-0.30}^{+0.21}$ & 0.242$\pm$0.019 & & & \\[2pt]
 & & & & & & & \\[1pt]
 6 & a$\Uparrow$ & 4.13988$\pm$2$\times10^{-5}$ & $13.68_{-0.04}^{+0.05}$ & $18.2_{-1.87}^{+1.61}$ & 0.396$\pm$0.045 & 0.549$\pm$0.060 & 74.3 & 94.7 \\[2pt]
  & b$\Uparrow$ & 4.14100$\pm$6$\times10^{-5}$ & $13.05_{-0.08}^{+0.08}$ & $6.12_{-1.06}^{+1.03}$ & 0.153$\pm$0.040 & & & \\[2pt]
 & & & & & & & \\[1pt]
 7 & a$\Uparrow$ & 4.61458$\pm$6$\times10^{-5}$ & $13.84_{-0.43}^{+0.46}$ & $25.4_{-12.4}^{+7.53}$ & 0.632$\pm$0.044 & 0.734$\pm$0.062 & 96.0 & 96.0 \\[2pt]
  & b$\Uparrow$ & 4.61490$\pm$0.00059 & $14.91_{-1.10}^{+0.73}$ & $5.27_{ -3.60}^{+3.68}$ & 0.102$\pm$0.043 & & & \\[2pt]
 & & & & & & & \\[1pt]
 8 & a$\Downarrow$ & 4.86294$\pm$0.00326 & $15.76_{-0.24}^{+0.28}$ & $18.1_{-0.67}^{+0.28}$ & 0.796$\pm$0.012 & 3.186$\pm$0.029 & 96.7 & 96.7 \\[2pt]
  & b$\Downarrow$ & 4.86387$\pm$0.09870 & $14.73_{-0.20}^{+0.15}$ & $21.1_{-0.46}^{+0.41}$ & 0.899$\pm$0.013 & & & \\[2pt]
  & c$\Downarrow$ & 4.86544$\pm$0.00031 & $13.27_{-0.01}^{+0.03}$ & $27.0_{ -6.10}^{+6.14}$ & 0.486$\pm$0.010 & & & \\[2pt]
  & d$\Downarrow$ & 4.86647$\pm$0.00013 & $13.84_{-0.39}^{+0.21}$ & $8.67_{-0.58}^{+0.81}$ & 0.324$\pm$0.009 & & & \\[2pt]
  & e$\Downarrow$ & 4.86732$\pm$0.00032 & $13.40_{-0.05}^{+0.06}$ & $27.4_{ -1.40}^{+1.46}$ & 0.465$\pm$0.011 & & & \\[2pt]
  & f$\Downarrow$ & 4.86869$\pm$5$\times10^{-5}$ & $13.14_{-0.15}^{+0.01}$ & $8.10_{-2.04}^{+1.16}$ & 0.212$\pm$0.014 & & & \\[2pt]
 & & & & & & & \\[1pt]
 9 & a$\Downarrow\Uparrow$ & 4.87902$\pm$0.00011 & $16.06_{-0.52}^{+1.07}$ & $14.1_{-0.33}^{+1.12}$ & 0.834$\pm$0.073 & 3.655$\pm$0.104 & 96.7 & 96.7 \\[2pt]
  & b$\Downarrow\Uparrow$ & 4.88060$\pm$0.00037 & $14.05_{-0.03}^{+0.27}$ & $32.9_{ -16.0}^{+16.5}$ & 1.036$\pm$0.060 & & & \\[2pt]
  & c$\Downarrow$ & 4.88203$\pm$0.00685 & $13.52_{-0.01}^{+0.01}$ & $74.5_{-10.9}^{+10.1}$ & 0.831$\pm$0.031 & & & \\[2pt]
  & d$\Downarrow$ & 4.88216$\pm$0.00029 & $16.17_{-0.90}^{+0.40}$ & $17.0_{-1.67}^{+0.21}$ & 0.953$\pm$0.030 & & & \\[2pt]
 & & & & & & & \\[1pt]
 \hline
 \hline
\end{tabular}

}

\begin{flushleft}

$\Downarrow$ $\hspace{0.8mm}$ denotes a component with a blended feature

$\Uparrow$ $\hspace{0.8mm}$ denotes a component polluted by a sky line or poor subtraction residual

 $^{\ast}$ $\hspace{1.2mm}$ denotes system which does not meet our 5$\sigma$ selection criteria

 $^{\ast}$ $^{\ast}$ denotes system which does not meet our 5$\sigma$ and 3$\sigma$ selection criteria

\end{flushleft}

\end{table*}

\begin{table*}
	\centering
	\caption{\mgii systems identified in ULAS J1319+0950 sightline. A system is defined as all components within 800 kms$^{-1}$ of the lowest redshift component. The Table lists $z$, $log$($N$) and $b$ which are the redshift, column density and doppler parameter for each component Voigt profile fit. The 3 and 5$\sigma$ selection criteria are defined in eq. \ref{eq:recovery}. No lower bound is presented for systems with $b$=1 kms$^{-1}$ as the minimum doppler parameter we allow for a Voigt Profile is 1 kms$^{-1}$.}
\label{tab:u1319Table}
\noindent\adjustbox{max width=\textwidth}{

\begin{tabular}{c c c c c c c c c}
   &    &      & 	  \Large{ULAS J1319+0959}  & 	    &  	   & 	   &   \\[3pt]
\hline
\hline
   &    &      & 	   & 	    &  	 comp  & 	 sys   & sys 5$\sigma$ & sys 3$\sigma$  \\[1pt]
 System & Component & z     & log(N)    & b     & EW$_{2796}$ & EW$_{2796}$ & recovery rate & recovery rate\\[1pt]
   &    &      & cm$^{-2}$    & kms$^{-1}$    &\AA   &\AA   & $\%$ & $\%$ \\[1pt]
\hline
 & & & & & & & \\[1pt]
 1$^{\ast}$ & a$\Downarrow$ & 2.26763$\pm$0.00010 & $12.41_{-0.24}^{+0.11}$ & $7.95_{-5.74}^{+7.50}$ & 0.078$\pm$0.005 & 0.078$\pm$0.005 & 37.9 & 70.6 \\[2pt]
 & & & & & & & \\[1pt]
 2 & a & 2.30368$\pm$3$\times10^{-5}$ & $12.73_{-0.13}^{+0.17}$ & $34.4_{-7.80}^{+12.7}$ & 0.193$\pm$0.007 & 0.298$\pm$0.008 & 96.7 & 96.7 \\[2pt]
  & b & 2.30391$\pm$1$\times10^{-5}$ & $12.77_{-0.07}^{+0.16}$ & $4.79_{-0.68}^{+0.57}$ & 0.104$\pm$0.004 & & & \\[2pt]
 & & & & & & & \\[1pt]
 3 & a & 2.40975$\pm$6$\times10^{-5}$ & $12.69_{-0.15}^{+0.06}$ & $29.3_{-8.14}^{+2.97}$ & 0.173$\pm$0.008 & 0.415$\pm$0.018 & 96.7 & 96.7 \\[2pt]
  & b & 2.41060$\pm$0.00105 & $12.30_{-0.19}^{+0.29}$ & $78.1_{-73.0}^{+1.75}$ & 0.145$\pm$0.014 & & & \\[2pt]
  & c & 2.41093$\pm$3$\times10^{-5}$ & $12.65_{-0.11}^{ }$ & $12.9_{-5.36}^{+0.28}$ & 0.097$\pm$0.007 & & & \\[2pt]
 & & & & & & & \\[1pt]
 4 & a & 3.28182$\pm$7$\times10^{-5}$ & $13.47_{-0.14}^{+0.07}$ & $5.69_{-0.59}^{+0.18}$ & 0.176$\pm$0.022 & 0.353$\pm$0.043 & 63.5 & 73.5 \\[2pt]
  & b & 3.28214$\pm$0.00050 & $12.69_{-0.19}^{+0.21}$ & $34.6_{-2.66}^{+9.17}$ & 0.176$\pm$0.037 & & & \\[2pt]
 & & & & & & & \\[1pt]
 5$^{\ast\ast}$ & a & 3.74477$\pm$0.00096 & $12.09_{-0.54}^{+0.73}$ & $1.00_{ }^{+0.05}$ & 0.024$\pm$0.013 & 0.057$\pm$0.017 & 20.1 & 28.5 \\[2pt]
  & b & 3.74494$\pm$0.00077 & $12.93_{-0.68}^{+0.33}$ & $1.08_{-0.07}^{+0.08}$ & 0.032$\pm$0.010 & & & \\[2pt]
 & & & & & & & \\[1pt]
 6$^{\ast\ast}$ & a$\Uparrow$ & 4.12256$\pm$0.00053 & $12.72_{-0.12}^{+0.13}$ & $13.0_{-10.7}^{+6.30}$ & 0.164$\pm$0.087 & 0.268$\pm$0.125 & 12.3 & 35.0 \\[2pt]
  & b$\Uparrow$ & 4.12306$\pm$0.00325 & $12.42_{-0.19}^{+0.13}$ & $26.1_{-13.9}^{+9.66}$ & 0.103$\pm$0.089 & & & \\[2pt]
 & & & & & & & \\[1pt]
 7 & a & 4.21618$\pm$4$\times10^{-5}$ & $12.81_{-0.10}^{+0.07}$ & $15.5_{-4.50}^{+2.73}$ & 0.189$\pm$0.009 & 0.222$\pm$0.011 & 55.2 & 55.2 \\[2pt]
  & b & 4.21681$\pm$0.00011 & $12.19_{-0.26}^{+0.27}$ & $2.03_{-1.02}^{+0.21}$ & 0.033$\pm$0.006 & & & \\[2pt]
 & & & & & & & \\[1pt]
 8 & a$\Uparrow$ & 4.56845$\pm$9$\times10^{-5}$ & $12.82_{-0.10}^{+0.10}$ & $27.6_{-9.63}^{+7.77}$ & 0.254$\pm$0.042 & 0.254$\pm$0.042 & 76.5 & 87.7 \\[2pt]
 & & & & & & & \\[1pt]
 9 & a$\Uparrow$ & 4.66297$\pm$6$\times10^{-5}$ & $13.20_{-0.21}^{+0.33}$ & $3.65_{-0.77}^{+0.20}$ & 0.123$\pm$0.069 & 0.309$\pm$0.074 & 61.1 & 64.2 \\[2pt]
  & b$\Uparrow$ & 4.66378$\pm$0.00023 & $12.59_{-0.14}^{+0.14}$ & $33.1_{ -2.70}^{+2.71}$ & 0.129$\pm$0.017 & & & \\[2pt]
  & c$\Uparrow$ & 4.66522$\pm$0.00010 & $14.15_{-0.89}^{+0.82}$ & $1.00 $ & 0.055$\pm$0.020 & & & \\[2pt]
 & & & & & & & \\[1pt]
 10$^{\ast\ast}$ & a & 5.37478$\pm$0.00011 & $13.53_{-0.73}^{+0.65}$ & $1.18_{-0.18}^{+0.13}$ & 0.048$\pm$0.009 & 0.082$\pm$0.012 & 25.9 & 42.1 \\[2pt]
  & b & 5.37518$\pm$0.00010 & $13.12_{-0.79}^{+0.81}$ & $1.00_{ }^{+0.18}$ & 0.034$\pm$0.007 & & & \\[2pt]
 & & & & & & & \\[1pt]
 11$^{\ast\ast}$ & a & 5.44109$\pm$0.00010 & $12.69_{-0.13}^{+0.12}$ & $14.8_{-9.47}^{+5.30}$ & 0.156$\pm$0.040 & 0.156$\pm$0.040 & 0.0 & 36.2 \\[2pt]
 & & & & & & & \\[1pt]
 12$^{\ast\ast}$ & a$\Uparrow$ & 5.92512$\pm$0.00029 & $14.20_{-0.52}^{+0.52}$ & $2.88_{-0.29}^{+0.17}$ & 0.122$\pm$0.088 & 0.287$\pm$0.130 & 0.00 & 6.28 \\[2pt]
  & b$\Uparrow$ & 5.92580$\pm$0.00032 & $13.10_{-0.08}^{+0.07}$ & $7.43_{-1.77}^{+1.13}$ & 0.165$\pm$0.096 & & & \\[2pt]
 & & & & & & & \\[1pt]
 \hline
 \hline
\end{tabular}
}
\begin{flushleft}

$\Downarrow$ $\hspace{0.8mm}$ denotes a component with a blended feature

$\Uparrow$ $\hspace{0.8mm}$ denotes a component polluted by a sky line or poor subtraction residual

 $^{\ast}$ $\hspace{1.2mm}$ denotes system which does not meet our 5$\sigma$ selection criteria

 $^{\ast}$ $^{\ast}$ denotes system which does not meet our 5$\sigma$ and 3$\sigma$ selection criteria

\end{flushleft}

\end{table*}

\section{False positive plots}

\begin{figure*}
	\begin{center}
		\includegraphics[ width=8.5cm]{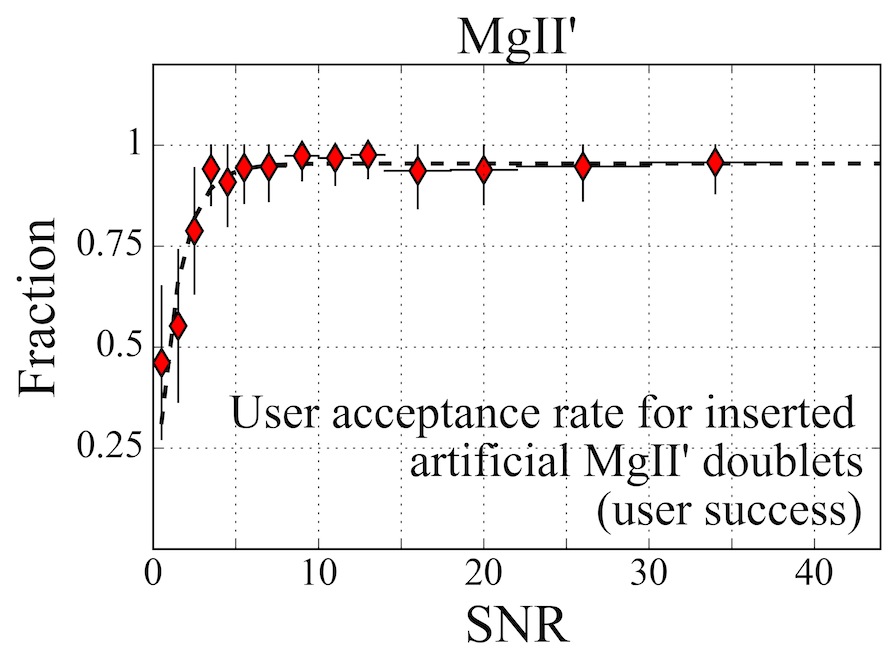} \includegraphics[ width=8.5cm]{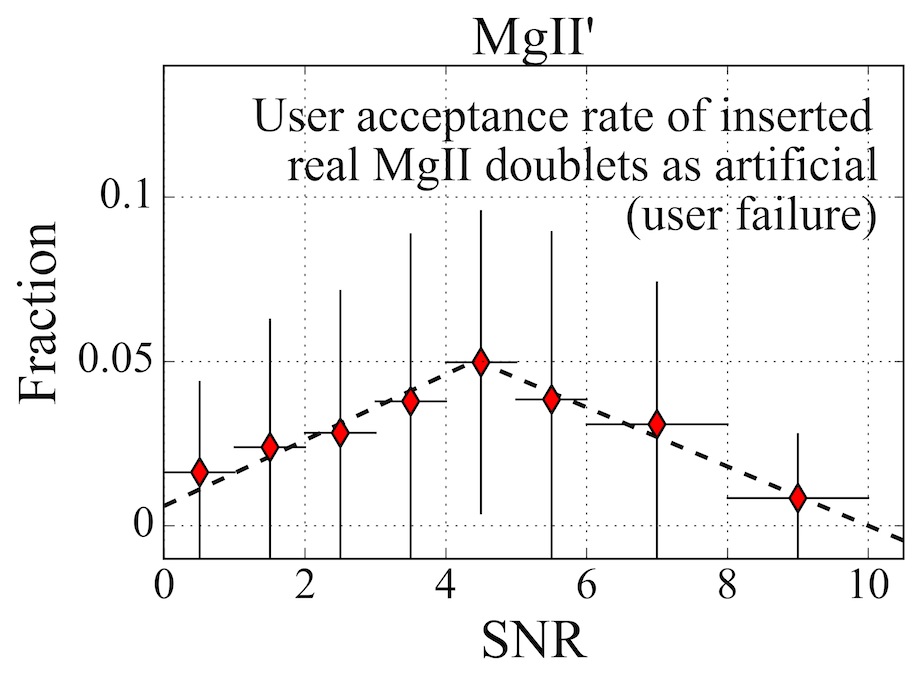}

	 	\caption{Plotted with red diamonds are the binned user success (top panel) and user failure (bottom panel). Plotted with dashed lines are the respective best fits (eqs. \ref{eq:pmg2} and \ref{eq:pfmg2}). The horizontal bounds denote the S/N bin considered and the vertical error bars correspond to the associated 95$\%$ Wilson confidence interval.}
 		\label{fig:falseacceptancestats}
	\end{center}
\end{figure*}

Here we present the user $success$ and $failure$ rate as a function of S/N as defined in equations \ref{eq:pmg2} and \ref{eq:pfmg2}. User $success$ represents the ability of the lead author to identify inserted \ion{Mg'}{II} absorbers as \ion{Mg'}{II} absorbers. User $failure$ represents the fraction of mis-identified \ion{Mg}{II} absorbers as \ion{Mg'}{II} when \mgii doublets are inserted. User $success$ and $failure$ can be seen in Figure \ref{fig:falseacceptancestats}.

\section{Completeness maps}

\begin{figure*}
	\begin{center}
	\includegraphics[ width=8.5cm]{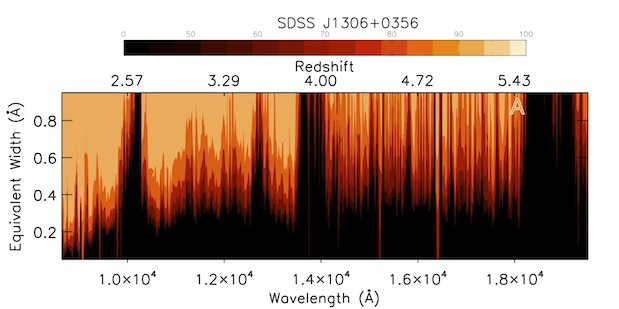}\includegraphics[ width=8.5cm]{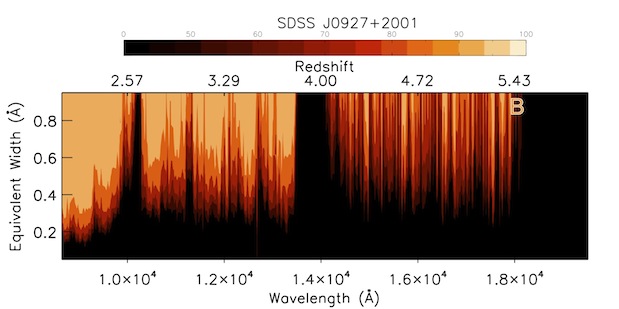}
	\includegraphics[ width=8.5cm]{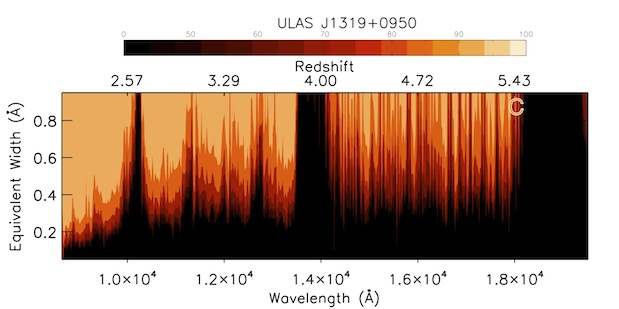}\includegraphics[ width=8.5cm]{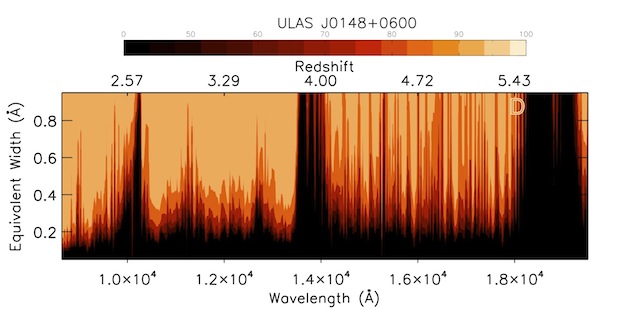}
 \caption{Completeness test results at the 3$\sigma$ level for each sightline. The $x$ and $y$ axis of each panel represent the wavelength (\AA) and $W_{2796}$ (\AA), respectively, of inserted systems. The top $x$ axis indicates the corresponding redshift of an inserted system. Their recovery rate, C($dW_{2796}$, $dz$), is denoted by the colour bar plotted in each panel. All recovery rates below 50$\%$ are shaded in black. Recovery rates for SDSS J1306+0356 are in panel $A$, SDSS J0927+2001 are in panel $B$, ULAS J1319+0950 are in panel $C$ and ULAS J0148+0600 are in panel $D$. The wavelength resolution of the recovery function, C($dW_{2796}$, $dz$), allows us to identify clean portions of the spectra in the near infrared as can be seen in panel $A$ at around $\sim$16660\AA$ $ where the recovery rate is above 50$\%$ for even the weakest of inserted systems. }
 \label{fig:completenesstest3sigma}
	\end{center}
\end{figure*}

Here we present the completeness (as defined in equation \ref{eq:recovery}) when searching for $real$ \mgii absorbers when considering a 3$\sigma$ selection criteria. It can be seen in Figure \ref{fig:completenesstest3sigma}.

\begin{figure*}
	\begin{center}
	\includegraphics[ width=8.5cm]{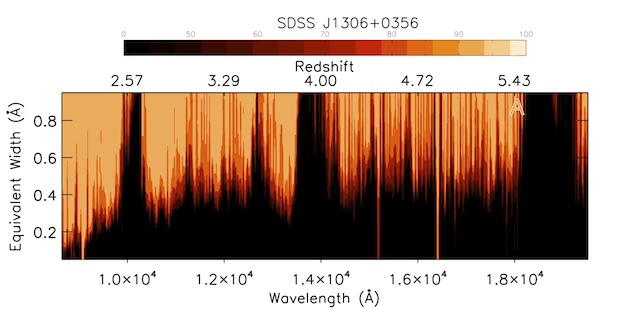}\includegraphics[ width=8.5cm]{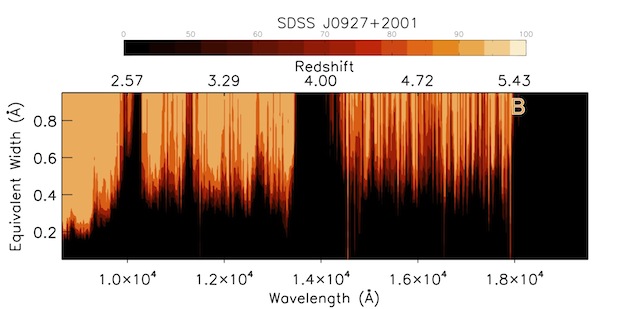}
	\includegraphics[ width=8.5cm]{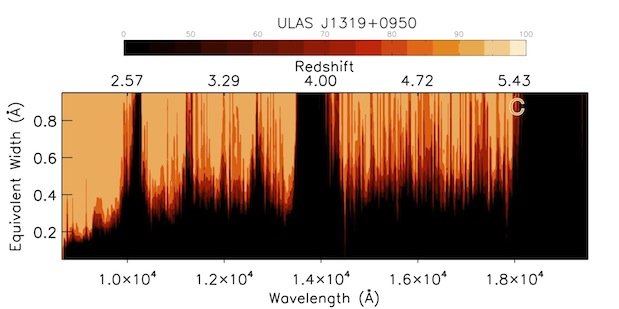}\includegraphics[ width=8.5cm]{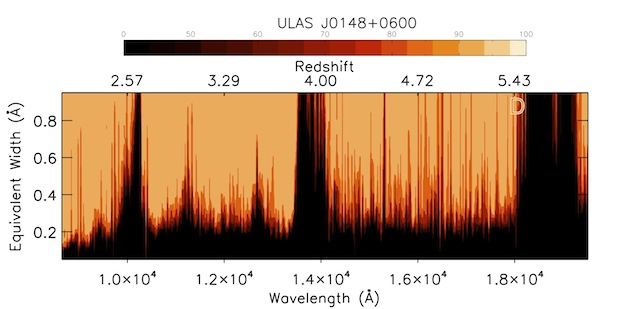}

 \caption{Completeness test results at the 5$\sigma$ level for each sightline as the result of a search for artificial \ion{Mg'}{II} doublets $\lambda\lambda$2796 2810. The $x$ and $y$ axis of each panel represent the wavelength (\AA) and $W'_{2796}$ (\AA), respectively, of inserted systems. The top $x$ axis indicates the corresponding redshift of an inserted system. Their recovery rate, C'($dW'_{2796}$, $dz$), is denoted by the colour bar plotted in each panel. All recovery rates below 50$\%$ are shaded in black. Recovery rates for SDSS J1306+0356 are in panel $A$, SDSS J0927+2001 are in panel $B$, ULAS J1319+0950 are in panel $C$ and ULAS J0148+0600 are in panel $D$. The wavelength resolution of the recovery function, C'($dW'_{2796}$, $dz$), allows us to identify clean portions of the spectra in the near infrared as can be seen in panel $A$ at around $\sim$16660\AA$ $ where the recovery rate is above 50$\%$ for even the weakest of inserted systems. }

 \label{fig:artificialcompletenesstest}
	\end{center}
\end{figure*}

\begin{figure*}
	\begin{center}
	\includegraphics[ width=8.5cm]{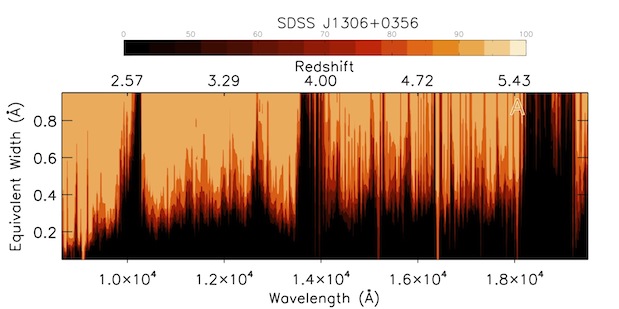}\includegraphics[ width=8.5cm]{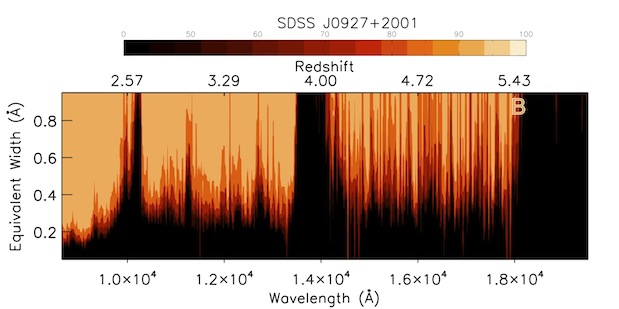}
	\includegraphics[ width=8.5cm]{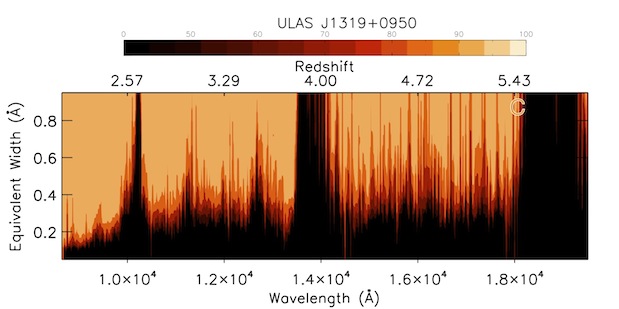}\includegraphics[ width=8.5cm]{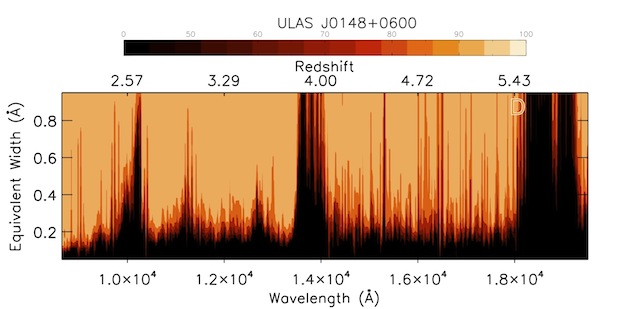}

 \caption{Completeness test results at the 3$\sigma$ level for each sightline as the result of a search for artificial \ion{Mg'}{II} doublets $\lambda\lambda$2796 2810. The $x$ and $y$ axis of each panel represent the wavelength (\AA) and $W'_{2796}$ (\AA), respectively, of inserted systems. The top $x$ axis indicates the corresponding redshift of an inserted system. Their recovery rate, C'($dW'_{2796}$, $dz$), is denoted by the colour bar plotted in each panel. All recovery rates below 50$\%$ are shaded in black. Recovery rates for SDSS J1306+0356 are in panel $A$, SDSS J0927+2001 are in panel $B$, ULAS J1319+0950 are in panel $C$ and ULAS J0148+0600 are in panel $D$. The wavelength resolution of the recovery function, C'($dW'_{2796}$, $dz$), allows us to identify clean portions of the spectra in the near infrared as can be seen in panel $A$ at around $\sim$16660\AA$ $ where the recovery rate is above 50$\%$ for even the weakest of inserted systems. }

 \label{fig:artificialcompletenesstest3s}
	\end{center}
\end{figure*}

The completeness (as defined in equation \ref{eq:recovery}) when searching for $artificial$ \ion{Mg'}{II} absorbers when considering a 3$\sigma$ and 5$\sigma$ selection criteria and can be seen in Figures \ref{fig:artificialcompletenesstest} and \ref{fig:artificialcompletenesstest3s} respectively.

\bsp	
\label{lastpage}
\end{document}